\documentclass[aps,prb,twocolumn,superscriptaddress,showpacs,longbibliography,notitlepage]{revtex4-1}
\usepackage[colorlinks=true,citecolor=blue,linkcolor=blue,breaklinks=true]{hyperref}
\usepackage{amssymb}
\usepackage{graphicx}
\usepackage{amsmath}
\usepackage{mathrsfs}
\usepackage{mathtools}
\usepackage[export]{adjustbox}
\usepackage{epsfig}
\usepackage{times}
\usepackage{color}
\usepackage{setspace}
\usepackage{calc}
\usepackage{natbib}
\usepackage{bm}
\DeclareMathAlphabet{\mathpzc}{OT1}{pzc}{m}{it}
\begin{document}

\newcommand {\beq} {\begin{equation}}
\newcommand {\eeq} {\end{equation}}
\newcommand {\bqa} {\begin{eqnarray}}
\newcommand {\eqa} {\end{eqnarray}}
\newcommand {\ba} {\ensuremath{b^\dagger}}
\newcommand {\Ma} {\ensuremath{M^\dagger}}
\newcommand {\psia} {\ensuremath{\psi^\dagger}}
\newcommand {\psita} {\ensuremath{\tilde{\psi}^\dagger}}
\newcommand{\lp} {\ensuremath{{\lambda '}}}
\newcommand{\A} {\ensuremath{{\bf A}}}
\newcommand{\Q} {\ensuremath{{\bf Q}}}
\newcommand{\kk} {\ensuremath{{\bf k}}}
\newcommand{\qq} {\ensuremath{{\bf q}}}
\newcommand{\kp} {\ensuremath{{\bf k'}}}
\newcommand{\rr} {\ensuremath{{\bf r}}}
\newcommand{\rp} {\ensuremath{{\bf r'}}}
\newcommand {\ep} {\ensuremath{\epsilon}}
\newcommand{\nbr} {\ensuremath{\langle ij \rangle}}
\newcommand {\no} {\nonumber}
\newcommand{\up} {\ensuremath{\uparrow}}
\newcommand{\dn} {\ensuremath{\downarrow}}
\newcommand{\rcol} {\textcolor{black}}
\newcommand{\bcol} {\textcolor{black}}
\newcommand{\bu} {\bold{u}}
\newcommand{\tr} {\text{tr}}

\begin{abstract}
Schwinger Keldysh field theory is a widely used paradigm to study
non-equilibrium dynamics of quantum many-body systems starting from a thermal state. We extend this formalism to describe
non-equilibrium dynamics of quantum systems starting from arbitrary
initial many-body density matrices. We show how this can be done for
both Bosons and Fermions, and for both closed and open quantum
systems, using additional sources coupled to bilinears of the fields at the initial time, calculating Green's functions in a theory with these sources, and then taking appropriate set of derivatives of these Green's functions w.r.t. initial sources to obtain physical observables. The set of derivatives depend on the initial density matrix. The physical correlators in a dynamics with arbitrary initial conditions do not satisfy Wick's theorem, even for non-interacting systems.
However our formalism constructs intermediate ``n-particle Green's functions'' which obey Wick's theorem and provide a prescription to obtain physical correlation functions from them. This allows us to obtain analytic answers for all physical many body correlation functions of a non-interacting system even when it is initialized to an arbitrary density matrix. We use these exact expressions to obtain an estimate of the violation of Wick's theorem, and relate it to presence of connected multi-particle initial correlations in the system. We illustrate this new formalism by calculating density and current profiles in many body Fermionic and Bosonic open quantum systems initialized to non-trivial density matrices. We have also shown how this formalism can be extended to interacting many body systems.

 \end{abstract}
\title{Non-Equilibrium Field Theory for Dynamics Starting from Arbitrary Athermal Initial Conditions}
\author{Ahana Chakraborty}\email{ahana@theory.tifr.res.in}
 \affiliation{Department of Theoretical Physics, Tata Institute of Fundamental
 Research, Mumbai 400005, India.}
\author{ Pranay Gorantla} 
\affiliation{Department of Theoretical Physics, Tata Institute of Fundamental
 Research, Mumbai 400005, India.}
\affiliation{Department of Physics, Princeton University, Washington Road,
Princeton, NJ 08544, USA}
\author{Rajdeep Sensarma}
 \affiliation{Department of Theoretical Physics, Tata Institute of Fundamental
 Research, Mumbai 400005, India.}

\date{\today}

\maketitle

The most general problem in non-equilibrium dynamics of quantum many
body systems can be stated in the following way: given a many body
Hamiltonian $H$, and an initial many body density matrix $\hat{\rho}_0$ at
$t=0$, one needs to find the evolution of the density matrix
$\hat{\rho}(t)$. This can then be used to calculate equal and unequal
time correlation functions in the system. The information of the full
many body density matrix can also be used to construct the reduced
density matrix of a  subsystem by tracing out remaining degrees of
freedom. This leads to calculation of non-local
information theoretic measures like entanglement entropy of the
subsystem~\cite{Entanglement_Review} with the rest of the degrees of freedom. In
case of an open quantum system, the evolution of $\hat{\rho}(t)$  is governed
by quantum master equations for Markovian dynamics~\cite{OQSBook,QME} and more
complicated equations with non-local memory kernels for non-Markovian
dynamics~\cite{Nakajima,Zwanzig,Ines_review,nori,Chakraborty1}. While a lot of progress has been made within this direct approach of solving the equation of motion of $\hat{\rho}(t)$, the method runs into the difficulty of dealing
with a Hilbert space growing exponentially with size of the
system. Several techniques~\cite{SCHOLLWOCK201196,Eisert,Tensor_Vidal} have been proposed in recent years to
reduce the size of the Hilbert space to be considered in the dynamics, 
with varying amount of success beyond one dimensional systems~\cite{2D-DMRG1,2D-DMRG2,Tensor_Vidal}.

Field theoretic techniques have been used extensively to obtain information about
quantum many-body systems, both in their ground state as well as in
thermal equilibrium at a finite temperature \cite{matsubara}. 
This approach can be extended to non-equilibrium situations by considering the time evolution of the density
matrix. The resulting Schwinger Keldysh (SK) field theory~\cite{Keldysh,*kamenev,*rammer_2007,kamenevbook}, which involves two sets of fields for each
space-time point, provides a path integral based approach to the
non-equilibrium dynamics of quantum many body systems. However, the
current formulation of SK field theory has a major
drawback: it can only efficiently deal with initial density matrices which are thermal 
(this includes ground states). In this case, the real time path integral is
extended into the Kadanoff-Baym contour~\cite{baymbook,*SciPost_Aron} along the imaginary time
axis. The SK field theory is also widely used in
describing steady states of quantum systems where the memory of the initial condition is assumed to be erased~\cite{keldysh_meir_time,Chakraborty1}. But several interesting questions in non-equilibrium dynamics
of many body systems, where dependence on initial conditions need to
be tracked explicitly, cannot even be posed within this formalism. This severely restricts the applicability of SK field
theory.
In this paper, we formulate a comprehensive action based field theoretic approach which can 
explicitly keep track of arbitrary initial conditions and their effect on the quantum
dynamics of Bosons and Fermions. This extends the domain of
applicability of SK field theory to a large class of problems
hitherto inaccessible to the field theoretic approaches.

Before we describe the new formalism, we would like to point out some important questions/problems in
non-equilibrium many body dynamics, where it is important to keep track
of the initial conditions explicitly. 
(i) Quantum computation works on the principle that different initial
conditions (inputs) will generically lead to different measurements
(outputs) in the system~\cite{nielsen_chuang_2010,*Qcomp_book}. It is obvious that ignoring initial conditions in problems related to implementation of quantum gates would lead to trivial results.
(ii) Discussion of approach to thermal equilibrium and
development of quantum chaos~\cite{chaos} in a many-body system requires studying dynamics
starting from an initial state far from equilibrium. In fact in a chaotic
system, one would expect the dynamics to be extremely sensitive to
initial conditions. (iii) Integrable systems~\cite{Rigol_integrable,Langen207} and many body localized systems~\cite{Nandkishore_MBL} retain memory of the initial state for long times and hence they 
do not thermalize. To capture this aspect, it is important to construct a description which explicitly takes the initial condition into account.
We would like to note that the only experimental
evidence ~\cite{Schreiber842,*Choi1547} for MBL is to measure the
residual memory of initial state in the long time dynamics.  (iv)
There are quantum systems whose long time behaviour changes
qualitatively depending on the initial condition, e.g. systems with
mobility edges \cite{BASKO20061126,Nandkishore_MBL} may or may not
thermalize depending on the state in which they are prepared. Cold atom
systems with strong non-linearity~\cite{Bodhaditya_bistable} have also
been found to reach qualitatively different steady states depending on
initial preparation. (v) An interesting class of problems related to
thermalization involves solving for the dynamics of open quantum
systems (OQS)\cite{OQSBook}, where a quantum system can exchange
energy/particles with a large reservoir/bath. In the open quantum
system set-up, it is interesting to study how the memory of the
initial state of the system is being retained in its subsequent
dynamics while the external dissipative effect from the baths tries to
erase it, as it approaches a thermal equilibrium/non-equilibrium
steady state. Interplay of multiple time scales, governing the
inherent dynamics of the system and the relaxation coming from the
external bath, make this problem particularly interesting. (vi) Recent
advances in ultra-fast spectroscopy has led to the study of transient
quantum transport\cite{transient1,*transient2,* transient3,*
  transient4,*transient5,*transient6,*transient7,*transient8} in
condensed matter systems, where the system is initialized to a highly
excited state and the change in its transport properties are
measured. The full counting statistics of charge and
spin in these systems~\cite{Esposito, Wang, Wang2} measure highly non-linear
response in these time-evolving systems. A proper investigation of
these properties also require a formalism to treat athermal initial conditions. (vii) Problems related to aging in quantum glasses also require a description of dynamics starting from non-equilibrium initial conditions. \cite{aging1,*aging2,*aging}
This is not an exhaustive list, but provides some context as to why such a formalism is important to
develop.


There have been two major streams of attempts in the past to include
arbitrary initial conditions within a field theoretic approach. The
first one starts from the Martin-Schwinger hierarchical equation \cite{kadanoff} for the one-particle Green's
function and then tries to include initial correlations in different
ways. In this case one assumes a Dyson
equation with a self energy structure, and then modifies the self
energy to satisfy initial boundary conditions \cite{Bonitz1,*Bonitz2}. There are two main
problems with this approach: (i) It assumes that a Dyson
equation for one-particle Green's function can be written in terms of
an irreducible self energy, which is itself a function of one particle Green's
functions, or with additive corrections representing initial correlations.
Since Wick's theorem is not valid in
a theory with arbitrary initial condition (as we will show from exact
expressions in our formalism), it is not clear under what condition
this can be done. (ii) Singling out the one particle correlation function does not
automatically provide a way to write down equations for higher order
correlation functions even in a non-interacting theory \cite{Zhang1,*Zhang2,Leeuwen} which will be evident from our formalism. The second
approach, due to Konstantinov and Perel \cite{KONSTANTINOV}, essentially states that since the
density matrix is a Hermitian operator with non-negative eigenvalues,
it can always be written as an exponential of {\it some}
many body Hamiltonian (which can be quite different from the
Hamiltonian which generates dynamics of the system) \cite{Wagner,*Stefanucci_GFF,*SciPost_Aron}. One can then use the old
Kadanoff-Baym contour, with the dynamics along the imaginary time contour governed by
this new ``Hamiltonian''. However, (i) for a given generic density matrix,
finding the ``imaginary time Hamiltonian''
requires a diagonalization in an exponentially large Hilbert space and
(ii) there is no guarantee that the resulting ``Hamiltonian'' will be
local or will only have few-body operators. Then the field theory
along the imaginary time contour becomes very hard to implement. Even for systems evolving in real time with a non-interacting Hamiltonian, the arbitrary non- thermal initial state maps the problem into a non-Gaussian field theory along the imaginary time axis of the Kadanoff-Baym contour.

In this paper we will develop a unified action based description of dynamics of
many Bosons/Fermions 
starting from arbitrary initial conditions. For this, we need to consider a SK
field theory in presence of a source, $\hat{u}$ which couples to bilinears of the initial
fields. We note that in contrast to the other approaches \cite{Markus,Leeuwen}, the additional term in the action, taking care of the initial correlations, is still quadratic and do not lead to high order vertices in this theory.
 This source is turned on only at the initial time, i.e. it acts
like an impulse. Different n-particle Green's
functions, $\hat{G}^{(n)}(u)$ are then calculated in this theory in presence of the source $\hat{u}$. The physical correlators, corresponding to dynamics
starting from a
particular $\hat{\rho}_0$, can then be obtained by taking a set of derivatives of the Green's functions with respect to $\hat{u}$ and then
setting $\hat{u}$ to zero. The particular set of derivatives to be taken depends on $\hat{\rho}_0$. We note that, in this formulation the calculation of the Green's functions are universal, i.e. they do not depend on particular  $\hat{\rho}_0$. The information of specific $\hat{\rho}_0$ is required solely to determine the set of derivatives (w.r.t $\hat{u}$) to be taken to obtain the physical correlators.

In
this formalism, we are able to construct a set of intermediate
quantities, $\hat{G}^{(n)}(u)$, which have the structure of ``n-particle Green's
functions'' and are derived from the action (with the source $\hat{u}$) in the
usual field theoretic way; i.e. Wick's theorem holds for these quantities. One can, for
example, construct a diagrammatic perturbation theory for $\hat{G}^{(n)}(u)$ using standard rules of SK field theory. The usual paradigms of obtaining interacting Green's functions in terms of self-energies and higher order vertex functions are valid for these quantities.
These are however {\it
  not} the physical n-particle correlators; we provide a
prescription to compute the physical correlators for different initial
density matrices from these intermediate quantities. The key
theoretical advance in this formalism is to prescribe a two step process: (i) construction of intermediate quantities where we
can apply the well studied structures and standard approximations of SK
quantum field theory, and (ii) a prescription to obtain physical
correlation functions from them. We would like to emphasize that the above statements are exact even for interacting open quantum systems and do not involve any ad-hoc approximation regarding the initial correlations. 

There are some other key advantages of
having an action based formalism: (i) {\it all} correlation functions can be
derived from a unified description by adding linear source fields $J$ to the action and then
taking appropriate derivatives w.r.t $J$. Hence they are all on the same theoretical
footing, as opposed to a focus on one particle correlators (ii) The general formalism keeps track of {\it
  all} ``n-particle initial correlations''. For non-interacting theories it leads to
exact answers for physical correlation functions, even for open quantum
systems. This is in itself non-trivial since there is no Wick's
theorem for physical correlators. This is an advantage from the
Konstantinov Perel (KP) formalism, where it is hard to get exact answers even
for non-interacting theories starting from arbitrary initial condition. (iii) For interacting theories,
it leads to exact expressions on which approximations have to be made
for practical calculations. In this case, this formalism provides the
most transparent way to understand and make useful approximations.
(iv) The action principle provides a way to integrate out degrees of
freedom and construct effective theories. Effective theories of
dynamics starting from arbitrary initial conditions is a completely
unexplored area where there may be new surprises. This may
lead to a renormalization group analysis \cite{Diehl,*Sangita} of non-equilibrium dynamics starting from non-trivial initial conditions.

In this paper, we will set up the general formalism, but focus mainly
on non-interacting systems (including open quantum systems), where we
can make exact statements. We will construct the intermediate quantities
for which a diagrammatic perturbation theory can be worked out in case
of an interacting system, and sketch how that can be done, but we will leave the question of the
different approximations and their validity in interacting systems for
a future work. We will now provide a guide map for the reader to
explore this paper. 
In section~\ref{sec:stdKFT}, we have briefly outlined the structure of the standard SK field theory formalism and set up the notation to be used in this paper. In the next section \ref{sec:genericKFT}, we have explained the main idea behind the extension of the SK formalism to include arbitrary initial condition and introduce the new ingredients of the field theory. In section \ref{sec:genericBoson}, we have explicitly worked them out for a system of Bosons starting from generic density matrix in Fock space. We first consider the pedagogical case of a single Bosonic mode starting from a density matrix diagonal in the number basis and derived the corresponding formalism. We then extend this to a multi-mode system starting with density matrix diagonal in the Fock basis. Finally, we consider the extension to arbitrary initial density matrices with off-diagonal elements in the Fock basis. In section \ref{sec:genericFermion}, we consider a Fermionic theory. A large part of the derivations of the Fermionic theory follow along lines similar to that of Bosonic theory. In this section, we mainly focus on the modifications required to convert the Bosonic theory to the Fermionic theory. In section \ref{sec:twoparticle}, we focus on calculating multi-particle physical correlators for a system of non-interacting Bosons and Fermions starting from arbitrary initial condition. We show how the Wick's theorem is violated by explicitly computing the corrections to the Wick reconstruction of the two particle physical correlators in terms of one particle physical correlators. In section \ref{sec:OQS} we extended the formalism to the case of a many body open quantum system. We also work out some examples of the above formalism to compute the evolution of densities and currents in many body open quantum systems. Finally, in section \ref{sec:interaction} we sketch the general structure of the interacting theory without going into the details of the approximation strategies.


\section{Brief Review of Standard Schwinger-Keldysh Field Theory}\label{sec:stdKFT}

\begin{figure}[t]
   \centering
  \includegraphics[width=0.48\textwidth]{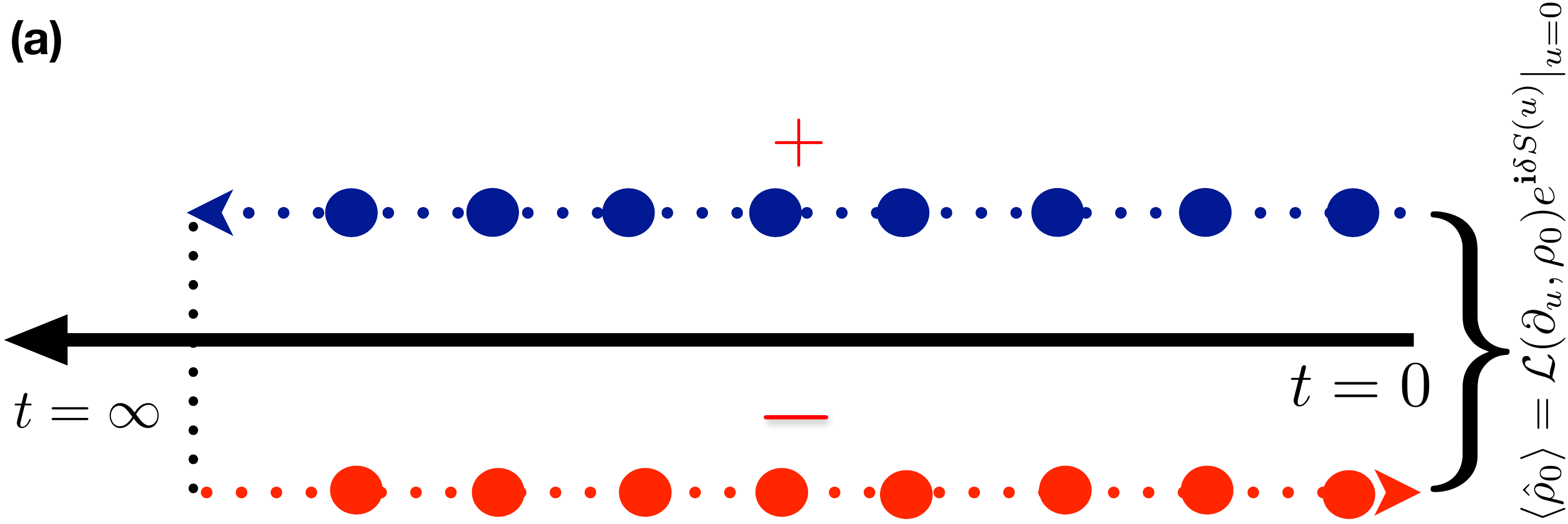}\\~ \\
  \includegraphics[width=0.48\textwidth]{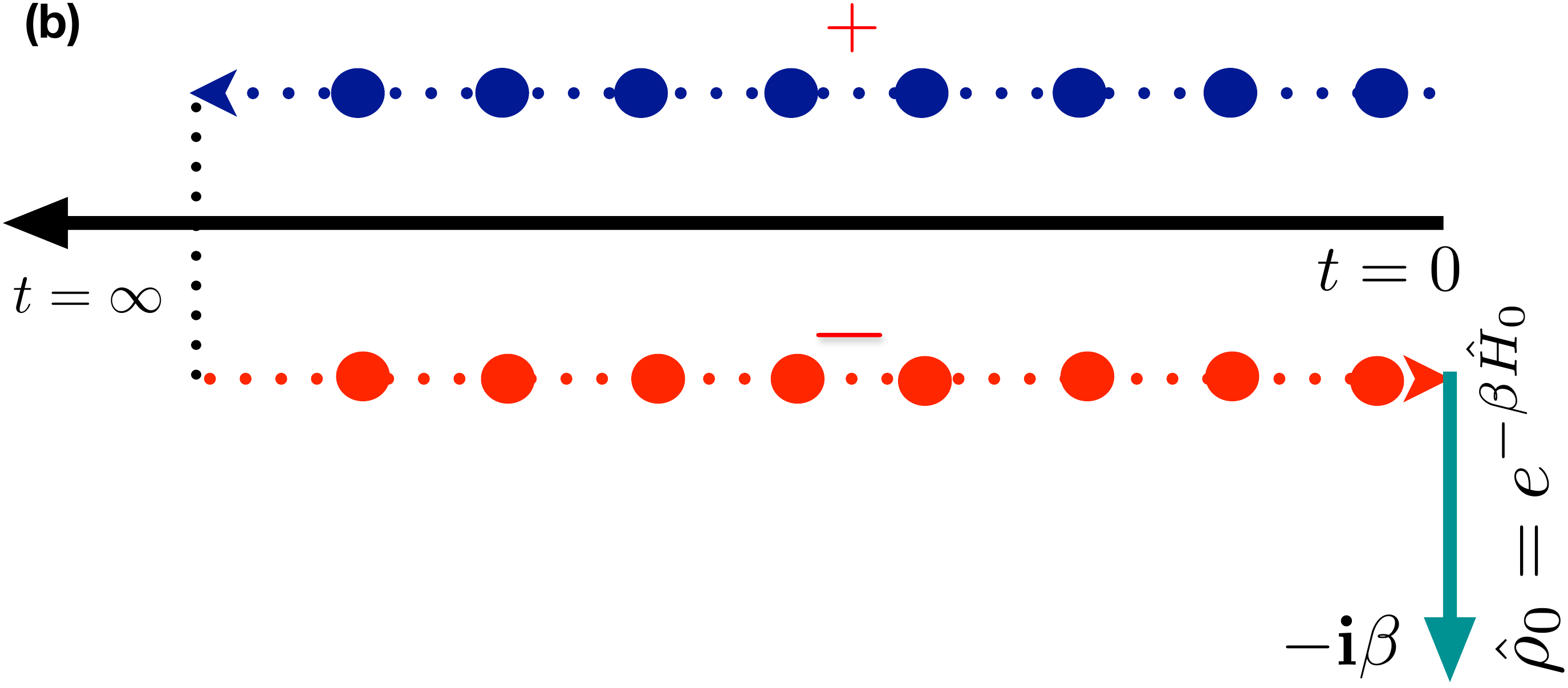}
  \caption{(a) Keldysh contours showing forward and backward propagation in time. In our formalism, the matrix element of the initial density matrix, $\hat{\rho}_0$ is written as the derivative of $exp[\mathbf{i}\delta S(u)]$, where $\delta S(u)$ is an added quadratic term in the action which couples to the initial bilinear source $\hat{u}$. The set of derivatives, ${\cal L}(\partial_u,\rho_0)$, to be taken, is completely dictated by initial $\hat{\rho}_0$.
(b) The Kadanoff Baym contour with an extension along the imaginary time axis, from $t=0$ to $t=-\mathbf{i}\beta$, in the Konstantinov Perel formalism. In the KP formalism, $\hat{\rho}_0 = exp(-\beta \hat{H}_0)$, with some many body operator $\hat{H}_0$, is written as a path integral along the imaginary axis.} 
   \label{fig:contour}
   \end{figure}

We start with a brief review of the standard SK field
theory \cite{kamenevbook}, both to set up notations and to provide context for our
extension of the formalism. The time evolution of a many body density
matrix is given by $\hat{\rho}(t)=U(t,0)\hat{\rho}_0U^\dagger(t,0)$,
where, for Hamiltonian dynamics of a closed quantum system, 
 the time evolution operator is $U(t,0)= {\cal T} [e^{-i\int_0^t dt'
  H(t')}]$. For an open quantum
system, $U$ is not an unitary operator in general. In SK field theory, each of $U$ and $U^\dagger$ is expanded
in a path/functional integral, resulting in the
Keldysh partition function for Bosons
\bqa
\label{eq:Z_1}
Z &=& Tr [U(\infty,0) \hat{\rho}_0
U^\dagger(\infty,0)]\\
\nonumber &=&\int D[\phi_+,\phi_-] e^{\mathbf{i}(S[\phi_+]-S[\phi_-])}\langle \phi_+(0)|\hat{\rho}_0|\phi_-(0)\rangle
\eqa
where the complex Bosonic fields $\phi_+$ and $\phi_-$ correspond to
the expansion of $U$ and $U^\dagger$ respectively, and $|\phi\rangle $ is a
many body Bosonic coherent state. Note that the time
  evolution operators, which result in the $e^{\pm \mathbf{i} S}$
  terms, shift the trace over final states to a trace over initial states.
The detailed form of $S$ is
not relevant for the present discussion. For Fermionic systems, a
similar expansion with Grassmann coherent states leads to 
\bqa
\label{eq:Z_2}
Z &=&\int D[\psi_+,\psi_-] e^{\mathbf{i} (S[\psi_+]-S[\psi_-])}\rcol{\langle  \psi_+(0)|\hat{\rho}_0|-\psi_-(0)\rangle}
\eqa
where $\psi_\pm$ are the Grassmann fields. Note the additional minus
sign in the matrix element, which comes from writing a trace in the
Fermionic Fock space as integrals over Grassmann fields~\cite{Negele_Orland}. This
will be important in the detailed discussion in Section \ref{sec:genericFermion}. Thus the
SK field theory is written in terms of doubled fields in a real time
formalism, with a path/functional integral over a contour shown in
Fig.~\ref{fig:contour}(a). It is clear that if the matrix element of
$\hat{\rho}_0$ can be written as an exponential of a low
order polynomial of the fields, one can obtain a standard action based
formalism for the dynamics. This can be achieved if $\hat{\rho}_0$ is
a thermal density matrix corresponding to a Hamiltonian $\hat{H}_0$ containing only a few body operators, i.e. $\hat{\rho}_0 = exp[-\beta \hat{H}_0]$ ($\hat{H}_0$
does not need to be generator of the real time dynamics; c.f. quantum
quench problems). In this case the matrix element can be written as an
Euclidean path integral, and the full $Z$ is a path integral over the
Kadanoff Baym contour shown in Fig.~\ref{fig:contour}(b), which
extends into the imaginary axis from $t=0$ to $t=-\mathbf{i} \beta$. We note that for a large class of $\hat{\rho}_0$, the above prescription does not work. We have already articulated the problem with the KP formalism, which tries to cast every $\hat{\rho}_0$ into the above mentioned formalism, even at the cost of having a $\hat{H}_0$ with arbitrary $n-$ particle interactions. Clearly a new formalism is required to treat the vast set of initial conditions, which do not lend themselves to a simple $\hat{H}_0$. 

Correlation functions are calculated in SK theory by coupling sources
$J_{\pm}$ linearly to the fields and taking appropriate derivatives
with respect to these sources. For one-particle Green's functions, the
doubled field approach leads to redundancies, i.e. the $4$ possible
Green's functions are not independent. To make this explicit, one works with symmetric and
anti-symmetric combination of the fields. For Bosons, these are called
``classical'' $\phi_{cl}=(\phi_++\phi_-)/\sqrt{2}$ and ``quantum'' fields,
$\phi_{q}=(\phi_+-\phi_-)/\sqrt{2}$. In this case, the quadratic
action has the form 
\beq
\label{eq:KS_Boson}
S= \int dt \int dt' [\phi^\ast_{cl}(t),\phi^\ast_q(t)]\left[\begin{array}{cc}
0 & G^{ -1}_A\\
G^{-1}_R & -\Sigma_K
\end{array}\right]\left[\begin{array}{c}
\phi_{cl}(t')\\
\phi_q(t')
\end{array}\right]
\eeq
We see that $S[\phi_q=0]=0$, a statement which holds true even when external baths and inter-particle interactions are present in the description. Here, $G_{R(A)}$ are the retarded (advanced)
Green's function, with $G_A=[G_R]^\dagger$, and $\Sigma_K$ is
anti-hermitian~\cite{kamenevbook}. This leads to the following structure in Green's
functions,
\beq
\nonumber \hat{G}(t,t')= \left[\begin{array}{cc}
G_K(t,t') & G_{R}(t,t')\\
G_{A}(t,t') & 0
\end{array}\right]
\eeq
where the Keldysh component $G_K$ is anti-hermitian. For Fermions, we
follow Larkin-Ovchinikov transformation,
$\psi_1=(\psi_++\psi_-)/\sqrt{2}$, $\psi^\ast_1=(\psi^\ast_+-\psi^\ast_-)/\sqrt{2}$,
$\psi_2=(\psi_+-\psi_-)/\sqrt{2}$,
and $\psi^\ast_2=(\psi^\ast_++\psi^\ast_-)/\sqrt{2}$ and get
\beq
\label{eq:KS_Fermion}
S= \int dt \int dt' [\psi^\ast_{1}(t),\psi^\ast_2(t)]\left[\begin{array}{cc}
G^{ -1}_R & -\Sigma_K  \\
0 &G^{-1}_A
\end{array}\right]\left[\begin{array}{c}
\psi_{1}(t')\\
\psi_2(t')
\end{array}\right]
\eeq
In this case, $S[\psi_1^\ast=0,\psi_2=0]=0$ and 
the Green's functions have a structure similar to that for
Bosons. Note that for non-interacting theories, $\hat{G}$ is obtained
simply by inverting the matrix in the microscopic action.

One can study the effects of interparticle interactions by adding terms to the Keldysh actions (eqns \ref{eq:KS_Boson}, \ref{eq:KS_Fermion}). For a pairwise interacting system, the added terms are quartic in the fields. For a generic interaction, the problem cannot be solved exactly, but a diagrammatic perturbation theory can be constructed with the matrix of propagators and vertices having $cl/q$ indices along with other quantum numbers. With these changes, standard field theoretic calculations, including non-perturbative resummation of the series can be undertaken in the usual way. The SK field theory then operationally becomes equivalent to the standard field theories with this added $2-$component structure.

It is important to note that the retarded Green's function $G_R(i,t;j,t')$ has the
physical interpretation of the probability amplitude of finding a
particle in state $i$ at time $t$ if it is already known to be in state $j$ at some earlier
time $t'$ without creating additional excitations in the system. For non-interacting systems, this is independent of the
initial conditions. For interacting systems, this amplitude does
depend on initial conditions, since probability amplitude of
scattering at intermediate times depend on the distribution functions,
which depends on initial conditions. On the contrary, the Keldysh
Green's function explicitly keeps track of the initial conditions
(e.g. it depends explicitly on the temperature of the initial
distribution for thermal cases).
\section{Structure of the new formalism for Arbitrary Initial Conditions} \label{sec:genericKFT}

In this section, we will describe the general structure of the formalism which
allows us to treat dynamics of a system of Bosons/Fermions starting
from an arbitrary initial density matrix. We will focus on the key modifications of the SK field theory required to achieve this, leaving the detailed derivation for later sections. 
We intend to highlight the fact that several properties, which are taken for granted in
standard field theories, do not hold in this case and the ways to get
around these difficulties.

We will develop our formalism for a system with large but finite number of degrees of freedom. We will consider the question of taking the continuum limit in terms of the physical correlation functions at the very end. In the new formalism,
the matrix element of $\hat{\rho}_0$ between coherent states in Eq.~\ref{eq:Z_1} and
Eq.~\ref{eq:Z_2} is written as a polynomial of the bi-linears of the initial fields. This can be exponentiated by adding to the standard Keldysh
action, a term $\delta S$, where functions of a source field $\hat{u}$,
couple to {\it bilinears of the fields only at $t=0$}. The polynomial
can then be retrieved by taking appropriate derivatives of
$exp[\mathbf{i} ~\delta S(u)]$ w.r.t $\hat{u}$ and setting $\hat{u}$ to zero [Fig. \ref{fig:contour}
(a)]. The additional initial source  is similar to the conjugate field in the full
counting statistics~\cite{*kamenev,Wang}, derivatives with respect to
which lead to moments of the number distribution. We have replaced the
arbitrary polynomial resulting from the matrix element of
$\hat{\rho}_0$ by its generating function in our formalism. We note
that our source
field is quite different from the additional field of
Ref.~\onlinecite{Wang}, where an integral with respect to the field
acts as a projector onto number states.
The detailed
derivation of the source function which achieves this will be slightly
different for Bosons and Fermions and depends on the structure of the
initial density matrix. These details will be filled in the next
sections, and are cataloged in Table \ref{table:Structure}. For both Bosons and Fermions, the new term can be seen
as an addition to the term $\Sigma^K$ in eq.~\ref{eq:KS_Boson} and
eq.~\ref{eq:KS_Fermion} and maintains the anti-hermiticity property of
$\Sigma^K$. This term can be thought of as a generalized impulse
potential felt by the system at the initial time.

The functional integral over the fields can be done first to obtain the partition function $Z(u)$ and the derivative w.r.t $\hat{u}$ can then be  taken on this quantity to get the physical partition function corresponding to $\hat{\rho}_0$.
On the top of this, sources $J$ which couple linearly to
the fields at all times $t>0$ can be added to this action, and the
functional integrals over the fields performed to yield the partition
function, $Z(J,u)$. Note that $u$ and $J$ couple differently to the
fields: $\hat{u}$ couples to bilinears only at $t=0$, while $J$ couples
linearly at all times. This implies that no cross derivative of any quantity w.r.t $\hat{u}$ and $J$ survives when all the source fields are set to zero. Then the Green's function in presence of $\hat{u}$ can
be calculated by taking appropriate derivatives of $Z(J,u)$ with
respect to $J$, and setting $J=0$. For a quadratic theory with action
 \beq
S(u)= \int dt \int dt' \Psi^\dagger (t) \hat{G}^{-1}(t,t',u) \Psi(t'),
\label{eq:KS_u}
\eeq
where $\Psi^\dagger (t) = [\phi^\ast_{cl}(t),\phi^\ast_q(t)]$ for Bosons and $\Psi^\dagger (t) = [\psi^\ast_{1}(t),\psi^\ast_2(t)]$ for Fermions, 
the physical one particle correlation function can be obtained by taking proper derivative of ${\cal N}(u)
\hat{G}(u)$, where the normalization ${\cal N}(u)=[\textrm{Det}~\{-\mathbf{i}
\hat{G}^{-1}(u)\}]^{-\zeta}$ comes from performing the Gaussian
integral, with $\zeta=\pm 1$ for Bosons (Fermions), and $\hat{G}(u)$
is the inverse of the matrix in equation \ref{eq:KS_u}. While $\hat{G}(u)$ is not
the physical one-particle correlation function, we will see that it is an important
intermediate construction, which has very useful properties and will
be used many times in developing the theory. We will call this object
the ``Green's function in presence of initial source $\hat{u}$'', since it is indeed the
Green's function for the saddle point equations of the action with the
initial bilinear source term. We stress once again that this is not
the physical one particle correlator of the system.

The physical one-particle correlator is now given by,
\beq
\hat{{\cal G}}_{\rho_0} = {\cal L}(\partial_u, \rho_0) [ {\cal N}(u) \hat{G}(u)]\vert_{u=0}
\eeq
where $ {\cal L}$ is a differential operator which depends on
$\hat{\rho}_0$ and encodes initial correlations. 
The different forms of $\delta S$, ${\cal N}(u)$ and
${\cal L}(\partial_u,\rho_0)$ for a large class of initial conditions for both
Bosons and Fermions are tabulated in Table \ref{table:Structure}. The detailed
derivations are given in later sections of this paper.
We can generalize the above procedure to the computation of a physical ``n-particle correlator'', i.e
\beq
\label{eq:physicalG}
\hat{{\cal G}^{(n)}}_{\rho_0} = {\cal L}(\partial_u, \rho_0) [ {\cal N}(u) \hat{G}^{(n)}(u)]\bigg\vert_{u=0}
\eeq
Note that the differential operator ${\cal L}$ and the normalization ${\cal N}(u) $ is the same for all order
correlation functions. $\hat{G}(u)$
and $\hat{G}^{(n)}(u)$ are derived from the action $S(u)$ using
standard SK field theoretic ways, i.e. initial conditions do not play
a role in the derivation. Thus, $\hat{G}^{(n)}(u)$ can be easily
written as a sum of products of $\hat{G}(u)$ using Wick's theorem. This relationship is violated by the application of the differential operator ${\cal L}(\partial_u, \rho_0)$, i.e. ${\cal G}^{(n)}_{\rho_0}$ can not be written as a sum of products of ${\cal G}_{\rho_0}$ even for a non-interacting theory. The absence of a Wick's theorem for physical correlators in a non-interacting theory is at the heart of all the complications in constructing physical correlators in interacting theory in terms of non-interacting correlators.

Our formalism bypasses this difficulty by constructing $\hat{G}_{int}(u)$ and $\hat{G}_{int}^{(n)}(u)$ for an interacting theory. These quantities are obtained by standard SK field theoretic techniques from an action $S(u)+S_{int}$ where $S_{int}$ represents the inter-particle interactions. The diagrammatic expansion of $\hat{G}_{int}(u)$, in terms of $\hat{G}(u)$ and the interaction vertices, follow the Feynman rules of the standard SK theory. The series can be resumed in terms of a self-energy $\Sigma[\hat{G}(u)]$ (for a perturbative expansion of $\Sigma$) or $\Sigma[\hat{G}_{int}(u)]$ (for a skeleton diagram expansion). Similarly, one can can construct $\hat{G}^{(n)}_{int}(u)$ in terms of $\hat{G}(u)$ and higher order vertex functions. All the knowledge from the standard SK field theory and different perturbative or non-perturbative approximations can be used to compute $\hat{G}_{int}(u)$ and $\hat{G}^{(n)}_{int}(u)$. We finally need to compute physical correlators, ${\cal G}^{(n)}_{ \rho_0,int}$ from $\hat{G}^{(n)}_{int}(u)$, which are once again related by eqn. \ref{eq:physicalG}, with $\hat{G}^{(n)}(u)$ replaced by $\hat{G}^{(n)}_{int}(u)$ and $\hat{{\cal G}}^{(n)}_{\rho_0}$ replaced by ${\cal G}^{(n)}_{\rho_0,int}$.

Our formalism thus breaks up the calculation of ``n-particle correlators'' in an interacting theory starting from arbitrary initial conditions into $2$ parts: (i) a universal calculation of $\hat{G}_{int}(u)$ and $\hat{G}^{(n)}_{int}(u)$ which does not depend on particular choice of $\hat{\rho}_0$ and uses standard SK field theoretic techniques with a $\hat{u}$ dependent bare Green's functions and (ii) obtaining ${\cal G}^{(n)}_{\rho_0,int}$ by applying ${\cal L}(\partial_u, \rho_0)$ on ${\cal N}(u) \hat{G}^{(n)}_{int}(u)$. All the dependence on $\hat{\rho}_0$ enters in the theory through the last step. We note that there is no approximation made in the construction of the theory, i.e. all statements made above are exact.
In the next sections, we provide a derivation of the theory outlined above, pointing out the details of how $\delta S$, ${\cal N}$ and ${\cal L}$ depend on
the statistics of the particles and the initial density matrix $\hat{\rho}_0$.

\begin{table*}[t!]
  \centering
   \begin{tabular}{|c||c||c||c||c|}
          \hline
System&Initial Density Matrix &  $\delta S(u)$ &${\cal N}(u)$ & ${\cal L}(\partial_u,\rho_0)$  \\
          \hline
  & Single mode :Diagonal $\hat{\rho}_0$  & $\mathbf{i}\phi^\ast_q(0)\phi_q(0) \frac{1+u}{1-u}$ &$\frac{1}{1-u}$  &  $\sum \limits_n \frac{1}{n!}c_n \partial_u^n $  \\
&$=\sum_n
c_n|n\rangle \langle n|$ & & & \\
          
&  & & & \\
& Multi-mode :Diagonal $\hat{\rho}_0$  & $\mathbf{i}\sum_\alpha 
\phi^\ast_q(\alpha,0)\phi_q(\alpha,0)\frac{1+u_\alpha}{1-u_\alpha}$ &$\frac{1}{\prod_\alpha (1-u_\alpha)}$  &  $\sum \limits_{\{n\}}c_{\{n\}}
\prod_\gamma \frac{\partial_{u_\gamma}^{n_\gamma}}{n_\gamma !} $  \\
Boson&$=\sum_{\{n\}}c_{\{n\}}| \{n\}\rangle\langle
\{n\}|$ & & & \\

&  & & & \\
& Multi-mode :Generic $\hat{\rho}_0$  &$ \mathbf{i}\sum \limits_{\alpha \beta} 
\phi^\ast_q(\alpha,0)\phi_q(\beta,0)[2 \left(1-\hat{u}\right)^{-1}-1]_{\alpha \beta} $&$\mathrm{Det}(1-\hat{u})^{-1}$  & $\sum \limits_{nm}c_{nm} \prod \limits_\alpha \frac{1}{\sqrt{n_\alpha ! m_\alpha
!}} \prod \limits_j \partial_{\alpha_j\beta_j}$   \\
&$=\sum \limits_{nm} c_{nm}|\{ n\}\rangle\langle \{m\}|$ & & & \\
&  & & & \\
    \hline     
    
    & Single mode :Diagonal $\hat{\rho}_0$  & $\mathbf{i}\psi^\ast_1(0)\psi_2(0) \frac{1-u}{1+u}$ &$1+u$  &  $c_0+c_1 \frac{\partial}{\partial u} $  \\
&$=\sum_{n=0,1}
c_n|n\rangle \langle n|$ & & & \\     
&  & & & \\  
& Multi-mode :Diagonal $\hat{\rho}_0$  & $\mathbf{i}\sum_\alpha 
\psi^\ast_1(\alpha,0)\psi_2(\alpha,0)\frac{1-u_\alpha}{1+u_\alpha}$ &$\prod_\alpha (1+u_\alpha)$  &  $\sum \limits_{\{n\}}c_{\{n\}}
\prod \limits_{\gamma \in \mathcal{A}} \partial_{u_\gamma}$  \\
Fermion&$=\sum_{\{n\}}c_{\{n\}}| \{n\}\rangle\langle
\{n\}|$ & & & \\

&  & & & \\
& Multi-mode :Generic $\hat{\rho}_0$  &$\mathbf{i}\sum \limits_{\alpha\beta} 
\psi^\ast_1(\alpha,0)\psi_2(\beta,0)[2 \left(1+\hat{u}\right)^{-1}-1]_{\alpha \beta}$  & $\mathrm{Det}(1+\hat{u})$ &  $ \sum_{nm}c_{nm} \prod \limits_j \partial_{\alpha_j\beta_j} $\\
&$=\sum \limits_{nm} c_{nm}|\{ n\}\rangle\langle \{m\}|$ & & & \\
&  & & & \\
  \hline     
   \end{tabular}
   \caption{ Modification in the structure of the Keldysh field theory to incorporate arbitrary initial density matrix, $\hat{\rho}_0$ for Bosonic and Fermionic systems: the matrix element of $\hat{\rho}_0$ is added as a quadratic term, $\delta S(u)$ in the action, where a function of the initial source $\hat{u}$ couples to the bilinears of the initial quantum fields, $\phi^\ast_q\phi_q$ for Bosons and $\psi_1^\ast\psi_2$ for Fermions. ${\cal N}(u)$ is the normalization of the partition function obtained from the modified action, $S+\delta S(u)$ and physical correlation functions are obtained by taking the set of derivatives, ${\cal L}(\partial_u,\rho_0)$, completely dictated by $\hat{\rho}_0$, of ${\cal N}(u) \hat{G}^{(n)}(u)$, where $\hat{G}^{(n)}(u)$ is the ``n-particle Green's function" in presence of the initial source $\hat{u}$. For the generic density matrix of a multi-mode system, $\hat{\rho}_0=\sum_{nm} c_{nm}|\{ n\}\rangle\langle \{m\}|$ with $N=\sum_\gamma n_\gamma =\sum_\gamma m_\gamma$, $~\partial_{\alpha_j \beta_j}$ denotes partial derivative with respect to $u_{\alpha_j \beta_j}$ which couples to the $j^{th}$ pair of the fields with indices $(\alpha_j, \beta_j)$. In case of Fermions, the set $\mathcal{A}$ denotes the set of occupied modes in the initial $\hat{\rho}_0$.}
    \label{table:Structure}
\end{table*}

\begin{widetext}
\section{Bosonic Field theory for Arbitrary
  Initial Conditions}\label{sec:genericBoson}

For pedagogical reasons, we will first derive the new formalism for
a closed system of a single non-interacting Bosonic mode (i.e. a
harmonic oscillator)  starting from a density matrix diagonal in number basis. While
dynamics of this system may seem trivial, we will see the general
structure mentioned in the previous section emerge in this simple
setting. Further, the derivation and the algebra in more complicated scenario, discussed in later subsections, follow along similar lines, and can be thought of as the extension of this basic theory.

\subsection{Single mode system}

We consider the dynamics of a single mode system described by the
Hamiltonian $H=\omega_0a^\dagger a$, where $\omega_0$ is the energy of the harmonic oscillator mode,
starting from an initial density matrix diagonal in the
number basis of $a^\dagger$, i.e.
\beq
\hat{\rho}_0=\sum_n
c_n|n\rangle \langle n|
\label{rho_0}
\eeq
 where $|n\rangle$ are number states, and $\sum_n c_n=1$.

The identity which enables us to exponentiate the matrix element
of $\hat{\rho}_0$ is
\bqa
\label{identity_u:sm}
\langle \phi| n\rangle\langle n|\phi^{'}\rangle &=&
\frac{(\phi^\ast\phi^{'})^n}{n!} =\left.\frac{1}{n!}\left [
  \frac{\partial}{\partial u}\right]^n e^{u
  \phi^\ast\phi^{'}}\right\vert_{u=0} \\
\nonumber \langle \phi_+(0)| \hat{\rho}_0|\phi_-(0)\rangle&=&
\sum_n \frac{c_n}{n!} \left [\frac{\partial}{\partial u}\right]^n e^{u
  \phi^\ast_+(0)\phi_-(0)}\bigg\vert_{u=0}.  
\eqa
where $|\phi\rangle$ are the harmonic oscillator coherent states.
One can thus exponentiate the initial matrix element in terms of a
source field $u$ coupling to the bilinear of the fields
$\phi^\ast_+\phi_-$ only at $t=0$, at the cost of taking multiple
derivatives with respect to this initial source. In the notation of
the previous section we have $\delta S(u)=-\mathbf{i}u \phi^\ast_+(0)\phi_-(0)$. The set of $u$ derivatives depend on $\hat{\rho}_0$ and in this particular case,
we have ${\cal L}(\partial_u,\rho_0)=\sum_n
(c_n/n!) \partial_u^n$.
Incorporating this in equation \ref{eq:Z_1}, we get the source dependent partition function,
%
\beq
\label{ZJu:sm}
Z(J,u)= \int D[\phi_+]D[\phi_-]e^{\mathbf{i}[\int_0^\infty dt \int_0^\infty dt'
  \phi^\dagger(t)\hat{G}^{-1}(t,t',u)\phi(t')+\int
  dt  J^\dagger(t)\phi(t) +h.c.]}
\eeq
%
where $\phi^\dagger(t)=[\phi^\ast_+(t),\phi^\ast_-(t)]$,
$J^\dagger(t)=[J^\ast_+(t),J^\ast_-(t)]$, and 
\begin{eqnarray*}
G^{-1}_{++}(t,t')&=&-G^{-1}_{--}(t,t')= \delta(t-t')
[i\partial_t -\omega_0]\\
\nonumber G^{-1}_{+-}(t,t',u) &=&-\mathbf{i} u \delta(t)\delta(t'),~~~G^{-1}_{-+}(t,t')=0
\end{eqnarray*}

Since we are working with a non-interacting system, one can easily show by
working with the time discretized version of the matrix $\hat{G}(u)$,
that $Det( -\mathbf{i}\hat{G}^{-1}) = (1-u)$ ~\cite{kamenev}. The gaussian
integrals over the fields then give
\beq
Z(J,u) =\frac{1}{1-u}e^{-\mathbf{i} \int_0^\infty dt \int_0^\infty dt'
  J^\dagger(t) \hat{G}(t,t',u)J(t')}
\label{ZJu1:sm}
\eeq
where the normalization factor ${\cal N}(u)=(1-u)^{-1}$ and $\hat{G}(u)$ is given by
\bqa
\nonumber G_{+-}(t,t',u)&=&\frac{-\mathbf{i}u}{1-u}e^{-i\omega_0(t-t')}\\
\nonumber G_{-+}(t,t',u)&=&\frac{-\mathbf{i}}{1-u}e^{-i\omega_0(t-t')}\\
\nonumber G_{++}(t,t',u)&=&\Theta(t-t') G_{-+}(t,t',u)+\Theta(t'-t)G_{+-}(t,t',u)\\
 G_{--}(t,t',u)&=&\Theta(t'-t) G_{-+}(t,t',u)+\Theta(t-t')
G_{+-}(t,t',u)
\label{gfn_u}
\eqa
We note that setting $u=0$ recovers the usual vacuum Green's functions
for the theory. Further, the physical partition function corresponding to $\hat{\rho}_0$ reduces to $Z_\rho=\sum_n c_n (1/n!) (\partial /\partial_u)^n Z(0,u)
\vert_{u=0}=\sum_n c_n =Tr \hat{\rho}_0$,
 where we have used $(\partial/\partial u)^n
 (1/1-u)\vert_{u=0}=n!$. These act as consistency checks for the Keldysh
partition function of a closed quantum system.

We take the derivatives of $Z(J,u)$ w.r.t the linear sources $J$ and set $J=0$, to define an
n-particle Green's function in
presence of the source $u$
\bqa
\nonumber \left.  \frac{\mathbf{i}^{n}~\partial^{2n}
    Z(J,u)}{\partial_{J(t_1)}..\partial_{J(t_n)}\partial_{J^\ast(t_{n+1})}...\partial_{J^\ast(t_{2n})}}\right\vert_{J=0} 
&=&\frac{1}{1-u}G ^{(n)}(t_1,...t_{2n},u)
\label{g2nu}
\eqa
Note that other than the normalization $(1-u)^{-1}$, which is kept
explicitly for its $u$ dependence, $G^{n}(u)$ is a standard ``n-particle
Green's function'' obtained from a field theory described by an action
$S+\delta S(u)$. We then take appropriate derivatives of $G^{(n)}(u)/(1-u)$ with respect
to $u$ to obtain the physical correlation function for the particular initial
density matrix $\hat{\rho}_0$ as
\beq
{\cal G}^{(n)}_\rho(t_1,...t_{2n})=\sum_n \frac{c_n}{n!}
\left[\frac{\partial}{\partial_u}\right]^n \frac{
  G^{(n)}(t_1,...t_{2n},u)}{1-u}\vert_{u=0} \nonumber
\eeq

 Focusing on the one particle Green's functions, we get $i {\cal
   G}_{+-}(t,t')=\sum_n nc_ne^{-i\omega_0(t-t')}$ and $\mathbf{i}{\cal G}_{-+}(t,t')=\sum_n (n+1)c_ne^{-\mathbf{i}\omega_0(t-t')}$.
At this point, it is useful to work in a rotated basis with the ``classical'' and ``quantum''
fields, $\phi_{cl}=(\phi_+ +\phi_-)/\sqrt{2}$ and
$\phi_q=(\phi_+-\phi_-)/\sqrt{2}$. In this new basis, $G_{qq}(t,t',u)=0$ and 
%
\bqa
G_R(t,t')&
=&-\mathbf{i}\Theta(t-t')e^{-i\omega_0(t-t')} \nonumber \\
 G_K(t,t',u)&=&-\mathbf{i}G^R(t,0)\frac{1+u}{1-u}G^A(0,t')=-\mathbf{i}\frac{1+u}{1-u}e^{-\mathbf{i}\omega_0(t-t')}
\label{gkusm}
\eqa
where $G_R$ is independent of the initial source $u$. 
It is easy to see that the physical retarded
one-particle correlator, ${\cal G}_{R \rho_0}(t,t')=G^R(t,t')$ is independent of the initial
density matrix (i.e. does not depend on $c_n$), while the Keldysh
propagator 
\bqa
\nonumber {\cal G}_{K\rho_0}(t,t')&=& -i\sum_n c_n
(2n+1)G^R(t,0)G^A(0,t')\\
\nonumber&=&-\mathbf{i} (2 \langle a^\dagger a\rangle_0+1) G^R(t,0)G^A(0,t')\\
&=&-\mathbf{i}(2 \langle a^\dagger a\rangle_0+1)  e^{-\mathbf{i}\omega_0 (t-t')}
\eqa 
carries the information of the initial distribution $\langle a^\dagger
a\rangle_0$.

We now construct a continuum action in Keldysh field theory, in the
$cl/q$ basis of the form 
\beq
 S=\int_0^\infty dt \int_0^\infty dt' \bar{\phi}(t) \left[\begin{array}{cc}
0& G_A^{-1}(t,t')\\
G_R^{-1}(t,t')& -\Sigma_K(t,t',u)
\end{array}\right] 
\phi(t') 
\label{Kactionu}
\eeq
with $\bar{\phi}(t) = \left[ \phi_{cl}^*(t),\phi_{q}^*(t) \right]$ and 
$G_R^{-1}(t,t')=\delta(t-t')[ \mathbf{i}\partial_t-\omega_0],
\rcol{\Sigma_K(t,t',u)=-\mathbf{i}(1+u)/(1-u) \delta(t)\delta(t')}.
$
This action $S(u)$ with the $u$ dependent part \rcol{$\delta
S(u)=\mathbf{i}\phi^\ast_q(0)\phi_q(0) (1+u)/(1-u)$ } correctly reproduces the Green's function in presence of the source $u$, i.e. $G_R(t,t')$ and $G_K(t,t',u)$.
From now on, this is the action we will start with and then add couplings to
baths or interparticle interactions, as the case may require, and work
out the dynamics of the system. We will finally take necessary $u$
derivatives to get the physical correlators with the correct initial
conditions.

To summarize, we have obtained a formalism similar to the one
described in the previous section for the dynamics of a single Bosonic
mode starting from a $\hat{\rho}_0= \sum_n c_n|n\rangle\langle n|$. As
shown in Table \ref{table:Structure}, 
\bqa
\nonumber \delta S(u)&=&\rcol{\mathbf{i}\phi^\ast_q(0)\phi_q(0) \frac{1+u}{1-u} }\\
{\cal N}(u)&=&(1-u)^{-1}~~~ and\\
\nonumber {\cal L}(\partial_u,\rho_0)&=&\sum_n \frac{c_n}{n!} \partial_u^n
\eqa
A special simplification takes place when the initial
density matrix has the form $\hat{\rho}_0=\rho^{\hat{n}}$;
i.e. $c_n=\rho^{n}$ for a real $\rho$. In this case ${\cal L}$ leads to
a Taylor series expansion, and as a result one can simply calculate
the physical correlators by setting $u=\rho$, rather than calculating
the derivatives. We note that the thermal density matrix is of this
form with $\rho=e^{-\omega_0/T}$, and hence the case of an initial
thermal distribution can be obtained by setting $u=e^{-\omega_0/T}$
rather than by taking derivatives with respect to $u$. For a time
independent Hamiltonian, this gives the same result which is obtained
for thermal states using usual infinitesimal regularization~\cite{kamenevbook}.

\subsection{Multi-mode systems with diagonal $\hat{\rho}_0$}
We now extend this formalism to a multi-mode Bosonic system starting from $\hat{\rho}_0$ which is diagonal in the occupation number basis in the Fock space. We will focus on a system with large but finite number of countable modes and develop this theory. We will comment on the continuum limit at the end of this section. Most of the algebra
will be similar to the single mode case, so we will point out the main differences in this case.
We consider a closed non-interacting system with $H=\sum_{\alpha,\beta}H_{\alpha\beta} a^\dagger_\alpha a_\beta$, 
where $\alpha,\beta$ denote one particle basis states. We consider an
initial density matrix diagonal in the Fock basis,
\beq
\label{rho_mm}
\hat{\rho}_0=\sum_{\{n\}}c_{\{n\}}| \{n\}\rangle\langle
\{n\}|,
\eeq 
where $|\{n\}\rangle =\prod_\alpha |n_\alpha\rangle$ is a
configuration in the Fock space with basis $\alpha$; e.g. if $\alpha$
indicates lattice sites, then the initial density matrix is diagonal
in the basis of local particle numbers. Note that we will {\it not} assume that the Hamiltonian is
diagonal in the basis $\alpha$ and hence our formalism can track non-trivial
dynamics of even in a closed non-interacting system. 

The first task is to find a way to exponentiate the matrix elements of
$\hat{\rho}_0$. Using the many body coherent states $|\phi\rangle$ we have
\bqa
\label{identity_u:mm}
\langle \phi| \{n\}\rangle\langle\{ n\}|\phi^{'}\rangle &=&
\prod_\alpha \frac{(\phi^\ast_\alpha\phi^{'}_\alpha)^{n_\alpha}}{n_\alpha!} =\left.\prod_\alpha\frac{1}{n_\alpha!}\left [
  \frac{\partial}{\partial {u_\alpha}}\right]^{n_\alpha} e^{\sum_\beta u_\beta
  \phi^\ast_\beta\phi^{'}_\beta}\right\vert_{\vec{u}=0}  \\
\nonumber \langle \phi_+(0)| \hat{\rho}_0|\phi_-(0)\rangle&=&
\sum_{\{n\}} c_{\{n\}}\left.\prod_\alpha\frac{1}{n_\alpha!}\left [
  \frac{\partial}{\partial {u_\alpha}}\right]^{n_\alpha} e^{\sum_\beta u_\beta
  \phi^\ast_{+\beta}(0)\phi_{-\beta}(0)}\right\vert_{\vec{u}=0} 
\eqa
An analysis similar to the single mode can
now be carried out, with the single source now extended to a vector $\vec{u}$. Working in the $\pm$
basis, the partition function can be written in a form similar to
eqn.~\ref{ZJu:sm} with the matrix structure in the space of quantum number $\alpha$. Here, $G^{-1}_{++}(\alpha,t;\beta,t')=\delta(t-t')
[\mathbf{i}\partial_t\delta_{\alpha\beta} -H_{\alpha\beta}]$,
$G^{-1}_{--}(\alpha,t;\beta,t')=-G^{-1}_{++}(\alpha,t;\beta,t')$, and
$G^{-1}_{-+}(\alpha,t;\beta,t')=0$. In equation \ref{identity_u:mm}, we see that the additional $\vec{u}$ dependent
action is given by 
$\delta S(u) =-\mathbf{i}\sum_\alpha u_\alpha
\phi^\ast_+(\alpha,0)\phi_-(\alpha,0)$, while the differential
operator used to obtain physical correlation functions ${\cal
  L}(\partial_u,\rho_0)=\sum_{\{n\}}c_{\{n\}}
\prod_\gamma \partial_{u_\gamma}^{n_\gamma}/n_\gamma !$

To continue the analysis
similar to the single mode case, we need to find expressions
for $Det(-\mathbf{i}\hat{G}^{-1})$, which gives the normalization factor
${\cal N}(u)$, and the Green's functions $\hat{G}(u)$. 
The detailed algebra
for analytic expressions of $Det(-\mathbf{i}\hat{G}^{-1})$ and $\hat{G}(u)$ are provided in
Appendix \ref{sec:AP:diagonal}. Here we quote the final answers for both of them. The determinant
is given by 

\beq
\label{Determinant}
Det[ -\mathbf{i} \hat{G}^{-1}] =Det[ -\mathbf{i} \hat{G}^{-1}(0)]\prod_\alpha 1-u_\alpha
\eeq
where $Det[ -\mathbf{i}G^{-1}(0)]$ is an $\vec{u}$ independent prefactor and can be
ignored as in usual field theory, while the $\vec{u}$ dependent
normalization ${\cal N}(u)=\prod_\alpha (1-u_\alpha)^{-1}$ has to be
kept in the calculations explicitly. 

Similarly, one can invert the matrix $\hat{G}^{-1}(u)$ to obtain (see
Appendix \ref{sec:AP:diagonal} for details) the $\vec{u}$ dependent Green's functions,

\bqa
\nonumber
G_{\mu\nu}(\alpha,t;\beta,t';\vec{u})&=&G^v_{\mu\nu}(\alpha,t;\beta,t') +\sum_\gamma
G^v_{\mu+}(\alpha,t;\gamma,0)\frac{\mathbf{i}~u_\gamma}{1-u_\gamma}G^v_{-\nu}(\gamma,0;\beta,t')
\eqa
where $\mu,\nu=\pm$. Here $\hat{G}^v$ are the Green's functions for the dynamics of a system starting from a vacuum state, and is obtained by setting $\vec{u}=0$ in $\hat{G}(u)$. Explicit expressions for $\hat{G}^v$ can be written in terms of the eigenvalues $E_a$ and the corresponding eigenvectors $\psi_a(\alpha)$ of the Hamiltonian: \rcol{$G^v_{-+}(\alpha,t;\beta,t')=-\mathbf{i}\sum_{a}
\psi^\ast_a(\beta)\psi_a(\alpha)e^{-iE_a (t-t')}$},
$G^v_{+-}(\alpha,t;\beta,t')=0$,
$G^v_{++}(\alpha,t;\beta,t')=\Theta(t-t')
G^v_{-+}(\alpha,t;\beta,t')$
and $G^v_{--}(\alpha,t;\beta,t')=\Theta(t'-t)
G^v_{-+}(\alpha,t;\beta,t')$.
The physical one-particle
correlator is then
given by 
%
\bqa
\nonumber {\cal G}_{ \mu\nu \rho_0}(\alpha,t;\beta,t')=
  G^v_{\mu\nu}(\alpha,t;\beta,t')+\mathbf{i} \sum_{\{n\}}c_{\{n\}}
  \sum_{\gamma}n_\gamma
  G^v_{\mu+}(\alpha,t;\gamma,0)G^v_{-\nu}(\gamma,0;\beta,t')
\eqa
Working in the {\it classical-quantum} basis, we find that 
 $G_R(\vec{u})=G_{R}^v={\cal G}_{R \rho_0}$,
 i.e. the retarded Green's function is independent of $\vec{u}$ and
 hence the physical retarded correlator is independent of the initial
 condition. Similarly we find 
\bqa
\label{Greens_B}
G_K(\alpha,t;\beta,t',\vec{u})
&=&-\mathbf{i} \sum_{\gamma}\frac{1+u_\gamma}{1-u_\gamma}
  G_{R}^v(\alpha,t;\gamma,0)G_{A}^v(\gamma,0;\beta,t')
\eqa
and the physical Keldysh correlator
%
\bqa
\nonumber {\cal G}_{K \rho_0}(\alpha,t;\beta,t')&=&-\mathbf{i}\sum_{\{n\}}c_{\{n\}} \sum_{\gamma}(2n_\gamma+1)
  G_{R}^v(\alpha,t;\gamma,0)G_{A}^v(\gamma,0;\beta,t')\\
  \label{eq:physicalGk_a}
&=&-\mathbf{i}\sum_\gamma (2\langle a^\dagger_\gamma a_\gamma \rangle_0+1)
  G_{R}^v(\alpha,t;\gamma,0)G_{A}^v(\gamma,0;\beta,t')
\eqa
%
where $\langle a^\dagger_\gamma a_\gamma \rangle_0$ is the occupancy of the mode $\gamma$ in the
initial density matrix. 

In this case all the correlation functions in the classical-quantum
basis can be obtained from a continuum Keldysh action of the same form
as in Eq.~\ref{Kactionu}, with
$
 G^{-1}_R(\alpha,t,\beta,t')=\delta(t-t')[ \mathbf{i}\partial_t \delta_{\alpha\beta}-H_{\alpha\beta}],
\rcol{\Sigma_K(\alpha,t,\beta,t',\vec{u})=-\mathbf{i}\delta_{\alpha\beta}(1+u_\alpha)/(1-u_\alpha)
\delta(t)\delta(t')}$.
One can now start with this action, add a bath or inter-particle
interactions, work out the correlators and take
appropriate derivatives to construct correlation functions in the physical
non-equilibrium system.

To summarize, for a many body  bosonic system with an initial density
matrix diagonal in the Fock basis, $\hat{\rho}_0=\sum_{\{n\}}c_{\{n\}}| \{n\}\rangle\langle
\{n\}|$, we have 
\bqa
\nonumber \delta S(u) &=&\rcol{\mathbf{i}\sum_\alpha 
\phi^\ast_q(\alpha,0)\phi_q(\alpha,0)\frac{1+u_\alpha}{1-u_\alpha}},\\
{\cal N}(u)&=&\prod_\alpha(1-u_\alpha)^{-1}~~~and\\
\nonumber {\cal  L}&=&\sum_{\{n\}}c_{\{n\}}
\prod_\gamma \frac{\partial_{u_\gamma}^{n_\gamma}}{n_\gamma !}
\eqa
We note that it is not easy to obtain the continuum limit of the
normalization ${\cal N}$ or the operator ${\cal  L}$ which is defined
w.r.t finite but large number of discrete modes. This stems from the
problem of defining a continuum limit of a many body density
matrix. However, it is clear from equation \ref{eq:physicalGk_a} that it is
straightforward to take the continuum limit of the physical
correlators obtained within this formalism by replacing the sum over
the modes by corresponding integrals.

We note once again that the case of a thermal initial density matrix
can be handled by getting rid of the derivatives and setting $u_a
= e^{-E_a/T}$ and matches with the answers from usual infinitesimal regularization.

\subsection{ Generic initial density matrix for multimode systems}
We now want to extend our formalism to the case of density matrices
which have off-diagonal matrix elements between occupation number
states. We will put the following restriction on the class of initial
density matrices: if the occupation number state $|\{ n\}\rangle$ and
$|\{m\}\rangle$ are connected by the initial density matrix, then
$\sum_\alpha n_\alpha=\sum_\alpha m_\alpha$, i.e. total particle
number in $|\{ n\}\rangle$ and
$|\{m\}\rangle$ are equal. The density matrix is thus block diagonal
in the fixed total particle number sectors of the Fock space. In this
case, we can again formulate the field theory in terms of an initial
source coupled to bilinears of the fields. We note that this
covers almost all density matrices where one can reasonably expect to
prepare the many body system. 

Let us consider an initial density matrix of the form 
\beq
\label{rho_off_mm}
 \hat{\rho}_0=\sum_{nm} c_{nm}|\{ n\}\rangle\langle \{m\}|
\eeq
where $c_{nm}=c^\ast_{mn}$ to maintain hermiticity of the density
matrix and $\sum_n c_{nn}=1$ for conservation of probabilities. The matrix element of $\hat{\rho}_0$ between initial
coherent states is given by
\beq
\nonumber \langle \phi|\hat{\rho}_0|\phi '\rangle=\sum_{nm} c_{nm}
\prod_\alpha
\frac{[\phi^\ast_\alpha]^{n_\alpha}[\phi '_\alpha]^{m_\alpha}}{\sqrt{n_\alpha
    ! m_\alpha !}}
\eeq
Now, if $\sum_\alpha n_\alpha=\sum_\alpha m_\alpha$, then one can
always pair up each $\phi^\ast_\alpha$ with a $\phi^{'}_\beta$ in the
above product. While this choice is not unique, we will proceed with a
particular pairing and show that our final answers for physical correlators are invariant
with respect to permutations leading to different pairings. 


In this case the exponentiation of the matrix element of
$\hat{\rho}_0$ is achieved by
\bqa
\nonumber
\prod_{\alpha}[\phi^\ast_{\alpha}]^{n_\alpha}[\phi '_{\alpha}]^{m_{\alpha}}&=&\prod_{j=1}^{N}{\phi}^\ast_{\alpha_j}\phi  '_{\beta_j}=\left.\prod_{j=1}^{N}\left[\frac{\partial}{\partial u_{\alpha_j \beta_j}}\right]e^{\sum_{\gamma \delta}u_{\gamma \delta}\phi^\ast_{\gamma}\phi '_{\delta}}\right\vert_{\hat{u}=0},\\
\langle \phi_+(0)| \hat{\rho}_0|\phi_-(0)\rangle &=& \sum_{nm}  \frac{c_{nm} }{\prod_\alpha\sqrt{n_\alpha ! m_\alpha !}}\left.\prod_{j}[\partial_{\alpha_j \beta_j}]e^{{\sum_{\gamma \delta}u_{\gamma \delta}\phi^\ast_{+\gamma}(0)\phi _{-\delta}(0)}}\right\vert_{\hat{u}=0},
\label{rho:mat-elem:umat}
\eqa
where, $(\alpha_j, \beta_j)$ are the mode indices of the fields forming the $j^{th}$ pair out of total $N=\sum_\alpha n_\alpha=\sum_\alpha m_\alpha$ pairs.
The vector source for the diagonal
density matrix  is now replaced by a matrix source $\hat{u}$ with
elements $u_{\alpha\beta}$ and $\partial_{\alpha_j \beta_j}$ indicate derivative with respect to $u_{\alpha_j \beta_j}$. Following algebra similar to the earlier
two cases, we find that we need to add a term to the Keldysh action 
$\delta S= -\mathbf{i}\sum_{\alpha\beta} u_{\alpha\beta}
\phi^\ast_+(\alpha,0)\phi_-(\beta,0)$, and the differential operator
used to obtain physical correlators is given by ${\cal
  L}(\partial_u,\rho_0)= \sum_{nm}c_{nm} \prod (n_\alpha ! m_\alpha
!)^{-1/2}\prod_j \partial_{\alpha_j\beta_j}$.

As before, we are interested in analytical expressions for $\mathrm{Det}
[-\mathbf{i}G^{-1}(\hat{u})]$ and the Green's functions $G(\hat{u})$, which are
given by 
\bqa
&&\mathrm{Det}[- \mathbf{i} G^{-1}(\hat{u})] =\mathrm{Det}[- \mathbf{i} G^{-1}(0)]\mathrm{Det}(1-\hat{u}) \nonumber \\
&&G_{\mu\nu}(\alpha,t;\beta,t';\hat{u})=G^v_{\mu\nu}(\alpha,t;\beta,t') +\mathbf{i}\sum_{\gamma \delta}
G^v_{\mu+}(\alpha,t;\gamma,0) [ \left(1-\hat{u}\right)^{-1}-1]_{\gamma \delta} G^v_{-\nu}(\delta,0;\beta,t')
\label{gfn_u:umat}
\eqa
Note that the derivation of these identities [See Appendix B for
derivation] follow a different route than those for the case of
diagonal initial density matrices.
We can now compute the physical Green's functions, ${\cal G}_{\rho_0}$, by taking appropriate functional derivatives with respect to $u_{\alpha \beta}$. We find that 
\bqa
\nonumber
{\cal G}_{\mu\nu \rho_0}(\alpha,t;\beta,t')=
  G^v_{\mu\nu}(\alpha,t;\beta,t')
  +\mathbf{i}\sum_{\gamma \delta}G_{\mu+}(\alpha,t;\gamma,0) \langle\hat{a}_{\delta}^{\dagger}\hat{a}_{\gamma}\rangle_{0} G_{-\nu}(\delta,0;\beta,t')
\eqa
where $\langle\hat{a}_{\delta}^{\dagger}\hat{a}_{\gamma}\rangle_{0}$
gives the initial one-particle correlations. 

After a Keldysh rotation to $cl/q$ basis, we find that $\hat{G}_R(\hat{u})=\hat{G}_R^v$.
The Keldysh Green's function, on the other hand, is given by 
\bqa
 G_K(\alpha,t;\beta,t',\hat{u})&=&
  G^v_{K}(\alpha,t;\beta,t') - \mathbf{i}  \sum_{\gamma \delta}
  G^v_{R}(\alpha,t;\gamma,0) [ 2\left(1-\hat{u}\right)^{-1}-1]_{\gamma \delta} G^v_{A}(\delta,0;\beta,t').
\eqa
The physical Green's functions are then given by,
\bqa
\nonumber
{\cal G}_{R\rho_0}(\alpha,t;\beta,t')&=&G^v_{R}(\alpha,t;\beta,t'),\\
\label{eq:physicalGK_off}
{\cal G}_{K\rho_0}(\alpha,t;\beta,t')&=& -\mathbf{i} \sum_{\gamma \delta}
  G^v_{R}(\alpha,t;\gamma,0) [2\langle\hat{a}_{\delta}^{\dagger}\hat{a}_{\gamma}\rangle_{0}+\delta_{\gamma \delta}] G^v_{A}(\delta,0;\beta,t').
\eqa
Once again, all the correlation functions in the classical-quantum
basis can be obtained from a continuum Keldysh action of the same form
as in Eq.~\ref{Kactionu}, with
$
 G^{-1}_R(\alpha,t,\beta,t')=\delta(t-t')[ \mathbf{i}\partial_t \delta_{\alpha\beta}-H_{\alpha\beta}],
\Sigma_K(\alpha,t,\beta,t',u)=\mathbf{i}[2 \left(1-\hat{u}\right)^{-1}-1]_{\alpha \beta}
\delta(t)\delta(t')$.
To summarize, for a many body  Bosonic system with an initial , $\hat{\rho}_0=\sum_{nm} c_{nm}|\{ n\}\rangle\langle \{m\}|$, we have 
\bqa
\nonumber \delta S(u) &=&-\mathbf{i}\sum_{\alpha \beta} 
\phi^\ast_q(\alpha,0)\phi_q(\beta,0)[2 \left(1-\hat{u}\right)^{-1}-1]_{\alpha \beta},\\
{\cal N}(u)&=&\mathrm{Det}(1-\hat{u})^{-1}~~~and\\
\nonumber {\cal
  L}(\partial_u,\rho_0)&=& \sum_{nm}c_{nm} \prod (n_\alpha ! m_\alpha
!)^{-1/2}\prod_j \partial_{\alpha_j\beta_j}
\eqa
This concludes the derivation of our new formalism which can treat the quantum dynamics of a Bosonic system starting from an arbitrary initial density matrix.

\section{Fermionic Field Theory for Arbitrary Initial Conditions}\label{sec:genericFermion}
In the previous sections, we have developed the Schwinger Keldysh path
integral based formalism to study the dynamics of a many body Bosonic
system starting from an arbitrary initial density matrix. In this
section, we will extend this newly developed formalism to a
Fermionic many body system. The basic structure of the theory follows
along a line similar to that proposed for Bosons, i.e. corresponding to the matrix element \rcol{$\langle \psi_+(0)| \hat{\rho}_0|-\psi_-(0)\rangle$} in eqn. \ref{eq:Z_2}, we
have to add a term $\delta S(u)$ to the standard Keldysh action, where
$\hat{u}$ is a source which couples to bilinears of the Grassmann fields only at
initial time. One can then calculate the Green's functions, $\hat{G}(u)$ from the action $S+\delta S(u)$ and the $\hat{u}$ dependent normalization ${\cal N}(u)$ by Gaussian integrals of the Grassmann fields. The physical correlation functions are then obtained by
applying appropriate set of derivatives ${\cal L}(\partial_u,\rho_0)$,
determined by the initial density matrix $\hat{\rho}_0$.
The derivation of $\delta S(u)$, ${\cal N}(u)$ and
${\cal L}(\partial_u,\rho_0)$ for a Fermionic theory for different initial conditions is very similar to that of Bosons, with some important changes. We will focus on the distinctions between Bosonic and Fermionic theory, instead of repeating the algebra similar to that in the previous sections.


To extend the new formalism for Fermions, we need to keep track of two major differences between Bosonic theories with complex fields and
Fermionic theories with Grassmann fields. The first one is that, in a Fermionic theory, the trace of an
operator, written as a functional integral over Grassmann fields, has an additional minus sign from that in the Bosonic
expression~\cite{Negele_Orland}, as seen in Eq.~\ref{eq:Z_2}. This is a characteristic of
all Fermionic theories. For example, for a diagonal
  density matrix in a single mode system, $\hat{\rho}_0=\sum_{n}c_{n}| n\rangle\langle
n|$, where $n =0,1$ for Fermionic systems, the matrix element
\bqa
\label{identity_u:mmf}
\nonumber \langle \psi_+(0)| \hat{\rho}_0|-\psi_-(0)\rangle&=&
\sum_{n} c_{n} [-\psi^\ast_{+}(0)\psi_{-}(0)]^{n}=\sum_{n} c_{n}[\partial_u]^n e^{-u\psi^\ast_{+}(0)\psi_{-}(0)}\vert_{u=0}
\eqa
Thus one can exponentiate the matrix element of the initial density
matrix in a way similar to that for Bosons,  with the additional minus sign absorbed by the transformation $u\rightarrow -u$. The second difference is that the Gaussian integration over Grassmann fields in the Fermionic partition function gives $Det[-\mathbf{i}\hat{G}^{-1}(u)]$ in the numerator as opposed to $1/Det[-\mathbf{i}\hat{G}^{-1}(u)]$ in the case of Bosons (eqn \ref{ZJu1:sm}).

We will consider a many body Fermionic system with Hamiltonian $H=\sum_{\alpha,\beta}H_{\alpha\beta} a^\dagger_\alpha a_\beta$ where $a^\dagger_\alpha $ creates a Fermion in mode $\alpha$ and an initial density matrix which is diagonal in Fock basis, given in equation \ref{rho_mm}, where the occupation numbers of the mode $\alpha$, $n_\alpha$, are restricted to be only $1$ or $0$ due to Pauli exclusion principle. In this case the matrix element of $\hat{\rho}_0$ is given by,
\beq
\label{eq:matrixelem_F}
\langle \psi_+(0)| \hat{\rho}_0|-\psi_-(0)\rangle =
\sum \limits_{\{n\}} c_{\{n\}}\prod_\alpha \left [
  \frac{\partial}{\partial_{u_\alpha}}\right]^{n_\alpha}  e^{-\sum_\beta u_\beta
  \psi^*_{+\beta}(0)\psi_{-\beta}(0)}\Bigg\vert_{\vec{u}=0} 
\eeq
where $\psi^*$ is the conjugate to the Grassmann field $\psi$. Using this, we obtain the Fermionic partition function $Z[J,u]$ in presence of both the sources: Grassmann source $J_{\pm}$ coupled linearly to $\psi^*_{\pm}$ and the real quadratic source $\vec{u}$ turned on at $t=0$ as,  
%
\beq
Z(J,u)= \int D[\psi_+]D[\psi_-]e^{\mathbf{i}[\int_0^\infty dt \int_0^\infty dt'
  \psi^\dagger(t)\hat{G}^{-1}(t,t',u)\psi(t')+\int
  dt  J^\dagger(t)\psi(t) +h.c.]}
\label{ZJu:sm_F}
\eeq
%
The inverse Green's function in the Fermionic action is the same as that in the Bosonic action (\ref{ZJu:sm}), except for the $+-$ component which is modified to 
$
G^{-1}_{+-}(\alpha,t;\beta, t',\vec{u}) =\mathbf{i} u_\alpha \delta_{\alpha\beta}
\delta(t)\delta(t')
$, i.e. $\delta S(u) = \mathbf{i}\sum_\beta u_\beta
  \psi^*_{+\beta}(0)\psi_{-\beta}(0)$.
We perform the Gaussian integration over the Grassmann fields to obtain,
\beq
\displaystyle Z[ J,u]=\prod_\alpha
(1+u_\alpha)e^{-\mathbf{i}\int_0^\infty dt\int_0^\infty dt' J^\dagger(\gamma,t)G(\gamma,t;\beta,t',\vec{u})J(\beta ,t')}
\eeq
A notable difference between the Fermionic partition function and the Bosonic one is that the determinant $Det[-\mathbf{i}G^{-1}]=\prod_\alpha(1+u_\alpha)$ appears in the numerator, leading to the normalization, ${\cal N}(u)=\prod_\alpha(1+u_\alpha)$. It is evident from equation \ref{eq:matrixelem_F}, that  ${\cal L}(\partial_u,\rho_0)=\sum_{\{n\}} c_{\{n\}}\prod_\alpha\left [
  \partial/\partial_{u_\alpha}\right]^{n_\alpha}=\sum_{\{n\}} c_{\{n\}}\prod_{\alpha \in \mathcal{A}} \partial/\partial_{u_\alpha}  $ where $\mathcal{A}$ denotes the set of modes occupied in the Fock state $|\{n\}\rangle$.
 We find that in the $+,-$ basis, the Fermionic Green's function $\hat{G}(u)$ can be obtained from the Bosonic ones by taking $\vec{u}\rightarrow - \vec{u}$. Working in the rotated basis $\psi_{1(2)}$, we obtain the retarded Green's function, $G_R(\alpha,t,\beta,t') =G_R^v(\alpha,t,\beta,t') $, again independent of $\vec{u}$, and the Keldysh Green's function,
\bqa
\label{Greens_B}
\nonumber G_K(\alpha,t;\beta,t',\vec{u})
=-\mathbf{i} \sum_{\gamma}\frac{1-u_\gamma}{1+u_\gamma}
  G_{R}^v(\alpha,t;\gamma,0)G_{A}^v(\gamma,0;\beta,t')
\eqa
%
The physical observables are obtained by applying ${\cal L}(\partial_u,\rho_0) $ on ${\cal N}(u)\hat{G}(u)$ and setting $\vec{u}=0$, i.e.
%
\bqa
{\cal G}_{R\rho}(\alpha,t;\beta,t')&=&G_{R}^v(\alpha,t;\beta,t')\\
\nonumber {\cal G}_{K\rho}(\alpha,t;\beta,t')&=&-\mathbf{i}\sum_{\{n\}}c_{\{n\}} \sum_{\gamma}(1-2n_\gamma)
  G_{R}^v(\alpha,t;\gamma,0)G_{A}^v(\gamma,0;\beta,t') \nonumber \\
  \label{eq1}
  &=&-\mathbf{i} \sum \limits_{\gamma}  (1-2\rcol{ \langle a^{\dagger}_\gamma a_\gamma \rangle _0 })  G_{R}^v(\alpha,t;\gamma,0)G_{A}^v(\gamma,0;\beta,t') 
\eqa
To continue working in the rotated $1(2)$ basis for Fermionic fields, we construct the Keldysh action in continuum in presence of the initial source $\vec{u}$. The retarded, advanced and Keldysh Fermionic propagators, $\hat{G}(\vec{u})$ can be obtained by inverting the kernels in the action \ref{action_12F}.
\bqa
\label{action_12F}
&&S=\int_0^\infty dt \int_0^\infty dt'\sum_{\alpha\beta} \psi^*(\alpha,t) \hat{G}^{-1}(\alpha,t;\beta,t',u)\psi(\beta,t') \\
&&\bar{G}^{-1}(\alpha,t,\beta,t')=\left[\begin{array}{cc}
G_R^{-1}(\alpha,t,\beta,t')& -\Sigma_K(\alpha,t,\beta,t',u)\\
0& G_A^{-1}(\alpha,t,\beta,t')
\end{array}\right]
\eqa
with
$\nonumber G_R^{-1}(\alpha,t,\beta,t')=\delta(t-t')[ \mathbf{i}\partial_t \delta_{\alpha\beta}-H_{\alpha\beta}]$ and $
\rcol{\nonumber\Sigma_K(\alpha,t,\beta,t',\vec{u})=-\mathbf{i}\delta_{\alpha\beta}\frac{1-u_\alpha}{1+u_\alpha}
\delta(t)\delta(t')}$.
To summarize, for a many body Fermionic system with an initial density
matrix diagonal in the Fock basis, $\hat{\rho}_0=\sum_{\{n\}}c_{\{n\}}| \{n\}\rangle\langle
\{n\}|$, we have 
\bqa
\nonumber \delta S(u) &=&\rcol{\mathbf{i}\sum_\alpha 
\psi^\ast_1(\alpha,0)\psi_2(\alpha,0)\frac{1-u_\alpha}{1+u_\alpha}},\\
{\cal N}(u)&=&\prod_\alpha(1+u_\alpha)~~~and\\
\nonumber {\cal  L}&=&\sum_{\{n\}}c_{\{n\}}
\prod_{\gamma \in \mathcal{A}} \partial_{u_\gamma}
\eqa
The Fermionic Green's functions satisfy a large number of constraints reflecting the fact that initial occupation numbers can not be greater than $1$. This leads to $(\partial /\partial u_\gamma)^n \mathcal{N}(u)\hat{G}(u)=0 ~\vert_{\vec{u}=0}$ for any $\gamma$ and $n \geq 2$. The non-interacting Green's functions derived above explicitly satisfy these conditions. We note that these relations are manifestations of Fermi statistics and should continue to hold for interacting systems as well as open quantum systems. The simplicity of the normalization factor $\mathcal{N}(u)$ allows us to write ${\cal G}_{\rho_0}=\sum_{\{n\}}c_{\{n\}}\prod_{\gamma \in \mathcal{A}} (1+\partial_{ u_{\gamma}}) \hat{G}(u) \vert_{\vec{u}=0}$. This compact relation is useful for practical computation of physical correlators for Fermionic systems.

This formalism can be generalized to the case of generic initial density matrix with off-diagonal elements in the Fock basis, given by eqn. \ref{rho_off_mm} in a way similar to that of Bosons with the modifications mentioned above. We will not go into the details, but provide the answers for the physical one particle correlators here,
\bqa
\nonumber
{\cal G}_{R\rho_0}(\alpha,t;\beta,t')&=&G^v_{R}(\alpha,t;\beta,t'),\\
\label{eq2}
{\cal G}_{K\rho_0}(\alpha,t;\beta,t')&=& -\mathbf{i} \sum_{\gamma \delta}
  G^v_{R}(\alpha,t;\gamma,0) [\delta_{\gamma \delta}-2\langle\hat{a}_{\delta}^{\dagger}\hat{a}_{\gamma}\rangle_{0}] G^v_{A}(\delta,0;\beta,t').
\eqa
Thus the initial off-diagonal density matrix for a system of Fermions leads to,
\bqa
\nonumber \delta S(u) &=&\rcol{\mathbf{i}\sum_\alpha 
\psi^\ast_1(\alpha,0)\psi_2(\beta,0)[2 \left(1+\hat{u}\right)^{-1}-1]_{\alpha \beta}},\\
{\cal N}(u)&=&\mathrm{Det}(1+\hat{u})~~~and\\
\nonumber {\cal
  L}(\partial_u,\rho_0)&=& \sum_{nm}c_{nm} \prod_j \partial_{\alpha_j\beta_j}
\eqa 
where we use similar notations as used in the Bosonic case.

\section{Two-particle Correlators and violation of Wick's
  theorem}\label{sec:twoparticle}

In standard field theories, Wick's theorem states that the expectation
of a multi-particle operator (i.e. a multi-particle correlation
function) in a non-interacting theory (gaussian action) can be
calculated as a product of single particle Green's functions, summed
over all possible pairings of the operators into bilinear forms. For an interacting theory, this
is the backbone of constructing a diagrammatic perturbation theory in
terms of single particle Green's functions and interaction vertices,
and various non-perturbative resummations that result from
this. Throughout this paper we have emphasized that the physical
correlators in a dynamics with arbitrary initial conditions are not
related by Wick's theorem, even for a non-interacting Hamiltonian. We will illustrate this
point in details in this section by considering physical two-particle
correlators in non-interacting Bosonic/Fermionic theories. In
fact, a major accomplishment of this formalism is to construct Green's
functions which satisfy Wick's theorem, and for which standard
approximations of field theories can be used. 

Our goal is not simply to establish a violation of Wick's theorem, but
to characterize and quantify the violation. To this end, we will work in the Keldysh rotated basis ($(cl,q)$ for Bosons and $(1,2)$ for Fermions), where the initial
condition dependence of the one particle correlators is more
streamlined. Any physical two particle correlator $\hat{{\cal G}}^{(2)}_{\rho_0}$ can be written in
terms of the corresponding ``two-particle Green's function in presence
of source", $\hat{G}^{(2)}(u)$ through Eq.~\ref{eq:physicalG}. To illustrate the violation, we will focus on a
multi-mode system starting from a density matrix diagonal in the Fock
basis  $\hat{\rho}_0=\sum_{\{n\}}c_{\{n\}}| \{n\}\rangle\langle
\{n\}|$; in this case, $\hat{{\cal G}}^{(2)}_{\rho_0}={\cal
  L}(\partial_u,\rho_0){\cal N}(u) \hat{G}^{(2)}(u)\vert_{u=0}$ with ${\cal  L}=\sum_{\{n\}}c_{\{n\}}
\prod_\gamma [\partial_{u_\gamma}^{n_\gamma}/n_\gamma !]$ and ${\cal N}(u)=\prod_\mu (1-\zeta u_\mu)^{-\zeta}$ where $\zeta=\pm
1$ for Bosons(Fermions).

As we have emphasized before, $\hat{G}^{(2)}(u)$ is related to the one
particle Green's functions $\hat{G}(u)$ through Wick's theorem,
i.e. $\hat{G}^{(2)}(u) =\sum_{(ab)}G_a(u)G_b(u)$, where $a,b = R/A/K$,
and $\sum_{(ab)}$ indicates sum over all allowed pairings. We will
now consider the action of ${\cal L}$ on ${\cal N}(u)G_a(u)G_b(u)$ for different combinations of $a,b$; the required sum over pairings can always be performed at
the end. Let us consider the action of ${\cal L}$ when both $a$ and
$b$ are either $R$ or $A$; i.e. we are considering a pair of retarded
or advanced Green's functions. In this case, $G_{R(A)}(u)$ is
independent of $u$, and ${\cal L}{\cal N}(u)\vert_{u=0}=1$ by
normalization of the density matrix; so this part of ${\cal
  G}^{(2)}_{\rho_0} = {\cal G}_{a,\rho_0}{\cal G}_{b,\rho_0}$,
i.e. this part of the physical 2-particle correlator can be written as a Wick
contraction over the physical retarded or advanced one-particle
correlators. We now consider the case where one, but not both of $a,b$ is the
Keldysh Green's function. In this case, $G_{R(A)}$ is
independent of $u$, ${\cal L}$ acts on ${\cal
  N}(u)G_K(u)$ to give ${\cal G}_{K,\rho_0}$, and once again Wick
contraction in terms of physical correlators work, i.e.  for this
part we also get ${\cal
  G}^{(2)}_{\rho_0} = {\cal G}_{R(A),\rho_0}{\cal G}_{K,\rho_0}$. 

The violation of Wick's theorem comes from the pairing where both
single particle Green's function are Keldysh propagators. For a
non-interacting system, 
\beq
G_K(\alpha,t,\beta,t',\vec{u})=-\mathbf{i}\sum_\gamma G_R(\alpha,t,\gamma,0)
G^\ast_R(\beta,t',\gamma,0)\frac{1+\zeta u_\gamma}{1-\zeta u_\gamma}.
\eeq
To show the structure of the violation, we consider the correlator, $\langle \phi^*_{cl}(\alpha,t)\phi^*_{cl}(\beta,t')\phi_{cl}(\gamma,t')\phi_{cl}(\delta,t) \rangle =\mathbf{i}^2 {\cal
  G}^{(2)}_{\rho_0}(\alpha,t,\beta,t',\gamma,t'\delta,t)$ for Bosons,
\bqa
{\cal
  G}^{(2)}_{\rho_0}(\alpha,t,\beta,t',\gamma,t'\delta,t)&=&\sum_{\{n\}}c_{\{n\}}\sum_{x,y}
\left[ G_R^\ast(\alpha,t,x,0)G_R(\gamma,t',x,0)
G_R^\ast(\beta,t',y,0)G_R(\delta,t,y,0)\right.\\
\nonumber &+&\left. G_R^\ast(\alpha,t,x,0)G_R(\delta,t,x,0)
G_R^\ast(\beta,t',y,0)G_R(\gamma,t',y,0)\right][ (2n_x+1)(2n_y+1)-2\delta_{x,y}n_x(n_x+1)] 
\eqa
Similarly, for Fermions we get,
\bqa
{\cal
  G}^{(2)}_{\rho_0}(\alpha,t,\beta,t',\gamma,t'\delta,t)&=&\sum_{\{n\}}c_{\{n\}}\sum_{x,y}
\left[ G_R^\ast(\alpha,t,x,0)G_R(\gamma,t',x,0)
G_R^\ast(\beta,t',y,0)G_R(\delta,t,y,0)\right.\\
\nonumber &+& \zeta \left. G_R^\ast(\alpha,t,x,0)G_R(\delta,t,x,0)
G_R^\ast(\beta,t',y,0)G_R(\gamma,t',y,0)\right][ (1-2n_x)(1-2n_y)-4\delta_{x,y}n_x] 
\eqa
For a single Fock state, where the $\sum_{\{n\}}$ is redundant, we
note that the first term with $(1+ \zeta 2n_x)(1+ \zeta 2n_y)$ can be written as
${\cal G}_{K\rho_0}{\cal G}_{K\rho_0}$, i.e. this part corresponds to
a Wick contraction with physical ${\cal G}_{K\rho_0}$. In this case the term
with $\delta_{x,y}$ contains the connected density correlations in the
initial state and leads to a violation of Wick's theorem. For a
generic diagonal density matrix, both terms lead to violation of Wick's
theorem, since even for $x\neq y$, the connected density correlations
in the initial state is non-zero. The expressions for Bosons and Fermions can be written
in a compact notation in terms of initial correlations in the system,
\bqa 
&&{\cal
  G}^{(2)}_{\rho_0}(\alpha,t,\beta,t',\gamma,t',\delta,t)=\sum_{x,y}
\left[ G_R(\alpha,t,x,0)G^\ast_R(\gamma,t',x,0)
G_R(\beta,t',y,0)G^\ast_R(\delta,t,y,0)\right.\\
\nonumber &+&\zeta \left. G_R(\alpha,t,x,0)G^\ast_R(\delta,t,x,0)
G_R(\beta,t',y,0)G^\ast_R(\gamma,t',y,0)\right][ \langle (1+\zeta
2\hat{n}_x)(1+\zeta 2\hat{n}_y)\rangle_0-2\delta_{x,y}\langle \hat{n}_x(\hat{n}_x+1)\rangle_0] 
\eqa
where $\hat{n}_x$ is the number operator in mode $x$, and $\langle
\rangle_0$ indicates expectation with the initial density matrix. Writing the above expression in terms of a Wick' theorem and a correction term, we have
\beq
{\cal
  G}^{(2)}_{\rho_0}(\alpha,t,\beta,t',\gamma,t'\delta,t)= {\cal
  G}_{K\rho_0}(\alpha,t,\gamma,t'){\cal
  G}_{K\rho_0}(\beta,t',\delta,t) + \zeta {\cal
  G}_{K\rho_0}(\alpha,t,\delta,t){\cal
  G}_{K\rho_0}(\beta,t',\gamma,t') + \delta {\cal
  G}^{(2)}
\eeq
where
\bqa
\delta {\cal
  G}^{(2)} =&&\sum \limits_{x,y}
\left[ G_R(\alpha,t,x,0)G^\ast_R(\gamma,t',x,0)
G_R(\beta,t',y,0)G^\ast_R(\delta,t,y,0)\right.\\
\nonumber && +\left. G_R(\alpha,t,x,0)G^\ast_R(\delta,t,x,0)
G_R(\beta,t',y,0)G^\ast_R(\gamma,t',y,0)\right] 2[\langle a^{\dagger}_x a^{\dagger}_y a_y a_x  \rangle_{0c} (2-\delta_{x,y}) ] 
\eqa
$\langle\rangle_{0c}$ indicates connected expectation value in the initial density matrix.
We
thus see that the violation of the Wick's theorem can be directly tied
to the presence of two particle connected correlations in the initial
state of the system.

The above calculation can easily be generalized to multi-particle
correlators. The Wick's theorem violating terms would come from having
multiple $G_K$ in the product decomposition and are proportional to
connected multi-particle correlations in the initial state. 

\section{Open Quantum systems with Arbitrary Initial Conditions}\label{sec:OQS}
In the previous sections, we have generalized the
Keldysh field theory to treat dynamics of closed quantum systems starting from
arbitrary initial conditions. In this section, we will extend this
formalism to study the dynamics of many particle open
  quantum systems (OQS)
coupled to external baths. We will then work out examples of a Bosonic
and a Fermionic OQS  undergoing non-unitary dynamics starting from
different initial conditions.

 The general problem of a system coupled to external
  baths can be treated using a Hamiltonian of the form
  $H=H_s+H_b+H_{sb}$, where $H_s$ and $H_b$ are the Hamiltonians of
  the system and the baths respectively, while $H_{sb}$ is a coupling
  between the system and the baths. Here, we will assume that both
  $H_s$ and $H_b$ are non-interacting Hamiltonians, whereas the system
  bath coupling $H_{sb}$ is linear in both the bath and system degrees
  of freedom, so that the combined system can be represented by a
  Gaussian theory. At $t=0$, 
the density matrix of the combined system, $\hat{\rho}_0 =
\hat{\rho}_{0S} \otimes \hat{\rho}^l_{0B}$, where $\hat{\rho}_{0S}$ is
an arbitrary density matrix of the system, which will be encoded by
using an initial bilinear source $\hat{u}$, similar to previous sections. Here $\hat{\rho}^l_{0B}$ is a
thermal density matrix for the $l^{th}$ bath with temperature $T_l$
and chemical potential $\mu_l$. We will assume that the system bath
coupling $H_{sb}$ is turned on through an infinitely rapid quench at
$t=0$.  This quench will break the time-translation invariance of the
full problem.

We will also assume that while the coupling to the baths
  changes the system dynamics for $t>0$, the baths themselves are not
  affected by the presence of the system. The bath Green's functions
  are then time-translation invariant and are given by the thermal
  Green's functions. These can be evaluated either by using standard
  infinitesimal regularization~\cite{kamenevbook} or by using a
  initial source field for the baths, and setting them to their
  thermal value. For $t>0$, we trace out the bath degrees of freedom
  and study the effective action of the system. Since the bath is
  non-interacting and the couplings are linear, this produces only
  quadratic terms in the effective action of the OQS, which can be
  written in the form of retarded and Keldysh self energies,
  $\Sigma^B_R$ and $\Sigma^B_K$ respectively. The matrix self-energy $\hat{\Sigma}^B$ has the structure,
\bqa
\hat{\Sigma}^B=\left[\begin{array}{cc}
0 & \Sigma^{ B}_A\\
\Sigma^{ B}_R & \Sigma^B_K
\end{array}\right] ~~\textrm{for ~Bosons~ and~~}
\hat{\Sigma}^B=\left[\begin{array}{cc}
\Sigma^{ B}_R & \Sigma^{ B}_K\\
 0& \Sigma^B_A
\end{array}\right]~~\textrm{for ~Fermions}.
\eqa
where $\Sigma_R^B=\Sigma_A^{B\dagger}$ incorporates the dissipative
effects of the bath, while $\Sigma_K^B$ incorporates the stochastic
fluctuations due to the bath. Since the bath Green's functions are
time translation invariant, it is easy to see that the self energies
have the following structure
\beq
\nonumber \Sigma^B_{R/K}(t,t')= \Theta(t) \Theta(t') \int \frac{d\omega}{2\pi}
\Sigma^B_{R/K}(\omega)e^{\mathbf{i} \omega t} 
\eeq
where $\Sigma^B_{R}(\omega)$ is related to the spectral density of the
baths $\mathcal{J}(\omega)$\cite{nori,Chakraborty1} ( a combination of
bath density of states and system bath coupling), while
$\Sigma^B_{K}(\omega)$ is related to both $\mathcal{J}(\omega)$ and
the thermal distributions in the baths. Note that we have suppressed
the quantum number indices for brevity here. The Dyson equation for
the retarded Green's function can be solved to get
\begin{eqnarray*}
G_R(t,t')&= G^O_R(t-t')&~ \text{for}~ t,t^{'} >0\\
&=G^O_R(t) G^0_R(-t') &~ \text{for}~ t>0,~ t^{'}<0\\
&=G^0_R(t-t') &~ \text{for}~ t, t^{'}<0
\end{eqnarray*}
where $G^0$ is the Green's function for the closed system and~\cite{nori} 
\beq
G^O_R(t-t') =\mathbf{i}\int \frac{d\omega}{\pi} e^{\mathbf{i} \omega
  t} \text{Im} [G^{0-1}_R(\omega) -\Sigma^B_R(\omega)]^{-1}
\label{eq:GOR}
\eeq
We note that the retarded Green's function of the OQS is still
independent of the source $\hat{u}$ and hence represents the physical
retarded Green's function, which is independent of $\hat{\rho}_{0S}$.
The information of $\hat{\rho}_{0S}$ is carried by the physical Keldysh correlation function,
%
%
\bqa
{\cal G}^O_{K\rho_0}(\alpha,t;\beta,t')&=& {\cal G}^s_{K\rho_0}(\alpha,t;\beta,t')
   + 
   \int_0^t
  dt_1\int_0^{t'} dt_2 G_{R}^O(\alpha,\gamma,t-t_1)\Sigma_K^B(\gamma,\delta,t_1-t_2)G_{A}^O(\delta,\beta,t_2-t')\nonumber \\
  \label{gkoqsB}
\eqa
where we have reinstated the quantum number of the modes and\rcol{
\beq
{\cal G}^s_{K\rho_0}(\alpha,t;\beta,t')=-\mathbf{i} \sum_{\gamma\delta}
G^O_R(\alpha,t;\gamma,0)[\delta_{\gamma\delta}+2 \zeta \langle
a^\dagger_\gamma a_\delta \rangle]G^O_A(\delta,0;\beta,t')
\eeq }
\end{widetext}
We note that $\hat{{\cal G}}^s_{K\rho_0}$ carries information about initial condition and is not a function of $(t-t')$. This, along with integration limits in the second term break the time translation invariance of the physical observables. 

 \begin{figure*}[t]
   \centering
    \includegraphics[width=0.27\textwidth]{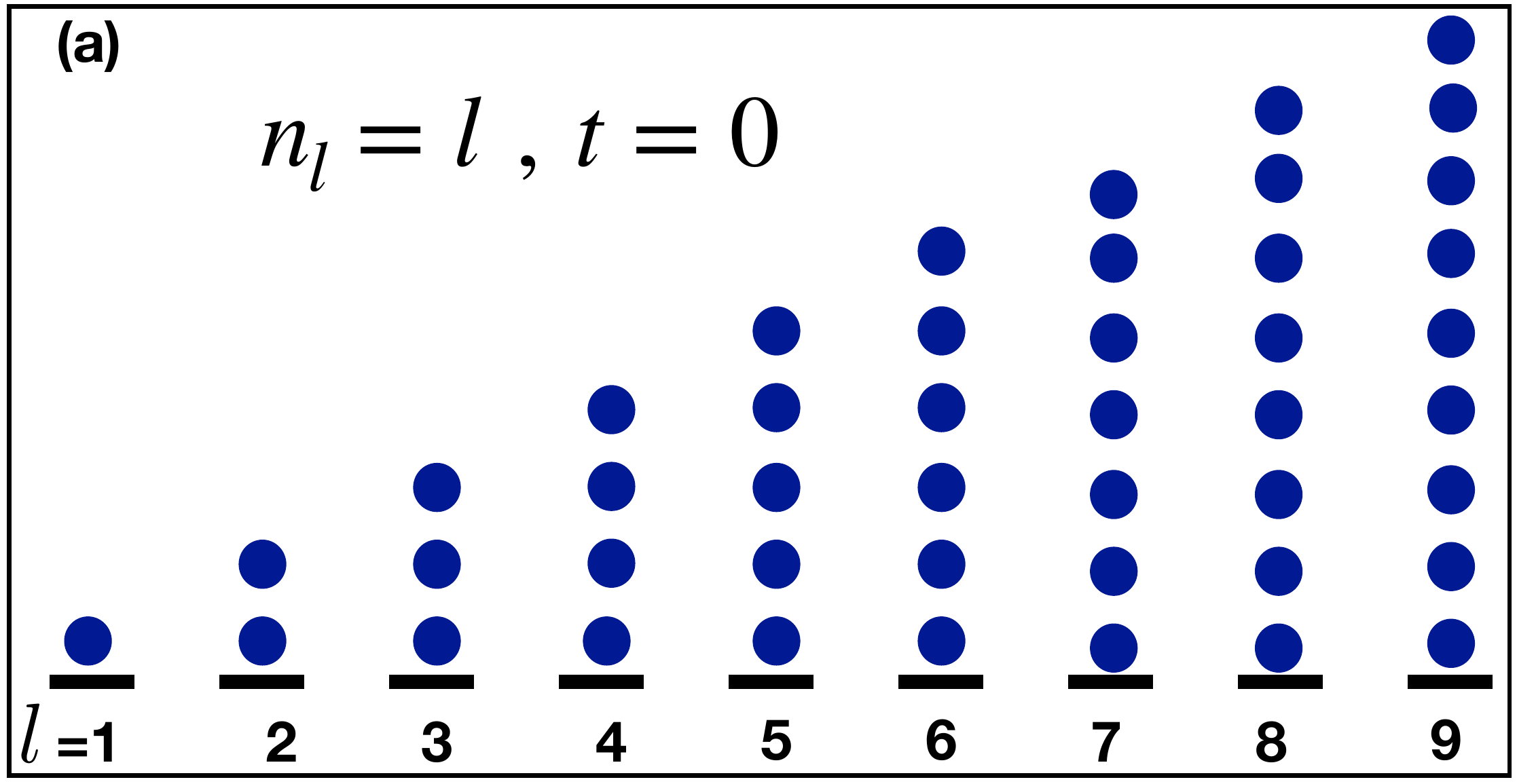}~
    \includegraphics[width=0.27\textwidth]{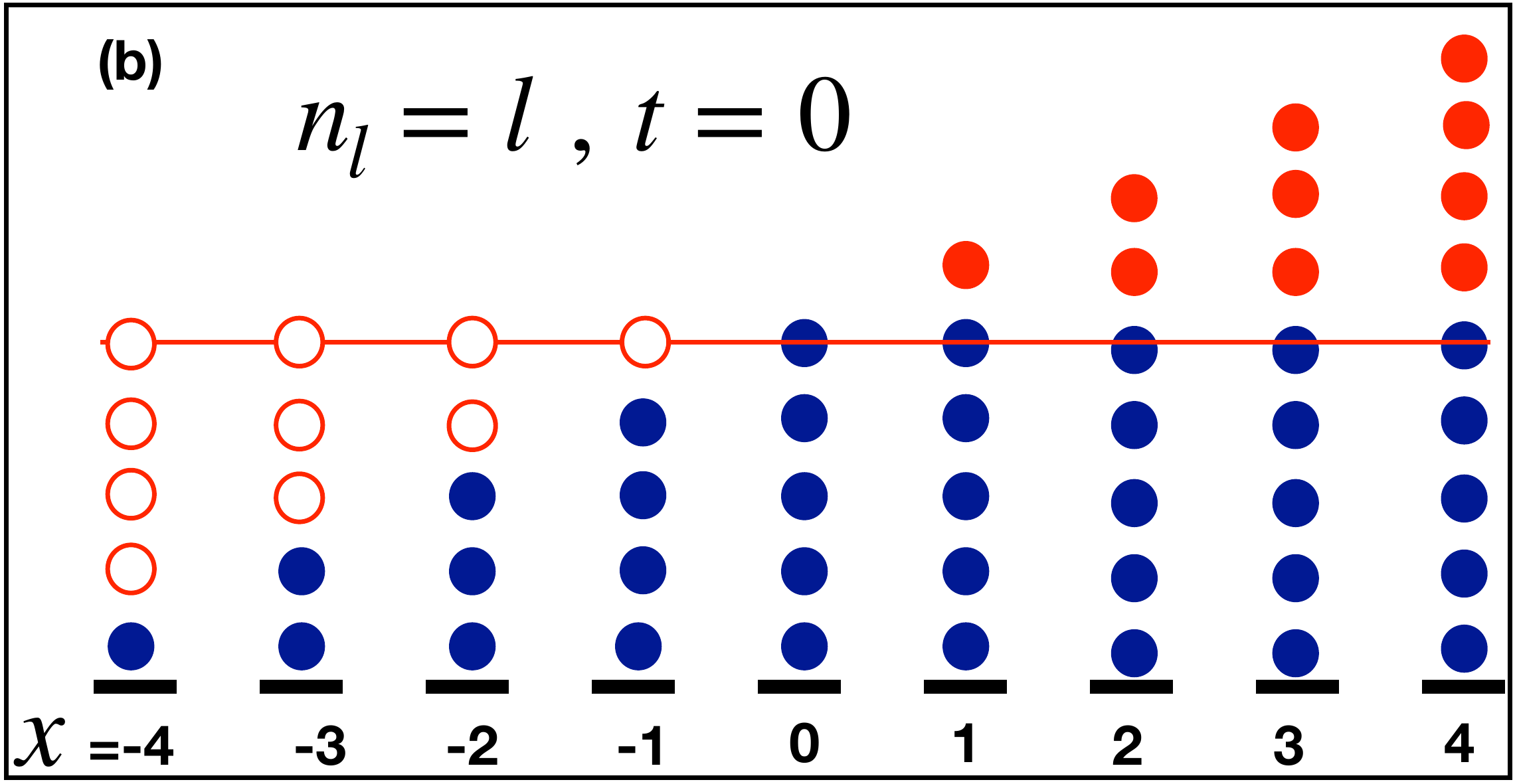}~~~
    \includegraphics[width=0.22\textwidth]{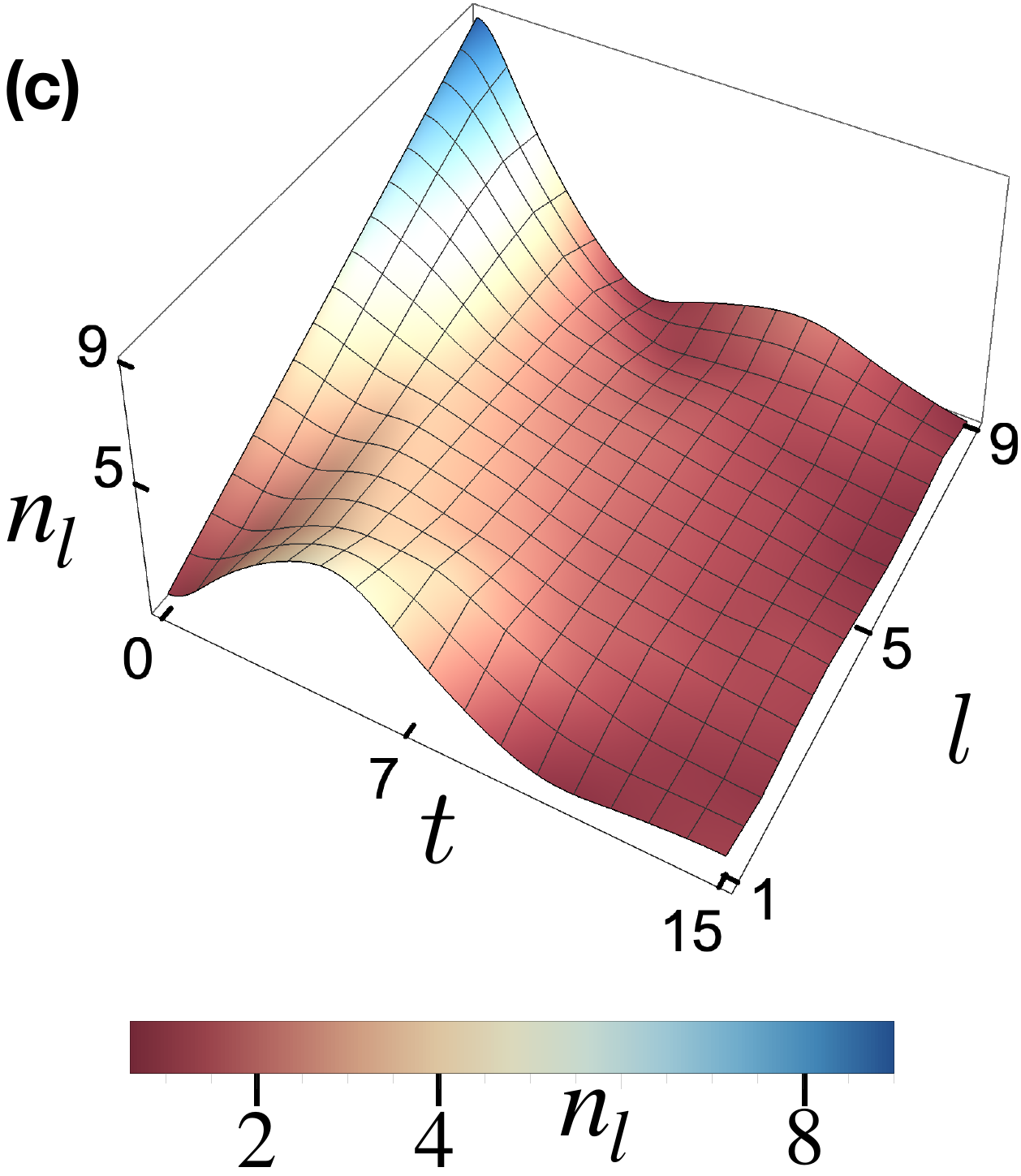}~~
    \includegraphics[width=0.22\textwidth]{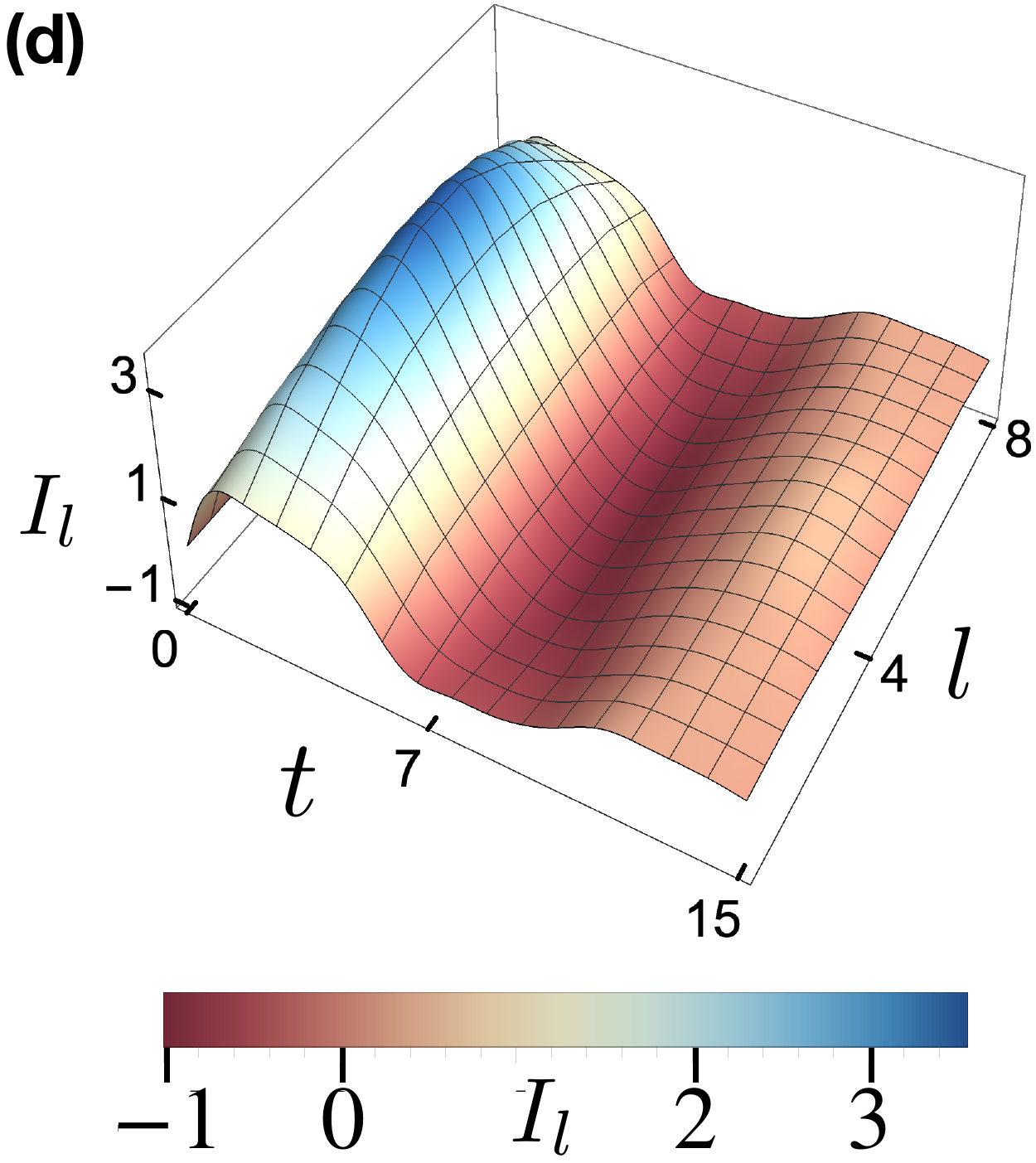}  
  \caption{Evolution of a linear chain of Bosons, starting from a Fock state. Each site $l$ is connected to a bath with temperature $T=g$ and chemical potential $\mu_l$, where $\mu_{l} = \mu_{1} + \nu (l-1)$ with $\mu_1 =-4.05 g$ and $ \nu = 0.75 g$. Here $g$ is the tunneling amplitude in the linear chain. (a) The initial Fock state in a $N=9$ site system where the $l^{th}$ site is occupied by $l$ particles. Each circle represents a particle. (b) The same Fock state with the origin shifted to the central site and local densities defined in terms of their deviations from the occupation of the central site, i.e $5$. The filled red circles indicate positive deviations while empty red circles indicate negative deviations. In terms of the deviations, the initial profile is anti-symmetric under reflection about the central site. (c) Color-plot of density, $n_l(t)$ and (d) current, $I_l(t)$ in the system as a function of site (link) number and time in under-damped regime with system bath coupling $\epsilon = 0.35 g$. The density profile executes a see-saw motion keeping the density of the central site almost constant at short times. The current shows a maximum at the center at short times. At long times, system settles to a density profile decreasing from left to right, governed by the chemical potential gradient in the baths. We use $g=1$ to set the unit of time, $t$ and $l$ is measured in units of lattice spacing.}
   \label{fig:diagonal}
   \end{figure*}
We now illustrate the potency of this formalism by studying the
dynamics of current and density profiles in Fermionic/ Bosonic OQS
initialized to specific $\hat{\rho}_{0S}$. We consider a system of
Bosons/Fermions hopping on a 1D lattice of $N$ sites with nearest 
neighbour tunneling amplitude $g$. Each site $l$ of the lattice is
coupled to the first site of a semi-infinite 1D Bosonic/Fermionic bath 
kept at fixed temperature $T_l$ and chemical potential $\mu_l$ with 
the same coupling strength $\epsilon$. The baths are modeled by a
hopping Hamiltonian with the hopping strength $t_B$. The total
Hamiltonian of the system ($H_{s}$) , the baths ($H_{b}$) and 
system bath interaction ($H_{sb}$) are then given by~\cite{Chakraborty1},
\begin{eqnarray}\label{model_Ham}
\nonumber H_{s}& =& -g\sum_{l=1}^{N}  a_{l}^{\dagger}a_{l+1} + h.c  ~~ and ~~ H_{sb} = \epsilon\sum_{l=1}^{N} a_{l}^{\dagger} b^{ (l)}_{1} + h.c \\
H_{b} &=& -t_B\sum_{l=1}^{N} \sum_{s=1}^{\infty}  b^{(l)\dagger }_{s} b^{ (l)}_{s+1}  +h.c. 
\end{eqnarray}
where $a_l$ is the annihilation operator (Bosons/Fermions) on the
$l^{th}$ site of the system, and $b^{(l)}_s$ is the annihilation
operator (Bosons/Fermions) of the $s^{th}$ site of 
the bath connected to $l^{th}$ site of the system. 
This kind of semi-infinite bath model yields a bath spectral 
function $\mathcal{J}(\omega)$, which has a square-root 
derivative singularity at the two band edges, $\omega = \pm 2t_B$,
\beq
 \mathcal{J}(\omega) = \Theta(4t_B^2-\omega^2)\frac{2}{t_{B}} \sqrt{1-\frac{\omega^2}{4t^2_{B}}}.
\label{Jomega_ourbath}
\eeq
This is a minimal model of non-Markovian dynamics of the OQS induced by non-analyticities in the bath spectral function~\cite{Chakraborty1}. The motivation for choosing this model is two-fold: (i) to show that our formalism can easily treat non-Markovian dynamics of OQS and (ii) this is an ideal case to study the effects of the initial condition, since the system retains memories over long timescales.
In this case [\onlinecite{Chakraborty1}], we have
\beq
\Sigma^B_{R}(\alpha,\beta,\omega )=\delta_{\alpha\beta} \left[ \frac{\epsilon^2\omega}{2t_B^2} - \mathbf{i}~\frac{\epsilon^2}{t_B}\left[1-\frac{(\omega+\mathbf{i} \eta )^2}{4t_B^2}\right]^{1/2}\right]\nonumber
\eeq
and hence the retarded Green's function is obtained to be,
\beq
{  G}_R^O(\alpha,\beta,\omega) =  (-1)^{\alpha+\beta} \frac{M_{\alpha-1}M_{N-\beta}}{g M_{N}} ~~~for~ \alpha<\beta  \nonumber
\eeq
and ${  G}_R^O(\alpha,\beta)={  G}_R^O(\beta,\alpha)$ for $\alpha>\beta$, where 
\beq
M_{\alpha}=\frac{\sinh[ (\alpha+1) \lambda]}{\sinh[\lambda]} ~~ with~~ \cosh \left[ \lambda \right]=\frac{1}{2g}[\omega-\Sigma^{R}(\omega)]. \nonumber
\eeq
From these analytical solutions, we obtain the retarded Green's
functions in time domain by performing the integral in
Eq.~\ref{eq:GOR}. Finally, the physical Keldysh Green's functions are
obtained by plugging ${  G}_R^O(\alpha,\beta,t-t')$ and 
\bqa
\nonumber \Sigma^B_{K}(\alpha,\beta,t-t')&=&-\mathbf{i}
\delta_{\alpha,\beta}~\epsilon^2/2\pi \int d\omega J(\omega)\\
\nonumber& &\left[ \coth
(\omega-\mu_\alpha)/2T_\alpha\right]^\zeta
exp[-\mathbf{i}\omega(t-t')]
\eqa
 back in eqn. \ref{gkoqsB}, where $\zeta=\pm 1$ for Bosons
 (Fermions).

The inherent non-Markovianness of the model is manifested as power law kernels, $\sim (t-t')^{-3/2}$ in $\Sigma^B_R(\alpha,\beta,t-t')$ and $\Sigma^B_K(\alpha,\beta,t-t')$. This leads to an initial exponential decay in ${  G}_R^O(\alpha,\beta,t-t')$, followed by a long time power law tail $\sim (t-t')^{-3/2}$, appearing at a time scale $\sim t_B/\epsilon^2$, which have been explored in great details in Ref [\onlinecite{Chakraborty1}]
. 
 
We first consider a linear chain of Bosons of $N=9$
  sites. The system is initialized in a Fock state where the first
  site has 1 particle, the second site has 2 particles .. the $l^{th}$
  site has $l$ particles, as shown in Fig. \ref{fig:diagonal}
  (a). This creates a positive density gradient from left to right in
  the initial state. We couple each site to a bath, with the chemical
  potential $\mu_{l} = \mu_{1} + \nu (l-1)$, keeping the temperature
  same for all baths. The chemical potential is set up in such a way
  that in the steady state, the system will have a positive density
  gradient from right to left, thus ensuring a non-trivial dynamics in
  this OQS. We choose the system bath coupling strength to be in the 
under-damped regime, i.e. $\epsilon/g = 0.35 <1$, so that we can 
study the interesting transient quantum dynamics of the OQS. 
The other parameters are chosen to be $t_B=2g,T_l=g,\mu_1 = -4.05g$ 
and $ \nu = 0.75 g$.

The time-dependent density at site $l$ and the current on the link between the sites $l$ and $l+1$ sites are given by, $n_l(t) =\zeta[ \mathbf{i} {\cal G}^O_{K\rho_0}(l,t,l,t)-1]/2$ and $I_l (t)= g ~Re[{\cal G}^O_{K\rho_0}(l,t,l+1,t)]$. 
The change in the density profile with time is plotted in
Fig. \ref{fig:diagonal}(c), while the change in current profile along the links of the
system is plotted in Fig. \ref{fig:diagonal}(d). At short times, we
find that the density at the central site, $\bar{n}$ does not change with time,
while the profile executes a see-saw type motion with the central site
as a fulcrum, i.e. the local density deviation from $\bar{n}$ increases in magnitude with distance from the central site and is
antisymmetric under reflection through this point. 
To understand the short time quantum dynamics of the system, it is enough to consider the dynamics of a closed system
with an odd ($2N+1$) number of sites (we will comment on the case of even
number of sites later). This description will be valid upto a time scale $\sim t_B/\epsilon^2$, when the effect of the bath starts to become prominent.
 In this case, it is useful to set the
origin at the central site, and denote the new co-ordinates by
$x$ ($-N \leq x \leq N$), so that the Hamiltonian has a reflection
symmetry about the origin ($x \rightarrow -x$). Further we consider the
deviation of the density from $\bar{n}$, $\delta n_x(t)$. The initial profile $\delta n_x(0)$ is antisymmetric under
reflection. This is shown in Fig. \ref{fig:diagonal}(c) in terms
of open (negative $\delta n_x(0)$)  and filled (positive $\delta
n_x(0)$) red circles.  Probability conservation implies that $\sum_y |G^R(x,t;y,0)|^2 =1$. Using this we get,
\beq
 \delta n_x(t) =\sum_y|G^R(x,t;y,0)|^2 \delta n_y(0)
\eeq
Here the retarded Green's function $G^R$ does not depend on the
initial conditions and exhibits the reflection symmetry of the
Hamiltonian, i.e. $G^R(x,t;y,0)=G^R(-x,t;-y,0)$, while $\delta
n_{-y}(0)=-\delta n_y(0)$. It is then easy to see that $\delta n_x(t)$
is antisymmetric under reflection, and hence $\delta n_x(t)$ is $0$
for the central site ($x=0$). This leads to a piling up of current in the middle at shown in figure \ref{fig:diagonal}(d). The maximum of the current at the center can be understood from the continuity equation $\partial n/\partial t \sim \nabla . \vec{j}$.We can get further insight for a large system, where the boundaries can be neglected. In this case, $|G^R(x,t,y,0)|^2$ is a function of $|x-y|$, and using the anti-symmetry of the initial profile, it can be shown that $\delta n_x(t) \sim x$, i.e it increases in magnitude linearly with the distance from the central site.
In presence of a series of baths
with a chemical potential gradient, the reflection symmetry is broken,
and at long times $\sim t_B/\epsilon^2$, the system gradually settles
down to a steady state behaviour. Note that for a system with even number
of sites, the reflection symmetry is about the center of a link. Sites
at the two ends of this central link will have a small but non-zero
change in density with mutually opposite signs at short times.

 \begin{figure*}[t]
   \centering
   \includegraphics[width=0.47\textwidth]{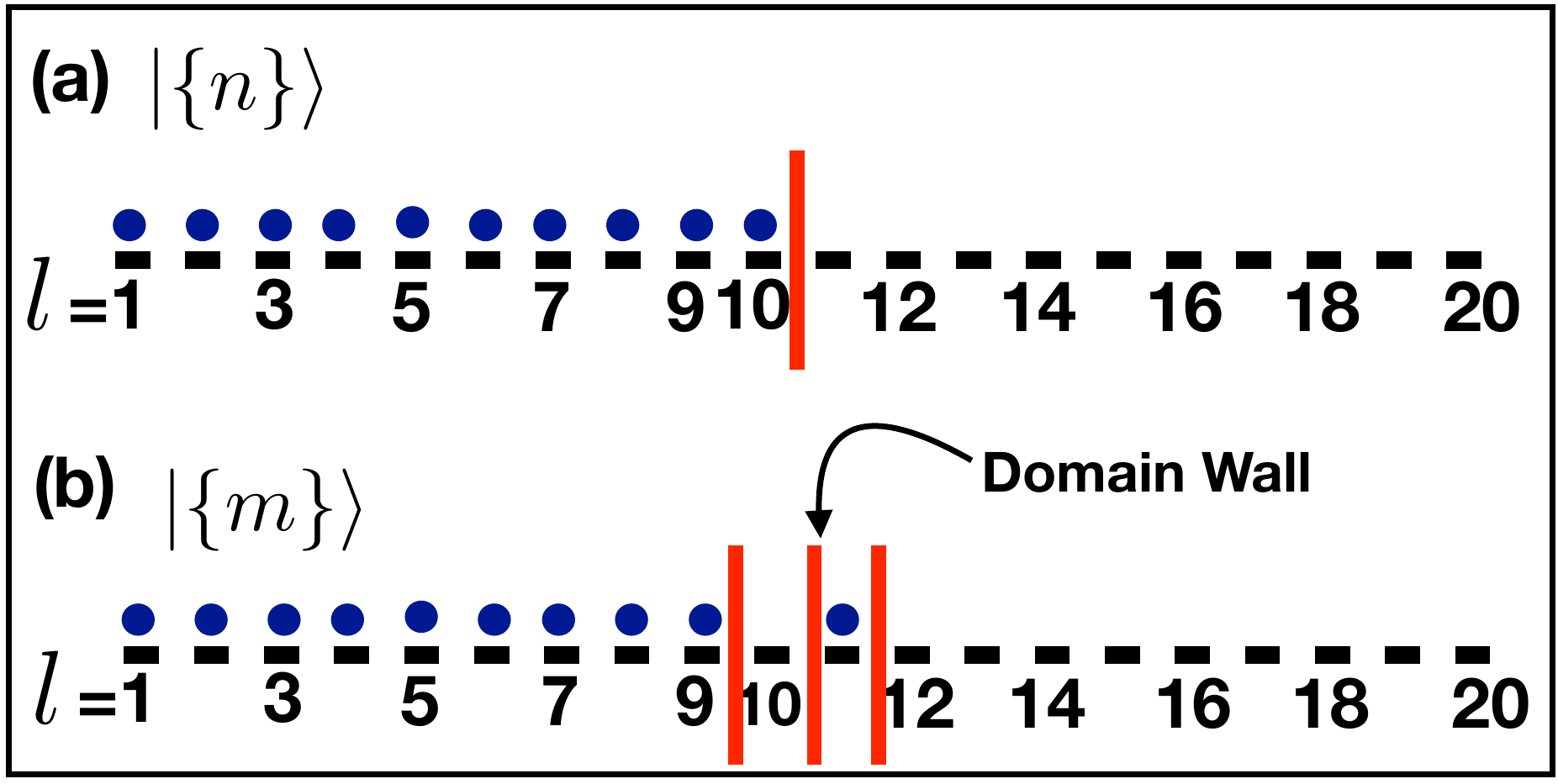}~~
   \includegraphics[width=0.22\textwidth]{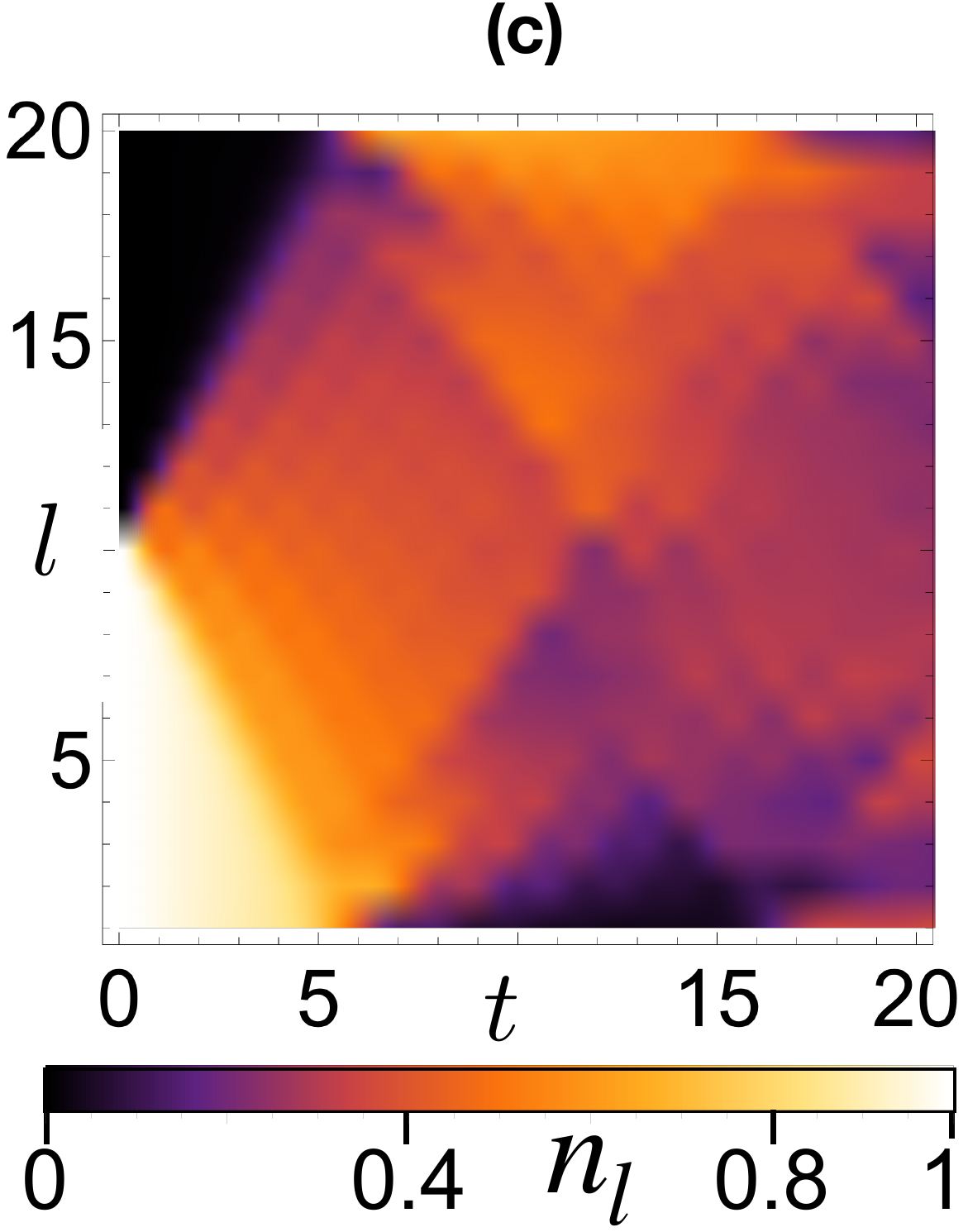}~~
  \includegraphics[width=0.25\textwidth]{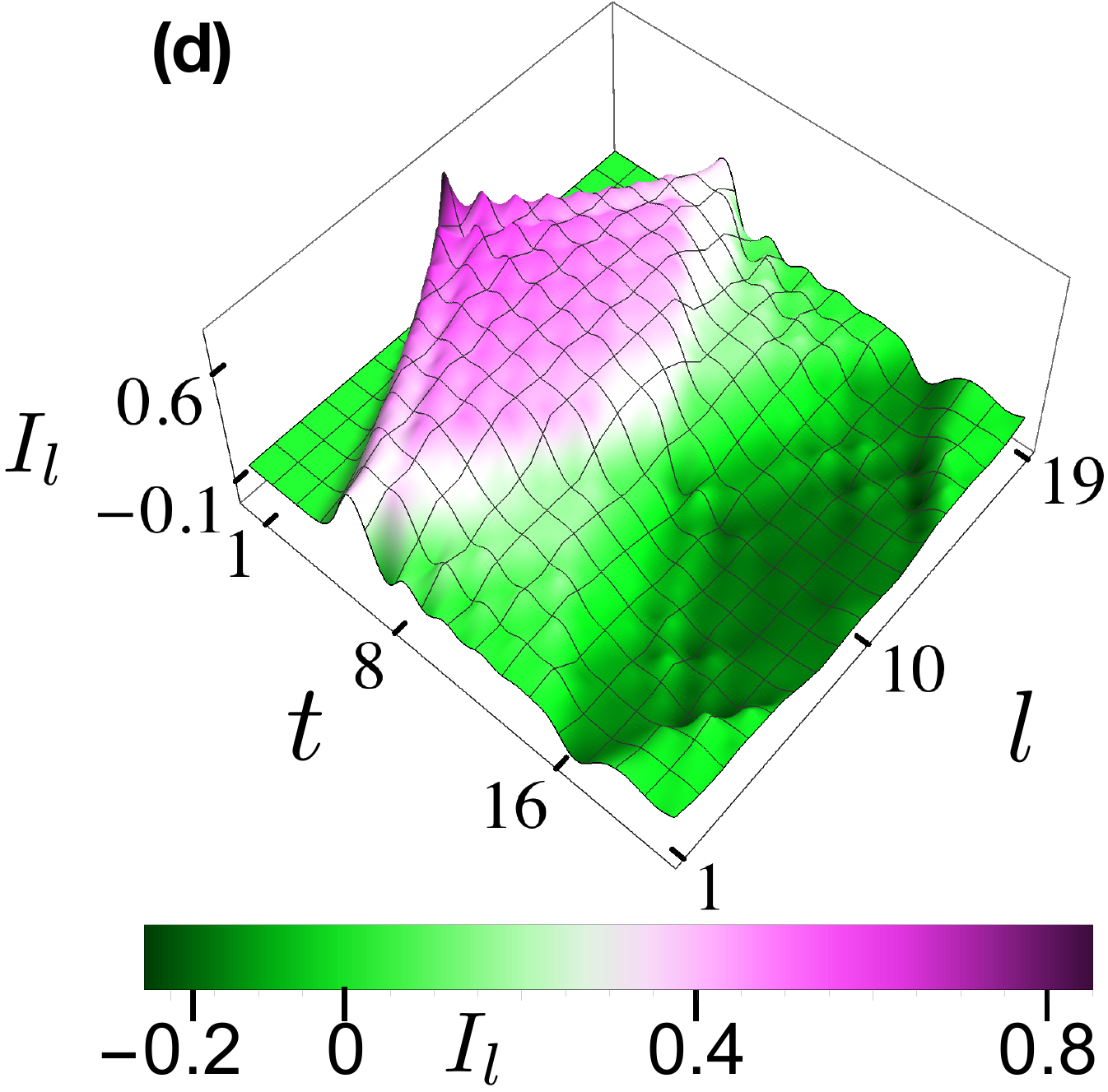}\\~\\
    \includegraphics[width=0.23\textwidth]{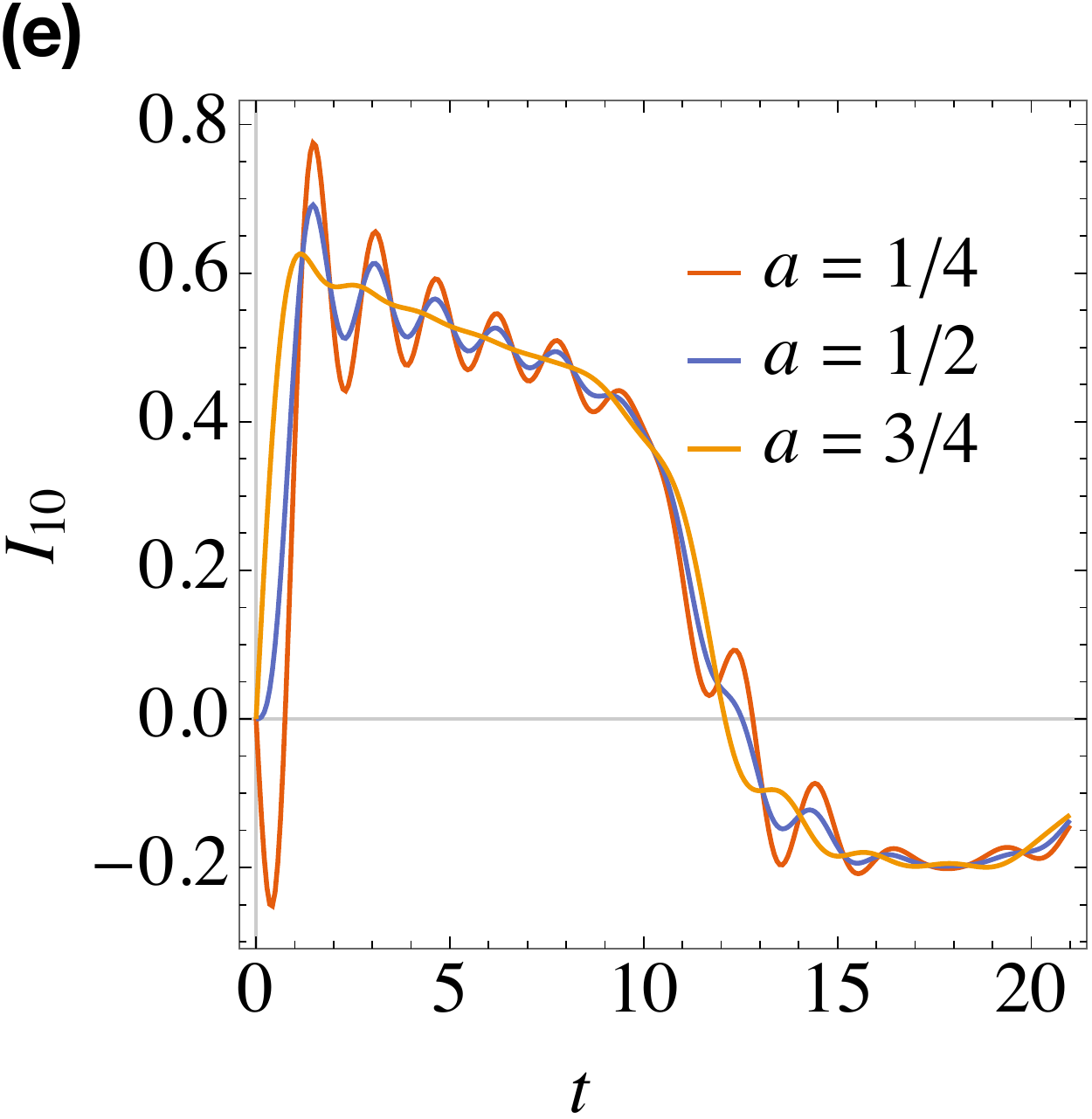}~~
    \includegraphics[width=0.23\textwidth]{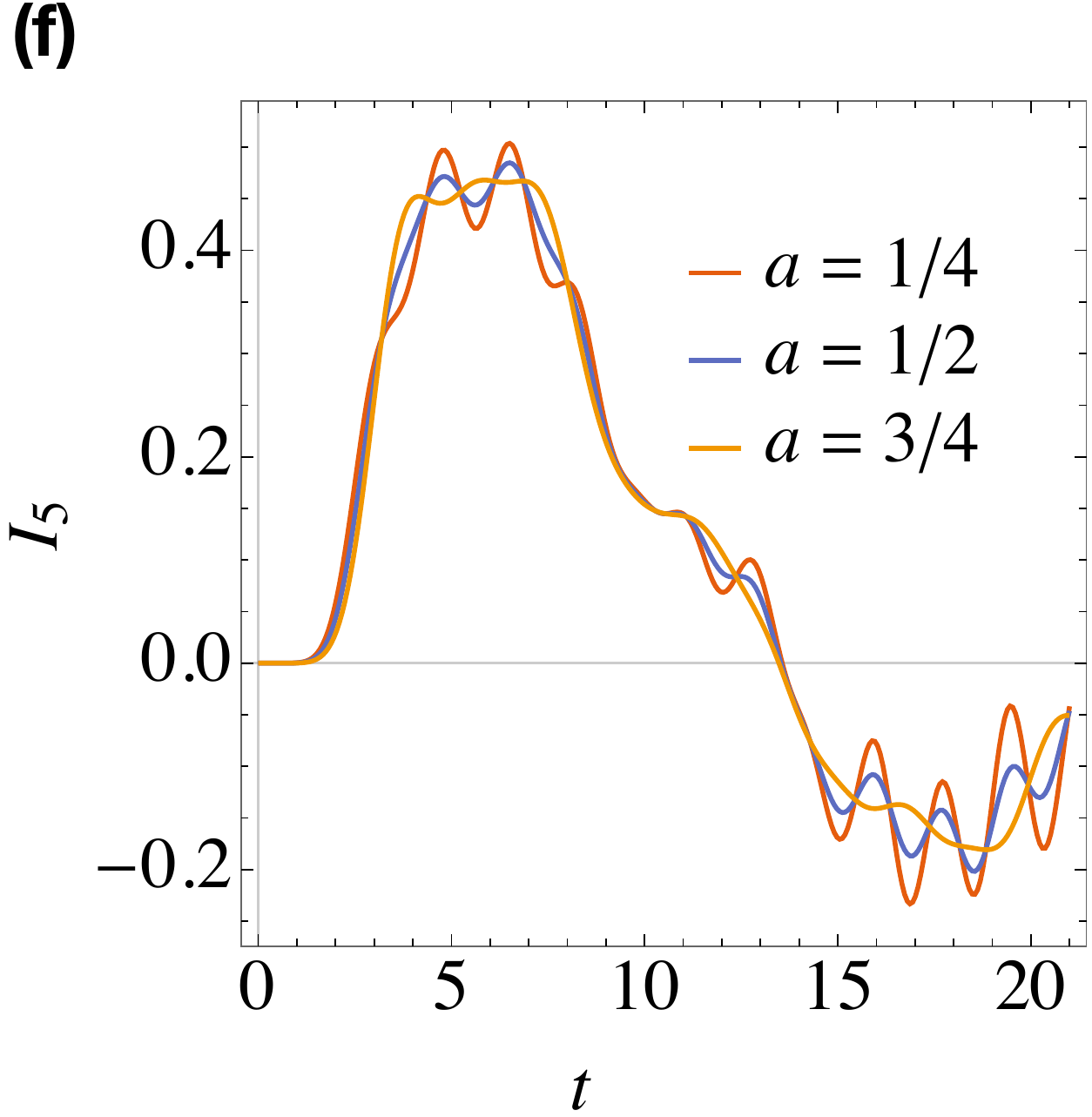}~~
    \includegraphics[width=0.23\textwidth]{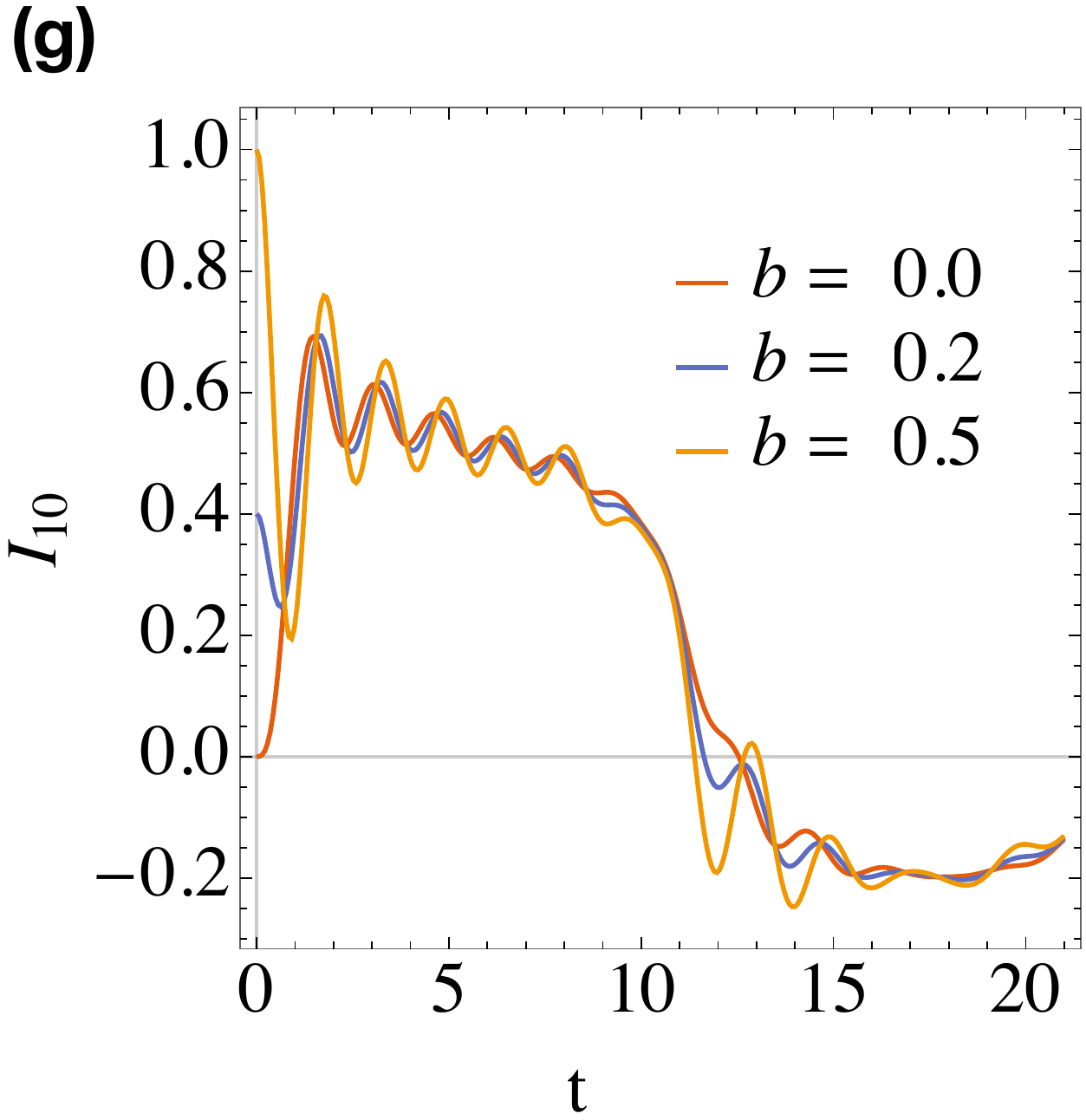}~~~~
    \includegraphics[width=0.23\textwidth]{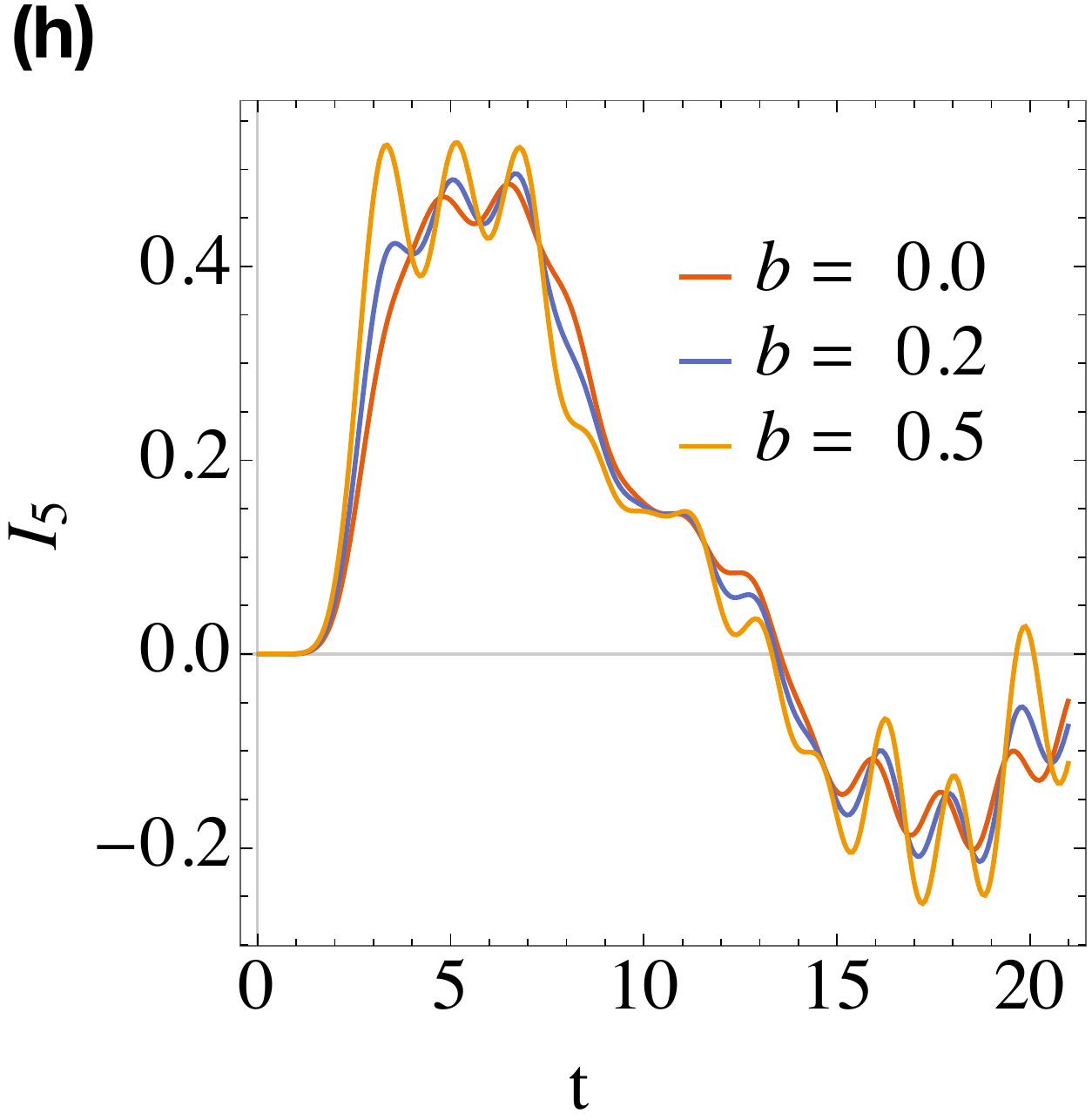}~~
 \caption{Dynamics of a Fermionic OQS starting from arbitrary initial condition. A $20$ site linear chain of spinless Fermions with nearest neighbour tunneling amplitude $g$ is coupled at each site to the baths at temperature $T_l=g$ and the chemical potential $\mu_l = -4.05 g$ through a coupling $\epsilon=0.2g$. (a) The initial state $| \{n \} \rangle$, with left half occupied, the right half empty and a domain wall at the center. (b) Another initial state $| \{m \} \rangle$, obtained from moving the rightmost particle in $| \{n \} \rangle$, one site to its right. Color plot of (c) for density and (d) for current as a function of site (link) number and time for a system starting with $| \{n \} \rangle \langle \{n \} |$. The diamonds are defined by ballistic motion of domain walls and their reflection from the edges. (e)-(h) Current as a function of time for off-diagonal $\hat{\rho}_{0S}=a\left|\left\{ n\right\} \right\rangle \left\langle \left\{ n\right\} \right|+\left(1-a\right)\left|\left\{ m\right\} \right\rangle \left\langle \left\{ m\right\} \right|+[\mathbf{i}b\left|\left\{ n\right\} \right\rangle \left\langle \left\{ m\right\} \right|+h.c.]$. (e) Current at the central link ($l=10$) for $b=0$ and $a=1/4,1/2,3/4$. (f) Same as (e) for a link far from the center ($l=5$). The initial oscillations at the central link are more pronounced for smaller $a$. (g) Current at the central link ($l=10$) and (h) current at the link far from the center $l=5$ for $\hat{\rho}_{0S}$ with $a=1/2$ and $b=0.0,0.2,0.5$. The initial current on the central link is controlled by the value of $b$. We use $g=1$ to set the unit of time, $t$ and $l$ is measured in units of lattice spacing.}
   \label{fig:off-diagonal}
   \end{figure*}
We next consider spinless Fermions hopping on a 1D lattice of $N=20$
sites. We first consider an initial Fock state, where the left half of
the lattice (sites $1$ to $10$) is occupied by particles, while the
right half of the system is empty, creating a domain wall in the
middle of the lattice, as shown in Fig \ref{fig:off-diagonal} (a). The
Fermionic bath parameters are fixed to $T_l = g$ and $\mu_l = -4.05 g$
and $\epsilon=0.2g$, i.e there is no inhomogeneity in the bath
parameters. At short times, the effect of the bath can be ignored and
the quantum dynamics can be understood by considering the domain wall
as a free particle. This particle splits coherently and
moves in either direction ballistically with a timescale $\sim g^{-1}$. The effect is seen both in the changes in the density
profile ( Fig. \ref{fig:off-diagonal} (c) ) and in the current profile
(Fig. \ref{fig:off-diagonal} (d)), which shows a sudden jump at a
site when the particle first passes through that site, creating the initial wedge
shaped profiles. The particle is coherently reflected back at the
boundary and rephases at a single point ~\cite{Preiss1229},
creating the diamond shape in the profile. Since the system is
underdamped, this cycle is repeated with associated sign change in the
current profile, as seen in Fig. \ref{fig:off-diagonal} (d). Beyond
the time scale $\sim t_B/\epsilon^2$, the presence of the bath governs
the dynamics; here the current goes to zero and the density profile
attains its steady uniform value dictated by the chemical potential in
the bath at long times, but the approach to the steady state is
governed by the power law of the non-Markovian bath.

We now consider the same Fermionic system initialized to a different density matrix. We consider 2 Fock states, with one state given by the domain wall profile shown in Fig. \ref{fig:off-diagonal} (a) (i.e. the initial state with the domain wall at the center). The second state is obtained from this state by hopping the particle at site $10$ to the site $11$, resulting in a configuration shown in Fig. \ref{fig:off-diagonal} (b). Let us call these states $| \left\{ n\right\} \rangle$ and $| \left\{ m\right\} \rangle$ respectively. We will consider a general $2 \times 2$ initial density matrix in the qubit space, spanned by these two states of the form, $\hat{\rho}_{0S}=a\left|\left\{ n\right\} \right\rangle \left\langle \left\{ n\right\} \right|+\left(1-a\right)\left|\left\{ m\right\} \right\rangle \left\langle \left\{ m\right\} \right|+[\mathbf{i}b\left|\left\{ n\right\} \right\rangle \left\langle \left\{ m\right\} \right|+h.c.]$. We note that the positivity of the eigenvalues of $\hat{\rho}_{0S}$ demands $\left|b\right|^{2}\le a\left(1-a\right)$. When $b$ is finite, the system has a non-zero current on the central link. We first consider the system with $b=0$, and plot the current on the central link as a function of time in Fig \ref{fig:off-diagonal} (e). In this case, the current is initially expected to rise for $a>1/2$, since there is more density at $l=10$ than at $l=11$, and to fall in value for $a<1/2$ (note that the current from right to left is considered to be positive in our notation). This is indeed observed as $a$ is varied from $1/4$ to $3/4$ in Fig. \ref{fig:off-diagonal} (e). We note that the amplitude of the oscillations of the current decreases with increases in $a$. In Fig. \ref{fig:off-diagonal} (f), we plot the current at a link $l=5$ far from the center. We see that the current rises after a finite time, as discussed in the previous case. We also find that changing $a$ from $1/2$ to $3/4$ causes minor changes in the current, i.e. the changes in the initial conditions mainly affect the dynamics in the center of the lattice. 

We now consider an off-diagonal $\hat{\rho}_{0S}$ with $a$ fixed to $1/2$, and change $b$ from $0$ to $1/2$. The current in the central link is plotted in Fig. \ref{fig:off-diagonal} (g), while the current in the link far away $(l=5)$ is plotted in Fig. \ref{fig:off-diagonal} (h). The key difference in seen in Fig. \ref{fig:off-diagonal} (g), where the current in the central link starts from a finite value, governed by $b$. The subsequent dynamics is almost independent of $b$ in all cases.

\section{Interacting Systems}\label{sec:interaction}

In the previous sections, we have built up a field theoretic formalism
to describe the dynamics of quantum many body systems starting from
arbitrary initial conditions. We have also extended this formalism to
the case of open quantum systems. However, till now, we have only looked at
non-interacting systems (quadratic or gaussian field theories), where
we can solve the problem exactly and the question of calculating a
correlator is reduced to evaluating one or a few integrals. In this
section we finally tackle the question of applying our
formalism to the dynamics of interacting quantum many body systems
starting from an arbitrary initial condition. 

In this case, we start by adding to the quadratic Keldysh action with
the initial bilinear source, $S(u)$, a term $S_{int}$,
representing the interaction between particles. We then consider the
field theory controlled by the action $S=S(u)+S_{int}$, and calculate
Green's functions $\hat{G}^{(n)}_{int}(u)$ in this
theory. $\hat{G}^{(n)}_{int}(u)$ has a diagrammatic expansion in terms of
the non-interacting Green's functions $\hat{G}(u)$ and the interaction
vertices of a standard SK field theory. The details of this
construction depends on the form of $S_{int}$, but the Feynman rules for
computing the diagrams are exactly similar to that of a SK field
theory, with $u$ dependent propagators $\hat{G}(u)$. 

The diagrammatic perturbation theory for the Green's functions work
well at short times, but one needs to resum the series or part of it
to all orders to obtain an accurate description of the long time
behaviour. This is a general characteristics of perturbation theories
and has nothing to do with arbitrary initial conditions. This is where
our formalism has an advantage: the standard resummation techniques
known in field theories apply to $G^{(n)}_{int}(u)$, while they do not apply
to the physical correlators  ${\cal G}^{(n)}_{int,\rho_0}= {\cal L}(\partial_u,\rho_0)
{\cal N}(u) G^{(n)}_{int}(u)\vert_{u=0}$. Focusing
on the one-particle Green's function, one can now write a Dyson
equation $\hat{G}_{int}(u)= [G^{-1}(u)-\Sigma(G(u))]^{-1}$, where the
irreducible self energy can be constructed diagrammatically in
perturbation theory. One can also use a skeleton expansion in terms of
$\Sigma(G_{int}(u))$, or resum a class of diagrams as in a RPA
expansion; in other words one can bring the full force of accumulated
knowledge of such approximation schemes to bear down on the problem of
calculating $\hat{G}_{int}(u)$. Similar constructions are possible for
higher order correlation functions in terms of higher order vertex
functions.

We will not go into any particular approximation in this paper since the
validity of different approximations are both model dependent and parameter
dependent. We will take this up in a future work. It may seem that
applying a large number of derivatives (equal to number of particles) through ${\cal L}$ will be a
daunting task in the case of thermodynamically large interacting systems, specially for
resumed approximations, where $G_{int}(u)$ may only be known
approximately, or even numerically. We will not provide a complete
solution to this problem here, but indicate a way forward. We will
consider the system to initially be in a single Fock state. The
generalization to arbitrary density matrices can be done suitably. For a
Fermionic system, starting in a Fock state $|\{ n\} \rangle$, where
the set of occupied modes are denoted by ${\cal A}$, we can write
\bqa
{\cal G}_{\rho_0}=  \prod_{\alpha \in {\cal A}}
[1+\partial_{u_\alpha}] G(u)\vert_{u=0} &=&G(0) +\sum_{\alpha \in {\cal
    A}}G^\alpha(0)  \nonumber \\
    &+&\frac{1}{2!} \sum_{\alpha \beta \in {\cal
    A},\alpha \neq\beta}G^{\alpha\beta}(0)+ ...\nonumber \\
\eqa
where $G^{ijk....n}= \partial_{u_i} ...\partial_{u_n} G(u)$. For a
Bosonic system, a similar derivative expansion can be written as 
\beq
{\cal G}_{\rho_0}=  \prod_{\alpha}
\sum_{m=0}^{n_\alpha}\frac{\partial^m_{u_\alpha}}{m!} G(u)\vert_{u=0} =G(0) +\sum_{\alpha \in {\cal
    A}}G^\alpha(0) + ...
\eeq
where ${\cal A}$ is the set of modes with at least $1$ particles.
For a thermodynamically large system, there are two possible
practical approximations to treat the derivative expansion: (i)
truncate the series or (ii) resum this series by assuming
factorization of correlation functions of higher order. We will not go
into the relative merits of these different approximation strategies,
and leave this as a topic of future studies on this subject. 

 \section{Conclusion}
In this paper, we have formulated a field theoretic description of dynamics of a quantum many body system (Bosons and Fermions) starting from an arbitrary initial density matrix. We have shown that the matrix element of the density matrix can be incorporated using a source which couples to the bilinears of the fields only at initial time, i.e by adding an impulse term to the original SK action. The Green's functions can be evaluated in this theory as a function of the addition source $\hat{u}$. The physical correlation functions can then be obtained by taking an appropriate set of derivatives of the Green's functions w.r.t the initial source and setting the sources to zero. The initial density matrix only governs the particular set of derivatives to be taken. Our formalism thus breaks up into two parts: (i) calculation of Green's functions in presence of a bilinear source, where the hierarchy of Green's functions satisfy Wick's theorem and the standard SK field theoretic techniques can applied to compute them, (ii) taking a particular set of derivatives, which depend on the initial conditions. We extend this formalism to open quantum systems and calculate evolution of density and current profile in Bosonic and Fermionic OQS. We calculate the exact expressions for physical one-particle and two-particle correlators in a non-interacting system and characterize the violation of Wick's theorem, relating it to the connected to particle correlations in the initial state. We have briefly sketched how our formalism can be extended to interacting systems. The biggest challenge that we have not addressed here are strategies to obtain reasonable approximation schemes which are controlled in particular limits. The issue of making conserving approximations which are valid at long times (i.e. no perturbation theory for physical correlators) is one of great importance which we hope to address in a future work.
\bibliographystyle{apsrev4-1} 
\bibliography{Keldysh_FockNI.bib}

\begin{thebibliography}{59}%
\makeatletter
\providecommand \@ifxundefined [1]{%
 \@ifx{#1\undefined}
}%
\providecommand \@ifnum [1]{%
 \ifnum #1\expandafter \@firstoftwo
 \else \expandafter \@secondoftwo
 \fi
}%
\providecommand \@ifx [1]{%
 \ifx #1\expandafter \@firstoftwo
 \else \expandafter \@secondoftwo
 \fi
}%
\providecommand \natexlab [1]{#1}%
\providecommand \enquote  [1]{``#1''}%
\providecommand \bibnamefont  [1]{#1}%
\providecommand \bibfnamefont [1]{#1}%
\providecommand \citenamefont [1]{#1}%
\providecommand \href@noop [0]{\@secondoftwo}%
\providecommand \href [0]{\begingroup \@sanitize@url \@href}%
\providecommand \@href[1]{\@@startlink{#1}\@@href}%
\providecommand \@@href[1]{\endgroup#1\@@endlink}%
\providecommand \@sanitize@url [0]{\catcode `\\12\catcode `\$12\catcode
  `\&12\catcode `\#12\catcode `\^12\catcode `\_12\catcode `\%12\relax}%
\providecommand \@@startlink[1]{}%
\providecommand \@@endlink[0]{}%
\providecommand \url  [0]{\begingroup\@sanitize@url \@url }%
\providecommand \@url [1]{\endgroup\@href {#1}{\urlprefix }}%
\providecommand \urlprefix  [0]{URL }%
\providecommand \Eprint [0]{\href }%
\providecommand \doibase [0]{http://dx.doi.org/}%
\providecommand \selectlanguage [0]{\@gobble}%
\providecommand \bibinfo  [0]{\@secondoftwo}%
\providecommand \bibfield  [0]{\@secondoftwo}%
\providecommand \translation [1]{[#1]}%
\providecommand \BibitemOpen [0]{}%
\providecommand \bibitemStop [0]{}%
\providecommand \bibitemNoStop [0]{.\EOS\space}%
\providecommand \EOS [0]{\spacefactor3000\relax}%
\providecommand \BibitemShut  [1]{\csname bibitem#1\endcsname}%
\let\auto@bib@innerbib\@empty
\bibitem [{\citenamefont {Eisert}\ \emph {et~al.}(2010)\citenamefont {Eisert},
  \citenamefont {Cramer},\ and\ \citenamefont {Plenio}}]{Entanglement_Review}%
  \BibitemOpen
  \bibfield  {author} {\bibinfo {author} {\bibfnamefont {J.}~\bibnamefont
  {Eisert}}, \bibinfo {author} {\bibfnamefont {M.}~\bibnamefont {Cramer}}, \
  and\ \bibinfo {author} {\bibfnamefont {M.~B.}\ \bibnamefont {Plenio}},\
  }\href {\doibase 10.1103/RevModPhys.82.277} {\bibfield  {journal} {\bibinfo
  {journal} {Rev. Mod. Phys.}\ }\textbf {\bibinfo {volume} {82}},\ \bibinfo
  {pages} {277} (\bibinfo {year} {2010})}\BibitemShut {NoStop}%
\bibitem [{\citenamefont {Breuer}\ and\ \citenamefont
  {Petruccione}(2002)}]{OQSBook}%
  \BibitemOpen
  \bibfield  {author} {\bibinfo {author} {\bibfnamefont {H.-P.}\ \bibnamefont
  {Breuer}}\ and\ \bibinfo {author} {\bibfnamefont {F.}~\bibnamefont
  {Petruccione}},\ }\href@noop {} {\emph {\bibinfo {title} {The theory of open
  quantum systems}}}\ (\bibinfo  {publisher} {Oxford University Press on
  Demand},\ \bibinfo {year} {2002})\BibitemShut {NoStop}%
\bibitem [{\citenamefont {Agarwal}(1969)}]{QME}%
  \BibitemOpen
  \bibfield  {author} {\bibinfo {author} {\bibfnamefont {G.~S.}\ \bibnamefont
  {Agarwal}},\ }\href {\doibase 10.1103/PhysRev.178.2025} {\bibfield  {journal}
  {\bibinfo  {journal} {Phys. Rev.}\ }\textbf {\bibinfo {volume} {178}},\
  \bibinfo {pages} {2025} (\bibinfo {year} {1969})}\BibitemShut {NoStop}%
\bibitem [{\citenamefont {Nakajima}(1958)}]{Nakajima}%
  \BibitemOpen
  \bibfield  {author} {\bibinfo {author} {\bibfnamefont {S.}~\bibnamefont
  {Nakajima}},\ }\href {\doibase 10.1143/PTP.20.948} {\bibfield  {journal}
  {\bibinfo  {journal} {Progress of Theoretical Physics}\ }\textbf {\bibinfo
  {volume} {20}},\ \bibinfo {pages} {948} (\bibinfo {year} {1958})}\BibitemShut
  {NoStop}%
\bibitem [{\citenamefont {Zwanzig}(1960)}]{Zwanzig}%
  \BibitemOpen
  \bibfield  {author} {\bibinfo {author} {\bibfnamefont {R.}~\bibnamefont
  {Zwanzig}},\ }\href {\doibase 10.1063/1.1731409} {\bibfield  {journal}
  {\bibinfo  {journal} {The Journal of Chemical Physics}\ }\textbf {\bibinfo
  {volume} {33}},\ \bibinfo {pages} {1338} (\bibinfo {year}
  {1960})}\BibitemShut {NoStop}%
\bibitem [{\citenamefont {de~Vega}\ and\ \citenamefont
  {Alonso}(2017)}]{Ines_review}%
  \BibitemOpen
  \bibfield  {author} {\bibinfo {author} {\bibfnamefont {I.}~\bibnamefont
  {de~Vega}}\ and\ \bibinfo {author} {\bibfnamefont {D.}~\bibnamefont
  {Alonso}},\ }\href {\doibase 10.1103/RevModPhys.89.015001} {\bibfield
  {journal} {\bibinfo  {journal} {Rev. Mod. Phys.}\ }\textbf {\bibinfo {volume}
  {89}},\ \bibinfo {pages} {015001} (\bibinfo {year} {2017})}\BibitemShut
  {NoStop}%
\bibitem [{\citenamefont {Zhang}\ \emph {et~al.}(2012)\citenamefont {Zhang},
  \citenamefont {Lo}, \citenamefont {Xiong}, \citenamefont {Tu},\ and\
  \citenamefont {Nori}}]{nori}%
  \BibitemOpen
  \bibfield  {author} {\bibinfo {author} {\bibfnamefont {W.-M.}\ \bibnamefont
  {Zhang}}, \bibinfo {author} {\bibfnamefont {P.-Y.}\ \bibnamefont {Lo}},
  \bibinfo {author} {\bibfnamefont {H.-N.}\ \bibnamefont {Xiong}}, \bibinfo
  {author} {\bibfnamefont {M.~W.-Y.}\ \bibnamefont {Tu}}, \ and\ \bibinfo
  {author} {\bibfnamefont {F.}~\bibnamefont {Nori}},\ }\href {\doibase
  10.1103/PhysRevLett.109.170402} {\bibfield  {journal} {\bibinfo  {journal}
  {Phys. Rev. Lett.}\ }\textbf {\bibinfo {volume} {109}},\ \bibinfo {pages}
  {170402} (\bibinfo {year} {2012})}\BibitemShut {NoStop}%
\bibitem [{\citenamefont {Chakraborty}\ and\ \citenamefont
  {Sensarma}(2018)}]{Chakraborty1}%
  \BibitemOpen
  \bibfield  {author} {\bibinfo {author} {\bibfnamefont {A.}~\bibnamefont
  {Chakraborty}}\ and\ \bibinfo {author} {\bibfnamefont {R.}~\bibnamefont
  {Sensarma}},\ }\href {\doibase 10.1103/PhysRevB.97.104306} {\bibfield
  {journal} {\bibinfo  {journal} {Phys. Rev. B}\ }\textbf {\bibinfo {volume}
  {97}},\ \bibinfo {pages} {104306} (\bibinfo {year} {2018})}\BibitemShut
  {NoStop}%
\bibitem [{\citenamefont {Schollwöck}(2011)}]{SCHOLLWOCK201196}%
  \BibitemOpen
  \bibfield  {author} {\bibinfo {author} {\bibfnamefont {U.}~\bibnamefont
  {Schollwöck}},\ }\href {\doibase https://doi.org/10.1016/j.aop.2010.09.012}
  {\bibfield  {journal} {\bibinfo  {journal} {Annals of Physics}\ }\textbf
  {\bibinfo {volume} {326}},\ \bibinfo {pages} {96 } (\bibinfo {year}
  {2011})},\ \bibinfo {note} {january 2011 Special Issue}\BibitemShut {NoStop}%
\bibitem [{\citenamefont {Eisert}(2013)}]{Eisert}%
  \BibitemOpen
  \bibfield  {author} {\bibinfo {author} {\bibfnamefont {J.}~\bibnamefont
  {Eisert}},\ }in\ \href@noop {} {\emph {\bibinfo {booktitle} {{Autumn School
  on Correlated Electrons: Emergent Phenomena in Correlated Matter Jülich,
  Germany, 23-27. September 2013}}}}\ (\bibinfo {year} {2013})\ \Eprint
  {http://arxiv.org/abs/1308.3318} {arXiv:1308.3318 [quant-ph]} \BibitemShut
  {NoStop}%
\bibitem [{\citenamefont {Evenbly}\ and\ \citenamefont
  {Vidal}(2015)}]{Tensor_Vidal}%
  \BibitemOpen
  \bibfield  {author} {\bibinfo {author} {\bibfnamefont {G.}~\bibnamefont
  {Evenbly}}\ and\ \bibinfo {author} {\bibfnamefont {G.}~\bibnamefont
  {Vidal}},\ }\href {\doibase 10.1103/PhysRevLett.115.180405} {\bibfield
  {journal} {\bibinfo  {journal} {Phys. Rev. Lett.}\ }\textbf {\bibinfo
  {volume} {115}},\ \bibinfo {pages} {180405} (\bibinfo {year}
  {2015})}\BibitemShut {NoStop}%
\bibitem [{\citenamefont {Stoudenmire}\ and\ \citenamefont
  {White}(2012)}]{2D-DMRG1}%
  \BibitemOpen
  \bibfield  {author} {\bibinfo {author} {\bibfnamefont {E.}~\bibnamefont
  {Stoudenmire}}\ and\ \bibinfo {author} {\bibfnamefont {S.~R.}\ \bibnamefont
  {White}},\ }\href {\doibase 10.1146/annurev-conmatphys-020911-125018}
  {\bibfield  {journal} {\bibinfo  {journal} {Annual Review of Condensed Matter
  Physics}\ }\textbf {\bibinfo {volume} {3}},\ \bibinfo {pages} {111} (\bibinfo
  {year} {2012})},\ \Eprint
  {http://arxiv.org/abs/https://doi.org/10.1146/annurev-conmatphys-020911-125018}
  {https://doi.org/10.1146/annurev-conmatphys-020911-125018} \BibitemShut
  {NoStop}%
\bibitem [{\citenamefont {Xiang}\ \emph {et~al.}(2001)\citenamefont {Xiang},
  \citenamefont {Lou},\ and\ \citenamefont {Su}}]{2D-DMRG2}%
  \BibitemOpen
  \bibfield  {author} {\bibinfo {author} {\bibfnamefont {T.}~\bibnamefont
  {Xiang}}, \bibinfo {author} {\bibfnamefont {J.}~\bibnamefont {Lou}}, \ and\
  \bibinfo {author} {\bibfnamefont {Z.}~\bibnamefont {Su}},\ }\href {\doibase
  10.1103/PhysRevB.64.104414} {\bibfield  {journal} {\bibinfo  {journal} {Phys.
  Rev. B}\ }\textbf {\bibinfo {volume} {64}},\ \bibinfo {pages} {104414}
  (\bibinfo {year} {2001})}\BibitemShut {NoStop}%
\bibitem [{\citenamefont {Altland}\ and\ \citenamefont
  {Simons}(2010)}]{matsubara}%
  \BibitemOpen
  \bibfield  {author} {\bibinfo {author} {\bibfnamefont {A.}~\bibnamefont
  {Altland}}\ and\ \bibinfo {author} {\bibfnamefont {B.~D.}\ \bibnamefont
  {Simons}},\ }\href@noop {} {\emph {\bibinfo {title} {Condensed matter field
  theory}}}\ (\bibinfo  {publisher} {Cambridge University Press},\ \bibinfo
  {year} {2010})\BibitemShut {NoStop}%
\bibitem [{\citenamefont {Keldysh}(1965)}]{Keldysh}%
  \BibitemOpen
  \bibfield  {author} {\bibinfo {author} {\bibfnamefont {L.}~\bibnamefont
  {Keldysh}},\ }\href {http://www.jetp.ac.ru/cgi-bin/dn/e_020_04_1018.pdf}
  {\bibfield  {journal} {\bibinfo  {journal} {JETP}\ }\textbf {\bibinfo
  {volume} {20}},\ \bibinfo {pages} {1018} (\bibinfo {year}
  {1965})}\BibitemShut {NoStop}%
\bibitem [{\citenamefont {Kamenev}\ and\ \citenamefont
  {Levchenko}(2009)}]{kamenev}%
  \BibitemOpen
  \bibfield  {author} {\bibinfo {author} {\bibfnamefont {A.}~\bibnamefont
  {Kamenev}}\ and\ \bibinfo {author} {\bibfnamefont {A.}~\bibnamefont
  {Levchenko}},\ }\href@noop {} {\bibfield  {journal} {\bibinfo  {journal}
  {Advances in Physics}\ }\textbf {\bibinfo {volume} {58}},\ \bibinfo {pages}
  {197} (\bibinfo {year} {2009})}\BibitemShut {NoStop}%
\bibitem [{\citenamefont {Rammer}(2007)}]{rammer_2007}%
  \BibitemOpen
  \bibfield  {author} {\bibinfo {author} {\bibfnamefont {J.}~\bibnamefont
  {Rammer}},\ }\href {\doibase 10.1017/CBO9780511618956} {\emph {\bibinfo
  {title} {Quantum Field Theory of Non-equilibrium States}}}\ (\bibinfo
  {publisher} {Cambridge University Press},\ \bibinfo {year}
  {2007})\BibitemShut {NoStop}%
\bibitem [{\citenamefont {Kamenev}(2011)}]{kamenevbook}%
  \BibitemOpen
  \bibfield  {author} {\bibinfo {author} {\bibfnamefont {A.}~\bibnamefont
  {Kamenev}},\ }\href@noop {} {\emph {\bibinfo {title} {Field theory of
  non-equilibrium systems}}}\ (\bibinfo  {publisher} {Cambridge University
  Press},\ \bibinfo {year} {2011})\BibitemShut {NoStop}%
\bibitem [{\citenamefont {Kadanoff}(1962)}]{baymbook}%
  \BibitemOpen
  \bibfield  {author} {\bibinfo {author} {\bibfnamefont {G.}~\bibnamefont
  {Kadanoff}, \bibfnamefont {Leo~P./Baym}},\ }\href@noop {} {\emph {\bibinfo
  {title} {Quantum Statistical Mechanics}}}\ (\bibinfo  {publisher} {Frontiers
  in Physics Series by W.A. Benjamin, Inc.},\ \bibinfo {year}
  {1962})\BibitemShut {NoStop}%
\bibitem [{\citenamefont {Aron}\ \emph {et~al.}(2018)\citenamefont {Aron},
  \citenamefont {Biroli},\ and\ \citenamefont {Cugliandolo}}]{SciPost_Aron}%
  \BibitemOpen
  \bibfield  {author} {\bibinfo {author} {\bibfnamefont {C.}~\bibnamefont
  {Aron}}, \bibinfo {author} {\bibfnamefont {G.}~\bibnamefont {Biroli}}, \ and\
  \bibinfo {author} {\bibfnamefont {L.~F.}\ \bibnamefont {Cugliandolo}},\
  }\href {\doibase 10.21468/SciPostPhys.4.1.008} {\bibfield  {journal}
  {\bibinfo  {journal} {SciPost Phys.}\ }\textbf {\bibinfo {volume} {4}},\
  \bibinfo {pages} {008} (\bibinfo {year} {2018})}\BibitemShut {NoStop}%
\bibitem [{\citenamefont {Jauho}\ \emph {et~al.}(1994)\citenamefont {Jauho},
  \citenamefont {Wingreen},\ and\ \citenamefont {Meir}}]{keldysh_meir_time}%
  \BibitemOpen
  \bibfield  {author} {\bibinfo {author} {\bibfnamefont {A.-P.}\ \bibnamefont
  {Jauho}}, \bibinfo {author} {\bibfnamefont {N.~S.}\ \bibnamefont {Wingreen}},
  \ and\ \bibinfo {author} {\bibfnamefont {Y.}~\bibnamefont {Meir}},\ }\href
  {\doibase 10.1103/PhysRevB.50.5528} {\bibfield  {journal} {\bibinfo
  {journal} {Phys. Rev. B}\ }\textbf {\bibinfo {volume} {50}},\ \bibinfo
  {pages} {5528} (\bibinfo {year} {1994})}\BibitemShut {NoStop}%
\bibitem [{\citenamefont {Nielsen}\ and\ \citenamefont
  {Chuang}(2010)}]{nielsen_chuang_2010}%
  \BibitemOpen
  \bibfield  {author} {\bibinfo {author} {\bibfnamefont {M.~A.}\ \bibnamefont
  {Nielsen}}\ and\ \bibinfo {author} {\bibfnamefont {I.~L.}\ \bibnamefont
  {Chuang}},\ }\href {\doibase 10.1017/CBO9780511976667} {\emph {\bibinfo
  {title} {Quantum Computation and Quantum Information: 10th Anniversary
  Edition}}}\ (\bibinfo  {publisher} {Cambridge University Press},\ \bibinfo
  {year} {2010})\BibitemShut {NoStop}%
\bibitem [{\citenamefont {Hoover}(1999)}]{Qcomp_book}%
  \BibitemOpen
  \bibfield  {author} {\bibinfo {author} {\bibfnamefont {W.~G.}\ \bibnamefont
  {Hoover}},\ }\href@noop {} {\emph {\bibinfo {title} {Time Reversibility,
  Computer Simulation, and Chaos}}},\ Vol.~\bibinfo {volume} {13}\ (\bibinfo
  {year} {1999})\BibitemShut {NoStop}%
\bibitem [{\citenamefont {Jensen}(1992)}]{chaos}%
  \BibitemOpen
  \bibfield  {author} {\bibinfo {author} {\bibfnamefont {R.~V.}\ \bibnamefont
  {Jensen}},\ }\href {http://dx.doi.org/10.1038/355311a0} {\bibfield  {journal}
  {\bibinfo  {journal} {Nature}\ }\textbf {\bibinfo {volume} {355}},\ \bibinfo
  {pages} {311 EP } (\bibinfo {year} {1992})}\BibitemShut {NoStop}%
\bibitem [{\citenamefont {Rigol}(2009)}]{Rigol_integrable}%
  \BibitemOpen
  \bibfield  {author} {\bibinfo {author} {\bibfnamefont {M.}~\bibnamefont
  {Rigol}},\ }\href {\doibase 10.1103/PhysRevLett.103.100403} {\bibfield
  {journal} {\bibinfo  {journal} {Phys. Rev. Lett.}\ }\textbf {\bibinfo
  {volume} {103}},\ \bibinfo {pages} {100403} (\bibinfo {year}
  {2009})}\BibitemShut {NoStop}%
\bibitem [{\citenamefont {Langen}\ \emph {et~al.}(2015)\citenamefont {Langen},
  \citenamefont {Erne}, \citenamefont {Geiger}, \citenamefont {Rauer},
  \citenamefont {Schweigler}, \citenamefont {Kuhnert}, \citenamefont
  {Rohringer}, \citenamefont {Mazets}, \citenamefont {Gasenzer},\ and\
  \citenamefont {Schmiedmayer}}]{Langen207}%
  \BibitemOpen
  \bibfield  {author} {\bibinfo {author} {\bibfnamefont {T.}~\bibnamefont
  {Langen}}, \bibinfo {author} {\bibfnamefont {S.}~\bibnamefont {Erne}},
  \bibinfo {author} {\bibfnamefont {R.}~\bibnamefont {Geiger}}, \bibinfo
  {author} {\bibfnamefont {B.}~\bibnamefont {Rauer}}, \bibinfo {author}
  {\bibfnamefont {T.}~\bibnamefont {Schweigler}}, \bibinfo {author}
  {\bibfnamefont {M.}~\bibnamefont {Kuhnert}}, \bibinfo {author} {\bibfnamefont
  {W.}~\bibnamefont {Rohringer}}, \bibinfo {author} {\bibfnamefont {I.~E.}\
  \bibnamefont {Mazets}}, \bibinfo {author} {\bibfnamefont {T.}~\bibnamefont
  {Gasenzer}}, \ and\ \bibinfo {author} {\bibfnamefont {J.}~\bibnamefont
  {Schmiedmayer}},\ }\href {\doibase 10.1126/science.1257026} {\bibfield
  {journal} {\bibinfo  {journal} {Science}\ }\textbf {\bibinfo {volume}
  {348}},\ \bibinfo {pages} {207} (\bibinfo {year} {2015})},\ \Eprint
  {http://arxiv.org/abs/http://science.sciencemag.org/content/348/6231/207.full.pdf}
  {http://science.sciencemag.org/content/348/6231/207.full.pdf} \BibitemShut
  {NoStop}%
\bibitem [{\citenamefont {Nandkishore}\ and\ \citenamefont
  {Huse}(2015)}]{Nandkishore_MBL}%
  \BibitemOpen
  \bibfield  {author} {\bibinfo {author} {\bibfnamefont {R.}~\bibnamefont
  {Nandkishore}}\ and\ \bibinfo {author} {\bibfnamefont {D.~A.}\ \bibnamefont
  {Huse}},\ }\href {\doibase 10.1146/annurev-conmatphys-031214-014726}
  {\bibfield  {journal} {\bibinfo  {journal} {Annual Review of Condensed Matter
  Physics}\ }\textbf {\bibinfo {volume} {6}},\ \bibinfo {pages} {15} (\bibinfo
  {year} {2015})},\ \Eprint
  {http://arxiv.org/abs/https://doi.org/10.1146/annurev-conmatphys-031214-014726}
  {https://doi.org/10.1146/annurev-conmatphys-031214-014726} \BibitemShut
  {NoStop}%
\bibitem [{\citenamefont {Schreiber}\ \emph {et~al.}(2015)\citenamefont
  {Schreiber}, \citenamefont {Hodgman}, \citenamefont {Bordia}, \citenamefont
  {L{\"u}schen}, \citenamefont {Fischer}, \citenamefont {Vosk}, \citenamefont
  {Altman}, \citenamefont {Schneider},\ and\ \citenamefont
  {Bloch}}]{Schreiber842}%
  \BibitemOpen
  \bibfield  {author} {\bibinfo {author} {\bibfnamefont {M.}~\bibnamefont
  {Schreiber}}, \bibinfo {author} {\bibfnamefont {S.~S.}\ \bibnamefont
  {Hodgman}}, \bibinfo {author} {\bibfnamefont {P.}~\bibnamefont {Bordia}},
  \bibinfo {author} {\bibfnamefont {H.~P.}\ \bibnamefont {L{\"u}schen}},
  \bibinfo {author} {\bibfnamefont {M.~H.}\ \bibnamefont {Fischer}}, \bibinfo
  {author} {\bibfnamefont {R.}~\bibnamefont {Vosk}}, \bibinfo {author}
  {\bibfnamefont {E.}~\bibnamefont {Altman}}, \bibinfo {author} {\bibfnamefont
  {U.}~\bibnamefont {Schneider}}, \ and\ \bibinfo {author} {\bibfnamefont
  {I.}~\bibnamefont {Bloch}},\ }\href {\doibase 10.1126/science.aaa7432}
  {\bibfield  {journal} {\bibinfo  {journal} {Science}\ }\textbf {\bibinfo
  {volume} {349}},\ \bibinfo {pages} {842} (\bibinfo {year} {2015})},\ \Eprint
  {http://arxiv.org/abs/http://science.sciencemag.org/content/349/6250/842.full.pdf}
  {http://science.sciencemag.org/content/349/6250/842.full.pdf} \BibitemShut
  {NoStop}%
\bibitem [{\citenamefont {Choi}\ \emph {et~al.}(2016)\citenamefont {Choi},
  \citenamefont {Hild}, \citenamefont {Zeiher}, \citenamefont {Schau{\ss}},
  \citenamefont {Rubio-Abadal}, \citenamefont {Yefsah}, \citenamefont
  {Khemani}, \citenamefont {Huse}, \citenamefont {Bloch},\ and\ \citenamefont
  {Gross}}]{Choi1547}%
  \BibitemOpen
  \bibfield  {author} {\bibinfo {author} {\bibfnamefont {J.-y.}\ \bibnamefont
  {Choi}}, \bibinfo {author} {\bibfnamefont {S.}~\bibnamefont {Hild}}, \bibinfo
  {author} {\bibfnamefont {J.}~\bibnamefont {Zeiher}}, \bibinfo {author}
  {\bibfnamefont {P.}~\bibnamefont {Schau{\ss}}}, \bibinfo {author}
  {\bibfnamefont {A.}~\bibnamefont {Rubio-Abadal}}, \bibinfo {author}
  {\bibfnamefont {T.}~\bibnamefont {Yefsah}}, \bibinfo {author} {\bibfnamefont
  {V.}~\bibnamefont {Khemani}}, \bibinfo {author} {\bibfnamefont {D.~A.}\
  \bibnamefont {Huse}}, \bibinfo {author} {\bibfnamefont {I.}~\bibnamefont
  {Bloch}}, \ and\ \bibinfo {author} {\bibfnamefont {C.}~\bibnamefont
  {Gross}},\ }\href {\doibase 10.1126/science.aaf8834} {\bibfield  {journal}
  {\bibinfo  {journal} {Science}\ }\textbf {\bibinfo {volume} {352}},\ \bibinfo
  {pages} {1547} (\bibinfo {year} {2016})},\ \Eprint
  {http://arxiv.org/abs/http://science.sciencemag.org/content/352/6293/1547.full.pdf}
  {http://science.sciencemag.org/content/352/6293/1547.full.pdf} \BibitemShut
  {NoStop}%
\bibitem [{\citenamefont {Basko}\ \emph {et~al.}(2006)\citenamefont {Basko},
  \citenamefont {Aleiner},\ and\ \citenamefont {Altshuler}}]{BASKO20061126}%
  \BibitemOpen
  \bibfield  {author} {\bibinfo {author} {\bibfnamefont {D.}~\bibnamefont
  {Basko}}, \bibinfo {author} {\bibfnamefont {I.}~\bibnamefont {Aleiner}}, \
  and\ \bibinfo {author} {\bibfnamefont {B.}~\bibnamefont {Altshuler}},\ }\href
  {\doibase https://doi.org/10.1016/j.aop.2005.11.014} {\bibfield  {journal}
  {\bibinfo  {journal} {Annals of Physics}\ }\textbf {\bibinfo {volume}
  {321}},\ \bibinfo {pages} {1126 } (\bibinfo {year} {2006})}\BibitemShut
  {NoStop}%
\bibitem [{\citenamefont {Labouvie}\ \emph {et~al.}(2016)\citenamefont
  {Labouvie}, \citenamefont {Santra}, \citenamefont {Heun},\ and\ \citenamefont
  {Ott}}]{Bodhaditya_bistable}%
  \BibitemOpen
  \bibfield  {author} {\bibinfo {author} {\bibfnamefont {R.}~\bibnamefont
  {Labouvie}}, \bibinfo {author} {\bibfnamefont {B.}~\bibnamefont {Santra}},
  \bibinfo {author} {\bibfnamefont {S.}~\bibnamefont {Heun}}, \ and\ \bibinfo
  {author} {\bibfnamefont {H.}~\bibnamefont {Ott}},\ }\href {\doibase
  10.1103/PhysRevLett.116.235302} {\bibfield  {journal} {\bibinfo  {journal}
  {Phys. Rev. Lett.}\ }\textbf {\bibinfo {volume} {116}},\ \bibinfo {pages}
  {235302} (\bibinfo {year} {2016})}\BibitemShut {NoStop}%
\bibitem [{\citenamefont {Yu}\ \emph {et~al.}(2016)\citenamefont {Yu},
  \citenamefont {Tang},\ and\ \citenamefont {Wang}}]{transient1}%
  \BibitemOpen
  \bibfield  {author} {\bibinfo {author} {\bibfnamefont {Z.}~\bibnamefont
  {Yu}}, \bibinfo {author} {\bibfnamefont {G.-M.}\ \bibnamefont {Tang}}, \ and\
  \bibinfo {author} {\bibfnamefont {J.}~\bibnamefont {Wang}},\ }\href {\doibase
  10.1103/PhysRevB.93.195419} {\bibfield  {journal} {\bibinfo  {journal} {Phys.
  Rev. B}\ }\textbf {\bibinfo {volume} {93}},\ \bibinfo {pages} {195419}
  (\bibinfo {year} {2016})}\BibitemShut {NoStop}%
\bibitem [{\citenamefont {My\"oh\"anen}\ \emph {et~al.}(2009)\citenamefont
  {My\"oh\"anen}, \citenamefont {Stan}, \citenamefont {Stefanucci},\ and\
  \citenamefont {van Leeuwen}}]{transient2}%
  \BibitemOpen
  \bibfield  {author} {\bibinfo {author} {\bibfnamefont {P.}~\bibnamefont
  {My\"oh\"anen}}, \bibinfo {author} {\bibfnamefont {A.}~\bibnamefont {Stan}},
  \bibinfo {author} {\bibfnamefont {G.}~\bibnamefont {Stefanucci}}, \ and\
  \bibinfo {author} {\bibfnamefont {R.}~\bibnamefont {van Leeuwen}},\ }\href
  {\doibase 10.1103/PhysRevB.80.115107} {\bibfield  {journal} {\bibinfo
  {journal} {Phys. Rev. B}\ }\textbf {\bibinfo {volume} {80}},\ \bibinfo
  {pages} {115107} (\bibinfo {year} {2009})}\BibitemShut {NoStop}%
\bibitem [{\citenamefont {Jin}\ \emph {et~al.}(2010)\citenamefont {Jin},
  \citenamefont {Tu}, \citenamefont {Zhang},\ and\ \citenamefont
  {Yan}}]{transient3}%
  \BibitemOpen
  \bibfield  {author} {\bibinfo {author} {\bibfnamefont {J.}~\bibnamefont
  {Jin}}, \bibinfo {author} {\bibfnamefont {M.~W.-Y.}\ \bibnamefont {Tu}},
  \bibinfo {author} {\bibfnamefont {W.-M.}\ \bibnamefont {Zhang}}, \ and\
  \bibinfo {author} {\bibfnamefont {Y.}~\bibnamefont {Yan}},\ }\href
  {http://stacks.iop.org/1367-2630/12/i=8/a=083013} {\bibfield  {journal}
  {\bibinfo  {journal} {New Journal of Physics}\ }\textbf {\bibinfo {volume}
  {12}},\ \bibinfo {pages} {083013} (\bibinfo {year} {2010})}\BibitemShut
  {NoStop}%
\bibitem [{\citenamefont {Zhou}\ \emph {et~al.}(2016)\citenamefont {Zhou},
  \citenamefont {Chen},\ and\ \citenamefont {Guo}}]{transient4}%
  \BibitemOpen
  \bibfield  {author} {\bibinfo {author} {\bibfnamefont {C.}~\bibnamefont
  {Zhou}}, \bibinfo {author} {\bibfnamefont {X.}~\bibnamefont {Chen}}, \ and\
  \bibinfo {author} {\bibfnamefont {H.}~\bibnamefont {Guo}},\ }\href {\doibase
  10.1103/PhysRevB.94.075426} {\bibfield  {journal} {\bibinfo  {journal} {Phys.
  Rev. B}\ }\textbf {\bibinfo {volume} {94}},\ \bibinfo {pages} {075426}
  (\bibinfo {year} {2016})}\BibitemShut {NoStop}%
\bibitem [{\citenamefont {Myöhänen}\ \emph {et~al.}(2008)\citenamefont
  {Myöhänen}, \citenamefont {Stan}, \citenamefont {Stefanucci},\ and\
  \citenamefont {van Leeuwen}}]{transient5}%
  \BibitemOpen
  \bibfield  {author} {\bibinfo {author} {\bibfnamefont {P.}~\bibnamefont
  {Myöhänen}}, \bibinfo {author} {\bibfnamefont {A.}~\bibnamefont {Stan}},
  \bibinfo {author} {\bibfnamefont {G.}~\bibnamefont {Stefanucci}}, \ and\
  \bibinfo {author} {\bibfnamefont {R.}~\bibnamefont {van Leeuwen}},\ }\href
  {http://stacks.iop.org/0295-5075/84/i=6/a=67001} {\bibfield  {journal}
  {\bibinfo  {journal} {EPL (Europhysics Letters)}\ }\textbf {\bibinfo {volume}
  {84}},\ \bibinfo {pages} {67001} (\bibinfo {year} {2008})}\BibitemShut
  {NoStop}%
\bibitem [{\citenamefont {Maciejko}\ \emph {et~al.}(2006)\citenamefont
  {Maciejko}, \citenamefont {Wang},\ and\ \citenamefont {Guo}}]{transient6}%
  \BibitemOpen
  \bibfield  {author} {\bibinfo {author} {\bibfnamefont {J.}~\bibnamefont
  {Maciejko}}, \bibinfo {author} {\bibfnamefont {J.}~\bibnamefont {Wang}}, \
  and\ \bibinfo {author} {\bibfnamefont {H.}~\bibnamefont {Guo}},\ }\href
  {\doibase 10.1103/PhysRevB.74.085324} {\bibfield  {journal} {\bibinfo
  {journal} {Phys. Rev. B}\ }\textbf {\bibinfo {volume} {74}},\ \bibinfo
  {pages} {085324} (\bibinfo {year} {2006})}\BibitemShut {NoStop}%
\bibitem [{\citenamefont {Yang}\ and\ \citenamefont
  {Zhang}(2016)}]{transient7}%
  \BibitemOpen
  \bibfield  {author} {\bibinfo {author} {\bibfnamefont {P.-Y.}\ \bibnamefont
  {Yang}}\ and\ \bibinfo {author} {\bibfnamefont {W.-M.}\ \bibnamefont
  {Zhang}},\ }\href {\doibase 10.1007/s11467-016-0640-z} {\bibfield  {journal}
  {\bibinfo  {journal} {Frontiers of Physics}\ }\textbf {\bibinfo {volume}
  {12}},\ \bibinfo {pages} {127204} (\bibinfo {year} {2016})}\BibitemShut
  {NoStop}%
\bibitem [{\citenamefont {Karlsson}\ \emph {et~al.}(2018)\citenamefont
  {Karlsson}, \citenamefont {van Leeuwen}, \citenamefont {Perfetto},\ and\
  \citenamefont {Stefanucci}}]{transient8}%
  \BibitemOpen
  \bibfield  {author} {\bibinfo {author} {\bibfnamefont {D.}~\bibnamefont
  {Karlsson}}, \bibinfo {author} {\bibfnamefont {R.}~\bibnamefont {van
  Leeuwen}}, \bibinfo {author} {\bibfnamefont {E.}~\bibnamefont {Perfetto}}, \
  and\ \bibinfo {author} {\bibfnamefont {G.}~\bibnamefont {Stefanucci}},\
  }\href {\doibase 10.1103/PhysRevB.98.115148} {\bibfield  {journal} {\bibinfo
  {journal} {Phys. Rev. B}\ }\textbf {\bibinfo {volume} {98}},\ \bibinfo
  {pages} {115148} (\bibinfo {year} {2018})}\BibitemShut {NoStop}%
\bibitem [{\citenamefont {Esposito}\ \emph {et~al.}(2009)\citenamefont
  {Esposito}, \citenamefont {Harbola},\ and\ \citenamefont
  {Mukamel}}]{Esposito}%
  \BibitemOpen
  \bibfield  {author} {\bibinfo {author} {\bibfnamefont {M.}~\bibnamefont
  {Esposito}}, \bibinfo {author} {\bibfnamefont {U.}~\bibnamefont {Harbola}}, \
  and\ \bibinfo {author} {\bibfnamefont {S.}~\bibnamefont {Mukamel}},\ }\href
  {\doibase 10.1103/RevModPhys.81.1665} {\bibfield  {journal} {\bibinfo
  {journal} {Rev. Mod. Phys.}\ }\textbf {\bibinfo {volume} {81}},\ \bibinfo
  {pages} {1665} (\bibinfo {year} {2009})}\BibitemShut {NoStop}%
\bibitem [{\citenamefont {Tang}\ and\ \citenamefont {Wang}(2014)}]{Wang}%
  \BibitemOpen
  \bibfield  {author} {\bibinfo {author} {\bibfnamefont {G.-M.}\ \bibnamefont
  {Tang}}\ and\ \bibinfo {author} {\bibfnamefont {J.}~\bibnamefont {Wang}},\
  }\href {\doibase 10.1103/PhysRevB.90.195422} {\bibfield  {journal} {\bibinfo
  {journal} {Phys. Rev. B}\ }\textbf {\bibinfo {volume} {90}},\ \bibinfo
  {pages} {195422} (\bibinfo {year} {2014})}\BibitemShut {NoStop}%
\bibitem [{\citenamefont {Tang}\ \emph {et~al.}(2014)\citenamefont {Tang},
  \citenamefont {Xu},\ and\ \citenamefont {Wang}}]{Wang2}%
  \BibitemOpen
  \bibfield  {author} {\bibinfo {author} {\bibfnamefont {G.-M.}\ \bibnamefont
  {Tang}}, \bibinfo {author} {\bibfnamefont {F.}~\bibnamefont {Xu}}, \ and\
  \bibinfo {author} {\bibfnamefont {J.}~\bibnamefont {Wang}},\ }\href {\doibase
  10.1103/PhysRevB.89.205310} {\bibfield  {journal} {\bibinfo  {journal} {Phys.
  Rev. B}\ }\textbf {\bibinfo {volume} {89}},\ \bibinfo {pages} {205310}
  (\bibinfo {year} {2014})}\BibitemShut {NoStop}%
\bibitem [{\citenamefont {Cugliandolo}\ \emph {et~al.}(2006)\citenamefont
  {Cugliandolo}, \citenamefont {Giamarchi},\ and\ \citenamefont
  {Doussal}}]{aging1}%
  \BibitemOpen
  \bibfield  {author} {\bibinfo {author} {\bibfnamefont {L.~F.}\ \bibnamefont
  {Cugliandolo}}, \bibinfo {author} {\bibfnamefont {T.}~\bibnamefont
  {Giamarchi}}, \ and\ \bibinfo {author} {\bibfnamefont {P.~L.}\ \bibnamefont
  {Doussal}},\ }\href {\doibase 10.1103/PhysRevLett.96.217203} {\bibfield
  {journal} {\bibinfo  {journal} {Phys. Rev. Lett.}\ }\textbf {\bibinfo
  {volume} {96}},\ \bibinfo {pages} {217203} (\bibinfo {year}
  {2006})}\BibitemShut {NoStop}%
\bibitem [{\citenamefont {Kennett}\ \emph {et~al.}(2001)\citenamefont
  {Kennett}, \citenamefont {Chamon},\ and\ \citenamefont {Ye}}]{aging2}%
  \BibitemOpen
  \bibfield  {author} {\bibinfo {author} {\bibfnamefont {M.~P.}\ \bibnamefont
  {Kennett}}, \bibinfo {author} {\bibfnamefont {C.}~\bibnamefont {Chamon}}, \
  and\ \bibinfo {author} {\bibfnamefont {J.}~\bibnamefont {Ye}},\ }\href
  {\doibase 10.1103/PhysRevB.64.224408} {\bibfield  {journal} {\bibinfo
  {journal} {Phys. Rev. B}\ }\textbf {\bibinfo {volume} {64}},\ \bibinfo
  {pages} {224408} (\bibinfo {year} {2001})}\BibitemShut {NoStop}%
\bibitem [{\citenamefont {Halimeh}\ \emph {et~al.}(2018)\citenamefont
  {Halimeh}, \citenamefont {Punk},\ and\ \citenamefont {Piazza}}]{aging}%
  \BibitemOpen
  \bibfield  {author} {\bibinfo {author} {\bibfnamefont {J.~C.}\ \bibnamefont
  {Halimeh}}, \bibinfo {author} {\bibfnamefont {M.}~\bibnamefont {Punk}}, \
  and\ \bibinfo {author} {\bibfnamefont {F.}~\bibnamefont {Piazza}},\ }\href
  {\doibase 10.1103/PhysRevB.98.045111} {\bibfield  {journal} {\bibinfo
  {journal} {Phys. Rev. B}\ }\textbf {\bibinfo {volume} {98}},\ \bibinfo
  {pages} {045111} (\bibinfo {year} {2018})}\BibitemShut {NoStop}%
\bibitem [{\citenamefont {Baym}\ and\ \citenamefont
  {Kadanoff}(1961)}]{kadanoff}%
  \BibitemOpen
  \bibfield  {author} {\bibinfo {author} {\bibfnamefont {G.}~\bibnamefont
  {Baym}}\ and\ \bibinfo {author} {\bibfnamefont {L.~P.}\ \bibnamefont
  {Kadanoff}},\ }\href {\doibase 10.1103/PhysRev.124.287} {\bibfield  {journal}
  {\bibinfo  {journal} {Phys. Rev.}\ }\textbf {\bibinfo {volume} {124}},\
  \bibinfo {pages} {287} (\bibinfo {year} {1961})}\BibitemShut {NoStop}%
\bibitem [{\citenamefont {Semkat}\ \emph {et~al.}(1999)\citenamefont {Semkat},
  \citenamefont {Kremp},\ and\ \citenamefont {Bonitz}}]{Bonitz1}%
  \BibitemOpen
  \bibfield  {author} {\bibinfo {author} {\bibfnamefont {D.}~\bibnamefont
  {Semkat}}, \bibinfo {author} {\bibfnamefont {D.}~\bibnamefont {Kremp}}, \
  and\ \bibinfo {author} {\bibfnamefont {M.}~\bibnamefont {Bonitz}},\ }\href
  {\doibase 10.1103/PhysRevE.59.1557} {\bibfield  {journal} {\bibinfo
  {journal} {Phys. Rev. E}\ }\textbf {\bibinfo {volume} {59}},\ \bibinfo
  {pages} {1557} (\bibinfo {year} {1999})}\BibitemShut {NoStop}%
\bibitem [{\citenamefont {Semkat}\ \emph {et~al.}(2000)\citenamefont {Semkat},
  \citenamefont {Kremp},\ and\ \citenamefont {Bonitz}}]{Bonitz2}%
  \BibitemOpen
  \bibfield  {author} {\bibinfo {author} {\bibfnamefont {D.}~\bibnamefont
  {Semkat}}, \bibinfo {author} {\bibfnamefont {D.}~\bibnamefont {Kremp}}, \
  and\ \bibinfo {author} {\bibfnamefont {M.}~\bibnamefont {Bonitz}},\ }\href
  {\doibase 10.1063/1.1286204} {\bibfield  {journal} {\bibinfo  {journal}
  {Journal of Mathematical Physics}\ }\textbf {\bibinfo {volume} {41}},\
  \bibinfo {pages} {7458} (\bibinfo {year} {2000})},\ \Eprint
  {http://arxiv.org/abs/https://doi.org/10.1063/1.1286204}
  {https://doi.org/10.1063/1.1286204} \BibitemShut {NoStop}%
\bibitem [{\citenamefont {Yang}\ \emph {et~al.}(2014)\citenamefont {Yang},
  \citenamefont {Lin},\ and\ \citenamefont {Zhang}}]{Zhang1}%
  \BibitemOpen
  \bibfield  {author} {\bibinfo {author} {\bibfnamefont {P.-Y.}\ \bibnamefont
  {Yang}}, \bibinfo {author} {\bibfnamefont {C.-Y.}\ \bibnamefont {Lin}}, \
  and\ \bibinfo {author} {\bibfnamefont {W.-M.}\ \bibnamefont {Zhang}},\ }\href
  {\doibase 10.1103/PhysRevB.89.115411} {\bibfield  {journal} {\bibinfo
  {journal} {Phys. Rev. B}\ }\textbf {\bibinfo {volume} {89}},\ \bibinfo
  {pages} {115411} (\bibinfo {year} {2014})}\BibitemShut {NoStop}%
\bibitem [{\citenamefont {Yang}\ \emph {et~al.}(2015)\citenamefont {Yang},
  \citenamefont {Lin},\ and\ \citenamefont {Zhang}}]{Zhang2}%
  \BibitemOpen
  \bibfield  {author} {\bibinfo {author} {\bibfnamefont {P.-Y.}\ \bibnamefont
  {Yang}}, \bibinfo {author} {\bibfnamefont {C.-Y.}\ \bibnamefont {Lin}}, \
  and\ \bibinfo {author} {\bibfnamefont {W.-M.}\ \bibnamefont {Zhang}},\ }\href
  {\doibase 10.1103/PhysRevB.92.165403} {\bibfield  {journal} {\bibinfo
  {journal} {Phys. Rev. B}\ }\textbf {\bibinfo {volume} {92}},\ \bibinfo
  {pages} {165403} (\bibinfo {year} {2015})}\BibitemShut {NoStop}%
\bibitem [{\citenamefont {van Leeuwen}\ and\ \citenamefont
  {Stefanucci}(2012)}]{Leeuwen}%
  \BibitemOpen
  \bibfield  {author} {\bibinfo {author} {\bibfnamefont {R.}~\bibnamefont {van
  Leeuwen}}\ and\ \bibinfo {author} {\bibfnamefont {G.}~\bibnamefont
  {Stefanucci}},\ }\href {\doibase 10.1103/PhysRevB.85.115119} {\bibfield
  {journal} {\bibinfo  {journal} {Phys. Rev. B}\ }\textbf {\bibinfo {volume}
  {85}},\ \bibinfo {pages} {115119} (\bibinfo {year} {2012})}\BibitemShut
  {NoStop}%
\bibitem [{\citenamefont {KONSTANTINOV}\ and\ \citenamefont
  {PEREL}(1961)}]{KONSTANTINOV}%
  \BibitemOpen
  \bibfield  {author} {\bibinfo {author} {\bibfnamefont {O.~V.}\ \bibnamefont
  {KONSTANTINOV}}\ and\ \bibinfo {author} {\bibfnamefont {V.~I.}\ \bibnamefont
  {PEREL}},\ }\href {http://www.jetp.ac.ru/cgi-bin/dn/e_012_01_0142.pdf}
  {\bibfield  {journal} {\bibinfo  {journal} {JETP}\ }\textbf {\bibinfo
  {volume} {12}},\ \bibinfo {pages} {142} (\bibinfo {year} {1961})}\BibitemShut
  {NoStop}%
\bibitem [{\citenamefont {Wagner}(1991)}]{Wagner}%
  \BibitemOpen
  \bibfield  {author} {\bibinfo {author} {\bibfnamefont {M.}~\bibnamefont
  {Wagner}},\ }\href {\doibase 10.1103/PhysRevB.44.6104} {\bibfield  {journal}
  {\bibinfo  {journal} {Phys. Rev. B}\ }\textbf {\bibinfo {volume} {44}},\
  \bibinfo {pages} {6104} (\bibinfo {year} {1991})}\BibitemShut {NoStop}%
\bibitem [{\citenamefont {Leeuwen}\ and\ \citenamefont
  {Stefanucci}(2013)}]{Stefanucci_GFF}%
  \BibitemOpen
  \bibfield  {author} {\bibinfo {author} {\bibfnamefont {R.~v.}\ \bibnamefont
  {Leeuwen}}\ and\ \bibinfo {author} {\bibfnamefont {G.}~\bibnamefont
  {Stefanucci}},\ }\href {http://stacks.iop.org/1742-6596/427/i=1/a=012001}
  {\bibfield  {journal} {\bibinfo  {journal} {Journal of Physics: Conference
  Series}\ }\textbf {\bibinfo {volume} {427}},\ \bibinfo {pages} {012001}
  (\bibinfo {year} {2013})}\BibitemShut {NoStop}%
\bibitem [{\citenamefont {Garny}\ and\ \citenamefont
  {M\"uller}(2009)}]{Markus}%
  \BibitemOpen
  \bibfield  {author} {\bibinfo {author} {\bibfnamefont {M.}~\bibnamefont
  {Garny}}\ and\ \bibinfo {author} {\bibfnamefont {M.~M.}\ \bibnamefont
  {M\"uller}},\ }\href {\doibase 10.1103/PhysRevD.80.085011} {\bibfield
  {journal} {\bibinfo  {journal} {Phys. Rev. D}\ }\textbf {\bibinfo {volume}
  {80}},\ \bibinfo {pages} {085011} (\bibinfo {year} {2009})}\BibitemShut
  {NoStop}%
\bibitem [{\citenamefont {Sieberer}\ \emph {et~al.}(2014)\citenamefont
  {Sieberer}, \citenamefont {Huber}, \citenamefont {Altman},\ and\
  \citenamefont {Diehl}}]{Diehl}%
  \BibitemOpen
  \bibfield  {author} {\bibinfo {author} {\bibfnamefont {L.~M.}\ \bibnamefont
  {Sieberer}}, \bibinfo {author} {\bibfnamefont {S.~D.}\ \bibnamefont {Huber}},
  \bibinfo {author} {\bibfnamefont {E.}~\bibnamefont {Altman}}, \ and\ \bibinfo
  {author} {\bibfnamefont {S.}~\bibnamefont {Diehl}},\ }\href {\doibase
  10.1103/PhysRevB.89.134310} {\bibfield  {journal} {\bibinfo  {journal} {Phys.
  Rev. B}\ }\textbf {\bibinfo {volume} {89}},\ \bibinfo {pages} {134310}
  (\bibinfo {year} {2014})}\BibitemShut {NoStop}%
\bibitem [{\citenamefont {Sarkar}\ \emph {et~al.}(2014)\citenamefont {Sarkar},
  \citenamefont {Sensarma},\ and\ \citenamefont {Sengupta}}]{Sangita}%
  \BibitemOpen
  \bibfield  {author} {\bibinfo {author} {\bibfnamefont {S.~D.}\ \bibnamefont
  {Sarkar}}, \bibinfo {author} {\bibfnamefont {R.}~\bibnamefont {Sensarma}}, \
  and\ \bibinfo {author} {\bibfnamefont {K.}~\bibnamefont {Sengupta}},\ }\href
  {http://stacks.iop.org/0953-8984/26/i=32/a=325602} {\bibfield  {journal}
  {\bibinfo  {journal} {Journal of Physics: Condensed Matter}\ }\textbf
  {\bibinfo {volume} {26}},\ \bibinfo {pages} {325602} (\bibinfo {year}
  {2014})}\BibitemShut {NoStop}%
\bibitem [{\citenamefont {Negele}\ and\ \citenamefont
  {Orland}(2002)}]{Negele_Orland}%
  \BibitemOpen
  \bibfield  {author} {\bibinfo {author} {\bibfnamefont {J.~W.}\ \bibnamefont
  {Negele}}\ and\ \bibinfo {author} {\bibfnamefont {H.}~\bibnamefont
  {Orland}},\ }\href@noop {} {\emph {\bibinfo {title} {Quantum Many Body
  Systems}}}\ (\bibinfo  {publisher} {Oxford University Press on Demand},\
  \bibinfo {year} {2002})\BibitemShut {NoStop}%
\bibitem [{\citenamefont {Preiss}\ \emph {et~al.}(2015)\citenamefont {Preiss},
  \citenamefont {Ma}, \citenamefont {Tai}, \citenamefont {Lukin}, \citenamefont
  {Rispoli}, \citenamefont {Zupancic}, \citenamefont {Lahini}, \citenamefont
  {Islam},\ and\ \citenamefont {Greiner}}]{Preiss1229}%
  \BibitemOpen
  \bibfield  {author} {\bibinfo {author} {\bibfnamefont {P.~M.}\ \bibnamefont
  {Preiss}}, \bibinfo {author} {\bibfnamefont {R.}~\bibnamefont {Ma}}, \bibinfo
  {author} {\bibfnamefont {M.~E.}\ \bibnamefont {Tai}}, \bibinfo {author}
  {\bibfnamefont {A.}~\bibnamefont {Lukin}}, \bibinfo {author} {\bibfnamefont
  {M.}~\bibnamefont {Rispoli}}, \bibinfo {author} {\bibfnamefont
  {P.}~\bibnamefont {Zupancic}}, \bibinfo {author} {\bibfnamefont
  {Y.}~\bibnamefont {Lahini}}, \bibinfo {author} {\bibfnamefont
  {R.}~\bibnamefont {Islam}}, \ and\ \bibinfo {author} {\bibfnamefont
  {M.}~\bibnamefont {Greiner}},\ }\href {\doibase 10.1126/science.1260364}
  {\bibfield  {journal} {\bibinfo  {journal} {Science}\ }\textbf {\bibinfo
  {volume} {347}},\ \bibinfo {pages} {1229} (\bibinfo {year} {2015})},\ \Eprint
  {http://arxiv.org/abs/http://science.sciencemag.org/content/347/6227/1229.full.pdf}
  {http://science.sciencemag.org/content/347/6227/1229.full.pdf} \BibitemShut
  {NoStop}%
\end{thebibliography}%
\begin{widetext}
\appendix
\newpage
\section{Calculation of $\mathcal{N}(u)$ and $\hat{G}(u)$ for the diagonal initial density matrix} \label{sec:AP:diagonal}

An important step in the formalism we have developed for a quantum many body system starting from $\hat{\rho}_0=\sum_{\{n\}}c_{\{n\}}| \{n\}\rangle\langle \{n\}|$ is to invert the kernel $\hat{G}^{-1}(\alpha,t,\beta,t',\vec{u})$ in the inverse Green's function analytically and obtain the closed form expression for the $\vec{u}$ dependent normalization, $\mathcal{N}(u)= Det[-\mathbf{i}\hat{G}^{-1}(u)]^{-\zeta}$ and also the Green's function, $\hat{G}(u)$ with the initial source $\vec{u}$, as they serve as the building blocks for the further steps of the many body formalism. In this appendix, we will work out the structure of calculating $\mathcal{N}(u)$ and $\hat{G}(u)$ from $\hat{G}^{-1}(\alpha,t,\beta,t',\vec{u})$ for a many body Bosonic system. 

To construct these
objects, it is useful to isolate the $\vec{u}$ dependent part in the action from the from part independent of the initial condition to write,
\beq
\hat{G}^{-1}(u_\alpha)=\hat{G}^{-1}(0)-\hat{\Delta}(u_\alpha)
\eeq
where $\hat{G}^{-1}(0)=\hat{G}^{-1}(\alpha,t;\beta,t')|_{\vec{u}=0}$ is the two component inverse Green's function when the system starts in
the vacuum state, and is obtained by setting
$u_\alpha=0$. It is evident from the text below equation \ref{identity_u:mm} of the main text, that the $\vec{u}$ dependent part $\Delta$ is finite only for the
$+-$ component, i.e. 
\beq
\Delta_{++}=\Delta_{--}=\Delta_{-+}=0~,~\Delta_{+-}(\alpha,t;\beta,t',\vec{u})=\mathbf{i}\delta_{\alpha\beta}\delta_t\delta_{t'}u_\alpha.
\eeq
Now, we will write,
\beq
Det[-\mathbf{i}\hat{G}^{-1}(\vec{u})] = e^{Tr[\log\{ -\mathbf{i} \hat{G}^{-1}(\vec{u}) \}]} \nonumber 
\eeq
 which leads to,
\bqa
\label{AP:trLog}
Tr[\log\{ -\mathbf{i} \hat{G}^{-1}(\vec{u}) \}] &=& Tr[\log\{ -\mathbf{i} \hat{G}^{-1}(0) \}] + Tr[\log\{1 - \hat{G}(0) \hat{\Delta}(\vec{u})\}] \nonumber \\
&=& Tr[\log\{ -\mathbf{i} \hat{G}^{-1}(0) \}]-Tr\left[\hat{G}(0) \hat{\Delta}(\vec{u})+ \frac{1}{2} \hat{G}(0) \hat{\Delta}(\vec{u})\hat{G}(0) \hat{\Delta}(\vec{u})+..\right]
\eqa

where,
\bqa
Tr\left[\hat{G}(0) \hat{\Delta}(\vec{u})\right] &=&Tr\left[\hat{G}_{-+}(0) \hat{\Delta}_{+-}(\vec{u})\right]  
= \mathbf{i}\sum \limits_{\alpha} G_{-+}(\alpha,0;\beta,0;0) u_\alpha 
= \sum \limits_{\alpha} u_\alpha.
\eqa
Here the vacuum Green's functions $\hat{G}(0)=\hat{G}^v$are given by
 \bqa
 G_{-+}^v(\alpha,t;\beta,t')&=&-\mathbf{i}\sum_{a}
\psi^\ast_a(\beta)\psi_a(\alpha)e^{-iE_a (t-t')}~,~
G_{+-}^v(\alpha,t;\beta,t')=0, \nonumber\\
G_{++}^v(\alpha,t;\beta,t')&=&\Theta(t-t')
G_{-+}^v(\alpha,t;\beta,t')~,~
and~ G_{--}^v(\alpha,t;\beta,t')=\Theta(t'-t)
G_{-+}^v(\alpha,t;\beta,t'),
\eqa
where
$E_a$ are the eigenvalues and $\psi_a(\alpha)$ are the corresponding
eigenvectors of the Hamiltonian of the multimode system. Using the orthogonality property of eigenmodes we get, $\mathbf{i}G_{-+}(\alpha,0;\beta,0;0)=\delta_{\alpha,\beta}$ at the initial time $t=t'=0$, Similarly,
\bqa
Tr\left[\hat{G}(0) \hat{\Delta}(\vec{u})\hat{G}(0) \hat{\Delta}(\vec{u})\right] &=&\frac{1}{2}Tr\left[\hat{G}_{-+}(\alpha,0;\beta,0;0) \mathbf{i} u_{\beta}\hat{G}_{-+}(\beta,0;\gamma,0;0) \mathbf{i} u_{\gamma}\right] 
=\frac{1}{2} \sum \limits_{\alpha} u^2_\alpha.
\eqa
Using similar argument for all terms in the expansion (equation \ref{AP:trLog}) and adding them up, we obtain,
\bqa
Tr[\log\{ -\mathbf{i} \hat{G}^{-1}(\vec{u}) \}] &=& Tr[\log\{ -\mathbf{i} \hat{G}^{-1}(0) \}]  + \sum \limits_\alpha\log \left(1-u_\alpha \right) \nonumber \\
Det[ -\mathbf{i} G^{-1}] &=&Det[ -\mathbf{i} G^{-1}(0)]\prod_\alpha 1-u_\alpha
\eqa 
which is quoted in equation \ref{Determinant} in the main text.

Now, we will show how to invert the kernel $\hat{G}^{-1}(u)$ to obtain closed form answer for $\hat{G}(u)$. We have,
\bqa
&&\hat{G}(\vec{u})
 =\left [\hat{G}^{-1}(0)-\hat{\Delta}(u_\alpha) \right] ^{-1} 
= \hat{G}(0)\left [1-\hat{G}(0)\hat{\Delta}(u_\alpha) \right] ^{-1} \nonumber \\
&&=\hat{G}(0)+\hat{G}(0)\hat{\Delta}(u_\alpha) \hat{G}(0) +\hat{G}(0)\hat{\Delta}(u_\alpha) \hat{G}(0) \hat{\Delta}(u_\alpha) \hat{G}(0)+ ... \nonumber 
\eqa
We will show here the structure of the above sum for one of the components, say $G_{++}(\alpha,t;\beta,t';\vec{u})$. The expansion of $\hat{G}_{++}(\vec{u})$ can be written as,
\bqa
G_{++}(\alpha,t;\beta,t';\vec{u}) &=& G_{++}^v(\alpha,t;\beta,t') + \mathbf{i}\sum \limits_{\gamma} G_{++}^v(\alpha,t;\gamma,0)u_\gamma  G_{-+}^v(\gamma,0;\beta,t') \nonumber \\
&+& \mathbf{i}^2 \sum \limits_{\gamma,\kappa} G_{++}^v(\alpha,t;\gamma,0)u_\gamma  G_{-+}^v(\gamma,0;\kappa,0)u_\kappa  G_{-+}^v(\kappa,0;\beta,t')+.. \nonumber \\
&=&G_{++}^v(\alpha,t;\beta,t') + \mathbf{i}\sum \limits_{\gamma} G_{++}^v(\alpha,t;\gamma,0)u_\gamma  G_{-+}^v(\gamma,0;\beta,t') + \mathbf{i}\sum \limits_{\gamma} G_{++}^v(\alpha,t;\gamma,0)u_\gamma^2  G_{-+}^v(\gamma,0;\beta,t') \nonumber  \\
&=& G_{++}^v(\alpha,t;\beta,t') + \mathbf{i} \sum \limits_{\gamma} \frac{u_\gamma}{1-u_\gamma} G_{++}^v(\alpha,t;\gamma,0)  G_{-+}^v(\gamma,0;\beta,t')
\eqa
Similar arguments will apply to the other components as well which will lead to equation for $G_{\mu\nu}(\alpha,t;\beta,t';\vec{u})$ in the main text.
%

\section{Generic density matrix}\label{ap:gen_formalism}

\subsection*{Green's functions\label{sec:ap-Green fn}}

The physical Green's function is given by 
\begin{equation}
{\cal G}^\rho=\sum_{nm}  \frac{c_{nm} }{\prod_\alpha\sqrt{n_\alpha ! m_\alpha !}}\left.\prod_{i}[\partial_{\alpha_i \beta_i}] Z[0,\hat{u}] G(\hat{u}) \right\vert_{\hat{u}=0},
\label{g_fn:mat}
\end{equation}
where $G(\hat{u})$ is given by eqn.~\ref{gfn_u:umat}. The first term in $G(\hat{u})$, the vacuum Green's function part, is independent of $\hat{u}$ and the just goes out of the above derivatives. For the second term, we need to evaluate  
\beq
\sum_{nm}  \frac{c_{nm} }{\prod_\alpha\sqrt{n_\alpha ! m_\alpha !}}\left.\prod_{i}[\partial_{\alpha_i \beta_i}] \frac{[(1-\hat{u})^{-1}-1]_{\gamma \delta}}{\det(1-\hat{u})^{-1}}\right\vert_{\hat{u}=0}.
\label{G_fn:umat:secondterm}
\eeq
We shall calculate the derivatives below. Again defining $A=1-\hat{u}$, we have 
\begin{alignat}{1}
d_{N}\cdots d_{1}\left(A^{-1}\det A^{-1}\right) & =A^{-1}d_{N}d_{N-1}\cdots d_{2}d_{1}\det A^{-1}\nonumber \\
 & \ \ \ +d_{1}A^{-1}d_{N}d_{N-1}\cdots d_{2}\det A^{-1}+d_{2}A^{-1}d_{N}d_{N-1}\cdots d_{3}d_{1}\det A^{-1}+\cdots\nonumber \\
 & \ \ \ +d_{2}d_{1}A^{-1}d_{N}d_{N-1}\cdots d_{3}\det A^{-1}+d_{3}d_{1}A^{-1}d_{N}d_{N-1}\cdots d_{4}d_{2}\det A^{-1}+\cdots\nonumber \\
 & \ \ \ +d_{3}d_{2}d_{1}A^{-1}d_{N}d_{N-1}\cdots d_{4}\det A^{-1}+\cdots\nonumber \\
 & \ \ \ +\cdots+d_{N}d_{N-1}\cdots d_{2}d_{1}A^{-1}\det A^{-1}.
\label{eq:ap-d^N (A^-1 det A^-1)}
\end{alignat}
After substituting $\hat{u}=0$, the first line of eqn.~\ref{eq:ap-d^N (A^-1 det A^-1)}
gives 
\begin{equation}
\left.[ A^{-1}]_{\gamma \delta}\prod_{i}[\partial_{\alpha_i \beta_i}]\det A^{-1}\right\vert_{\hat{u}=0}=\delta_{\gamma \delta} \langle \{m\} | \{n\} \rangle \prod_\alpha n_\alpha !\ .
\end{equation}
which cancels the contribution from $[-1]_{\gamma \delta}$ part in eqn.~\ref{gfn_u:umat} 
exactly. To get an intuition for how to deal with the rest of the terms, let us focus on the first term in the second line of eqn.~\ref{eq:ap-d^N (A^-1 det A^-1)},  
\begin{equation}
\left. [\partial_{\alpha_{1}\beta_{1}}A^{-1}] _{\gamma \delta}\prod_{i\ne1} [\partial_{\alpha_i \beta_i}] \det A^{-1}\right\vert_{\hat{u}=0}
=\delta_{\gamma \alpha_{1}}\delta_{\beta_{1} \delta}\sum_P \langle P(\vec\beta) | \vec\alpha \rangle.
\end{equation}
where $P$ is now understood to be a permutation on $N$ labels that fixes $\beta_1$, that is, $P(\beta_1)=\beta_1$, and the matrix element has only $\alpha_i$'s and $\beta_i$'s for $i=2,\ldots,N$. The other terms in the sum can now be determined using the symmetry arguments using before in calculating the partition function. We could start our analysis with the states $|\alpha'_1 \cdots \alpha'_N\rangle =|Q(\alpha_1) \cdots Q(\alpha_N)\rangle $ and $|\beta'_1 \cdots \beta'_N\rangle =|R(\beta_1) \cdots R(\beta_N)\rangle $, where $Q$ and $R$ any permutations, because nothing physical depends on this choice. Using these new labels, the above equation would give us 
\beq
\left. [\partial_{\alpha'_{1}\beta'_{1}}A^{-1}] _{\gamma \delta}\prod_{i\ne1} [\partial_{\alpha'_i \beta'_i}] \det A^{-1}\right\vert_{\hat{u}=0}
= \delta_{\gamma \alpha'_{1}}\delta_{\beta'_{1} \delta}\sum_{P}\langle P(\vec\beta') | \vec\alpha' \rangle 
= \delta_{\gamma \alpha_{k}}\delta_{\beta_{l} \delta}\sum_{P}\langle P(Q(\vec\beta))  | R(\vec\alpha) \rangle
= \delta_{\gamma \alpha_{k}}\delta_{\beta_{l} \delta}\sum_{P}\langle P(\vec\beta)  | \vec\alpha \rangle,
\eeq
where, there exist $k$ and $l$ such that $Q(\alpha_1)=\alpha_k$, $R(\beta_1)=\beta_l$, and the vectors $\vec\alpha$ and $\vec\beta$ do not contain $\alpha_k$ and $\beta_l$ respectively. Also, $P$ runs over all permutations that fix $\beta_l$.

It is not hard to convince oneself that all the terms in eqn.~\ref{eq:ap-d^N (A^-1 det A^-1)} are of the above form for different choices of $Q$ and $R$. For example, the second term in the second line, $d_{2}A^{-1}d_{N}d_{N-1}\cdots d_{3}d_{1}\det A^{-1}$, corresponds to $k=l=2$, whereas the sum 
\bqa
&=& d_2 d_1 A^{-1} d_N \cdots d_3 \det A^{-1}+ d_2 d_3 d_1 A^{-1} d_N \cdots d_4 \det A^{-1} + d_2 d_4 d_1 A^{-1} d_N \cdots d_5 d_3 \det A^{-1} + \cdots  \\
&& +\ d_2 d_3 d_4 d_1 A^{-1} d_N \cdots d_5 \det A^{-1} + d_2 d_3 d_5 d_1 A^{-1} d_N \cdots d_6 d_4 \det A^{-1} + d_2 d_4 d_5 d_1 A^{-1} d_N \cdots d_6 d_3 \det A^{-1} + \cdots,\nonumber
\eqa
[that is, the sum of all terms with $d_1$ and $d_2$ being the first and last derivatives to act on $A^{-1}$] corresponds to $k=2$ and $l=1$.
Hence, the LHS of eqn.~\ref{eq:ap-d^N (A^-1 det A^-1)} reduces to the following expression in occupation number basis,  
\begin{alignat}{1}
\prod_{i} [\partial_{\alpha_i \beta_i}] \frac{[ A^{-1}-1]_{\gamma \delta}}{\det A}  & =\sum_{kl}\delta_{\gamma \alpha_{k}}\delta_{\beta_{l} \delta}\sum_{P}\langle P(\vec\beta)  | \vec\alpha \rangle
=\sum_{P}\langle P(\vec\beta) | \hat a ^\dagger _\delta \hat a _\gamma | \vec\alpha \rangle = \langle \{m\} | \hat a ^\dagger _\delta \hat a _\gamma | \{ n \} \rangle \prod_\gamma \sqrt{n_\gamma ! m_\gamma !}\ .
\end{alignat}
The expression in eqn.~\ref{G_fn:umat:secondterm} can then be written as an expectation value in the initial density matrix, 
\beq
\sum_{nm}  \frac{c_{nm} }{\prod_\alpha\sqrt{n_\alpha ! m_\alpha !}}\left.\prod_{i}[\partial_{\alpha_i \beta_i}] \frac{[(1-\hat{u})^{-1}-1]_{\gamma \delta}}{\det(1-\hat{u})^{-1}}\right\vert_{\hat{u}=0}
=\sum_{nm}  c_{nm} \langle \{m\} | \hat a ^\dagger _\delta \hat a _\gamma | \{ n \} \rangle =\tr (\hat\rho_0 \hat a ^\dagger _\delta \hat a _\gamma) \equiv \langle \hat a ^\dagger _\delta \hat a _\gamma \rangle_0.
\eeq
Substituting this result into eqn.~\ref{g_fn:mat} gives the physical Green's functions in the main text,
\bqa
{\cal G}_{K\rho_0}(\alpha,t;\beta,t')= -\mathbf{i} \sum_{\gamma \delta}
  G^v_{R}(\alpha,t;\gamma,0) [2\langle\hat{a}_{\delta}^{\dagger}\hat{a}_{\gamma}\rangle_{0}+\delta_{\gamma \delta}] G^v_{A}(\delta,0;\beta,t').
\eqa
\end{widetext}


\end{document}